\documentclass[%
superscriptaddress,
preprint,
 amsmath,amssymb,
 aps,
pra,
]{revtex4-2}

\usepackage{graphicx}
\usepackage{dcolumn}
\usepackage{bm}
\usepackage{braket}
\usepackage{comment} 
\usepackage{hyperref}
\usepackage{siunitx}

\usepackage{xcolor}
\newcommand{\red}[1]{\textcolor{black}{#1}}

\usepackage{nicematrix}
\NiceMatrixOptions{cell-space-limits = 1pt}

\usepackage{mathtools}

\usepackage{amsthm}

\newtheorem*{theorem*}{Theorem}

\theoremstyle{definition}

\theoremstyle{remark}

\newtheorem*{example*}{Example}
\newtheorem*{algorithm*}{Algorithms}

\usepackage{etoolbox}

\makeatletter
\patchcmd{\frontmatter@abstract@produce}
  {\vskip200\p@\@plus1fil
   \penalty-200\relax
   \vskip-200\p@\@plus-1fil}
  {}
  {}
  {}
\makeatother

\begin{document}

\title{Unitary control of partially coherent waves. II. \\ Transmission or reflection}

\author{Cheng Guo}
\email{guocheng@stanford.edu}
\affiliation{Ginzton Laboratory and Department of Electrical Engineering, Stanford University, Stanford, California 94305, USA}

\author{Shanhui Fan}
\email{shanhui@stanford.edu}
\affiliation{Ginzton Laboratory and Department of Electrical Engineering, Stanford University, Stanford, California 94305, USA}%

\date{\today}

\begin{abstract}
Coherent control of wave transmission and reflection is crucial for applications in communication, imaging, and sensing. However, many practical scenarios involve partially coherent waves rather than fully coherent ones. We present a systematic theory for the unitary control of partially coherent wave transmission and reflection. For a linear time-invariant system with an incident partially coherent wave, we derive analytical expressions for the range of attainable total transmittance and reflectance under arbitrary unitary transformations. We also introduce an explicit algorithm to construct a unitary control scheme that achieves any desired transmission or reflection within the attainable range. As applications of our theory, we establish conditions for four novel phenomena: partially coherent perfect transmission, partially coherent perfect reflection, partially coherent zero transmission, and partially coherent zero reflection. We also prove a theorem that relates the degree of coherence of the incident field, quantified by the majorization order, to the resulting transmission and reflection intervals. Furthermore, we demonstrate that reciprocity (or energy conservation) imposes direct symmetry constraints on bilateral transmission (or transmission and reflection) of partially coherent waves under unitary control. Our results provide fundamental insights and practical guidelines for using unitary control to manipulate the transmission and reflection of partially coherent waves. This theory applies to various wave systems, including electromagnetic and acoustic waves.
\end{abstract}
\maketitle


\section{introduction}\label{sec:introduction}

Transmission and reflection are fundamental wave phenomena~\cite{planck1991,chen2005,zhang2007,howell2016,fan2017,cuevas2018b,li2021e}. Controlling these phenomena is crucial for various applications, including communication~\cite{miller2013c,miller2019,seyedinnavadeh2023}, imaging~\cite{sebbah2001a,kittel2005,ntziachristos2010a,cizmar2012,kang2015,guo2018,guo2018a,yoon2020b,wang2020p,long2021,bertolotti2022,wang2022,long2022}, and sensing~\cite{aulbach2011,sarma2015,mounaix2016,jeong2018a,muraviev2018,tan2020,liu2020s}. Approaches to controlling wave transmission and reflection can be classified into two categories: structural design and wave manipulation. In the structural design approach, desired transmission and reflection behaviors are achieved by directly designing the transmitting or reflecting media. For instance, recent advances in nanophotonics have enabled the creation of novel photonic structures with transmission and reflection properties that differ significantly from traditional media~\cite{greffet2002a, guo2012b, dezoysa2012, yu2013, pelton2015, thompson2018, baranov2019a, guo2021a,guo2023c}. Photonic structures can be designed with negative permittivity and permeability to achieve perfect lensing~\cite{veselago1968,pendry2000,smith2000,shelby2001,fan2005,valentine2008} or with wavevector-dependent transmission and reflection to perform novel tasks such as analog optical computing~\cite{guo2018,guo2018a,wang2020p,zhu2021,long2021,wang2022}, compressing free space~\cite{guo2020a,reshef2021,long2022,long2023}, and generating light bullets~\cite{guo2021c}.

In the wave manipulation approach, desired transmission and reflection behaviors are achieved by manipulating the external waves interacting with the media~\cite{popoff2014,liew2016,mounaix2016}. A significant advancement in the wave manipulation approach has been the development of wavefront shaping techniques, particularly using spatial light modulators (SLMs)~\cite{vellekoop2007,yu2017e}. SLMs can modulate the phase of reflected light, transforming a coherent input field into a tailored wavefront. This results in the desired transmission or reflection patterns of waves interacting with a complex medium. This technique, known as coherent control~\cite{popoff2014,liew2016,mounaix2016}, has greatly enhanced our ability to manipulate wave transmission and reflection, achieving novel phenomena such as reflectionless scattering modes~\cite{sweeney2020a,stone2021,horodynski2022}.

\red{Initial work on coherent control via SLMs focused on manipulating a single coherent incident wave. Recently, motivated by various applications, this approach has been extended to simultaneously control multiple coherent incident waves~\cite{berdague1982,luo2014,su2021}.} The feasibility of multimode control is now emerging with programmable unitary photonic devices such as Mach-Zehnder interferometer meshes~\cite{reck1994,miller2013c,miller2013a,miller2013b,carolan2015,miller2015,clements2016,ribeiro2016,wilkes2016,annoni2017,miller2017d,perez2017,harris2018,pai2019} and multiplane light conversion systems~\cite{morizur2010,labroille2014,tanomura2022,kupianskyi2023,taguchi2023,zhang2023b}. These devices can perform arbitrary unitary transformations and hold significant potential for applications in quantum computing~\cite{carolan2015,carolan2020,elshaari2020,wang2020aw,chi2022,madsen2022,pelucchi2022}, machine learning~\cite{shen2017,prabhu2020,zhang2021f,ashtiani2022,bandyopadhyay2022,ohno2022,chen2023,pai2023}, and optical communications~\cite{clements2016,annoni2017,burgwal2017,melati2017,choutagunta2020,buddhiraju2021}. By converting between different sets of orthogonal incident modes, these devices can achieve advanced multimode control of wave behaviors. This type of control is termed \emph{unitary control}~\cite{guo2023b}, as it is mathematically described by a unitary transformation of the input wave space. It can have broad applications in scenarios where the transmitting and reflecting media cannot be altered~\cite{vellekoop2008,popoff2010,aulbach2011,kim2012a,shi2012,yu2013a,gerardin2014,pena2014a,popoff2014,davy2015,shi2015a,bender2020c}.

The concept of unitary control has been explored for coherent waves to manipulate multimode absorption~\cite{guo2023b} and transmission~\cite{guo2024}. However, many practical applications, such as microscopy and astronomy, involve the transmission and reflection of partially coherent waves~\cite{mandel1995,goodman2000}, since many wave sources are inherently partially coherent. To develop a theory of unitary control for manipulating the transmission or reflection of partially coherent waves, one needs to consider the interplay between the properties of the structure and the coherence properties of the incident waves. Such a theory represents a significant advancement beyond the theory of unitary control for coherent waves.

In this paper, we develop a systematic theory for the unitary control of the transmission or reflection of partially coherent waves. Our theory addresses two fundamental questions: (i) Given an object and an incident partially coherent wave, what is the range of all attainable total transmittance and reflectance under unitary control? (ii) How can we achieve a given total transmittance or reflectance via unitary control? The first question addresses the capabilities and limitations of unitary control over transmission and reflection, while the second focuses on implementation. We provide comprehensive answers to both questions. As applications of our theory, we establish the conditions for four new phenomena: partially coherent perfect transmission, partially coherent perfect reflection, partially coherent zero transmission, and partially coherent zero reflection. We also examine how the degree of coherence, measured by the majorization order, affects the attainable transmission and reflection, and prove that majorized coherence implies nested transmission and reflection intervals. Furthermore, we investigate the symmetry constraints on the unitary control of bilateral transmission and reflection for partially coherent waves. We show that reciprocity enforces direct constraints on transmission, while energy conservation enforces direct constraints on both transmission and reflection.

This paper is the second in a series on the unitary control of partially coherent waves. In the first paper~\cite{guo2024a}, we investigated the unitary control of the absorption of partially coherent waves. In this work, we further extend the unitary control method to manipulate the transmission and reflection of partially coherent waves. We have adopted the same mathematical notations (see Ref.~\cite{guo2024a} Sec.~II) and similar proof techniques throughout this series of papers. Throughout this paper, we will refer to Ref.~\cite{guo2024a} as Paper~1.

The rest of this paper is organized as follows. In Sec.~\ref{sec:theory}, we develop a general theory of unitary control over partially coherent wave transmission and reflection. In Sec.~\ref{sec:applications}, we discuss the physical applications of our theory. We conclude in Sec.~\ref{sec:conclusion}.

\section{Theory}\label{sec:theory}

\subsection{Partially coherent waves}\label{subsec:partially_coherent_waves}

Let $H$ be an $n$-dimensional Hilbert space of waves. A partially coherent wave is represented by a density matrix~\cite{landau1981,oneill2003,wolf2003,delimabernardo2017,zhang2019m,korotkova2022} $\rho$, also known as a coherency matrix~\cite{wolf1985,goodman2000,yamazoe2012,okoro2017} in optics. $\rho$ is positive semidefinite. The trace of $\rho$ corresponds to the total power, which we assume to be normalized:
\begin{equation}\label{eq:rho_normalization}
\operatorname{tr} \rho = 1.
\end{equation}
The coherence properties are encoded in the eigenvalues of $\rho$, known as the coherence spectrum:
\begin{equation}
\bm{\lambda}^{\downarrow}(\rho) =  (\lambda^{\downarrow}_{1}(\rho),\dots,\lambda^{\downarrow}_{n}(\rho)).   
\end{equation}

\subsection{Partially coherent wave transmission or reflection}

\begin{figure}[hbtp]
    \centering
    \includegraphics[width=0.5\textwidth]{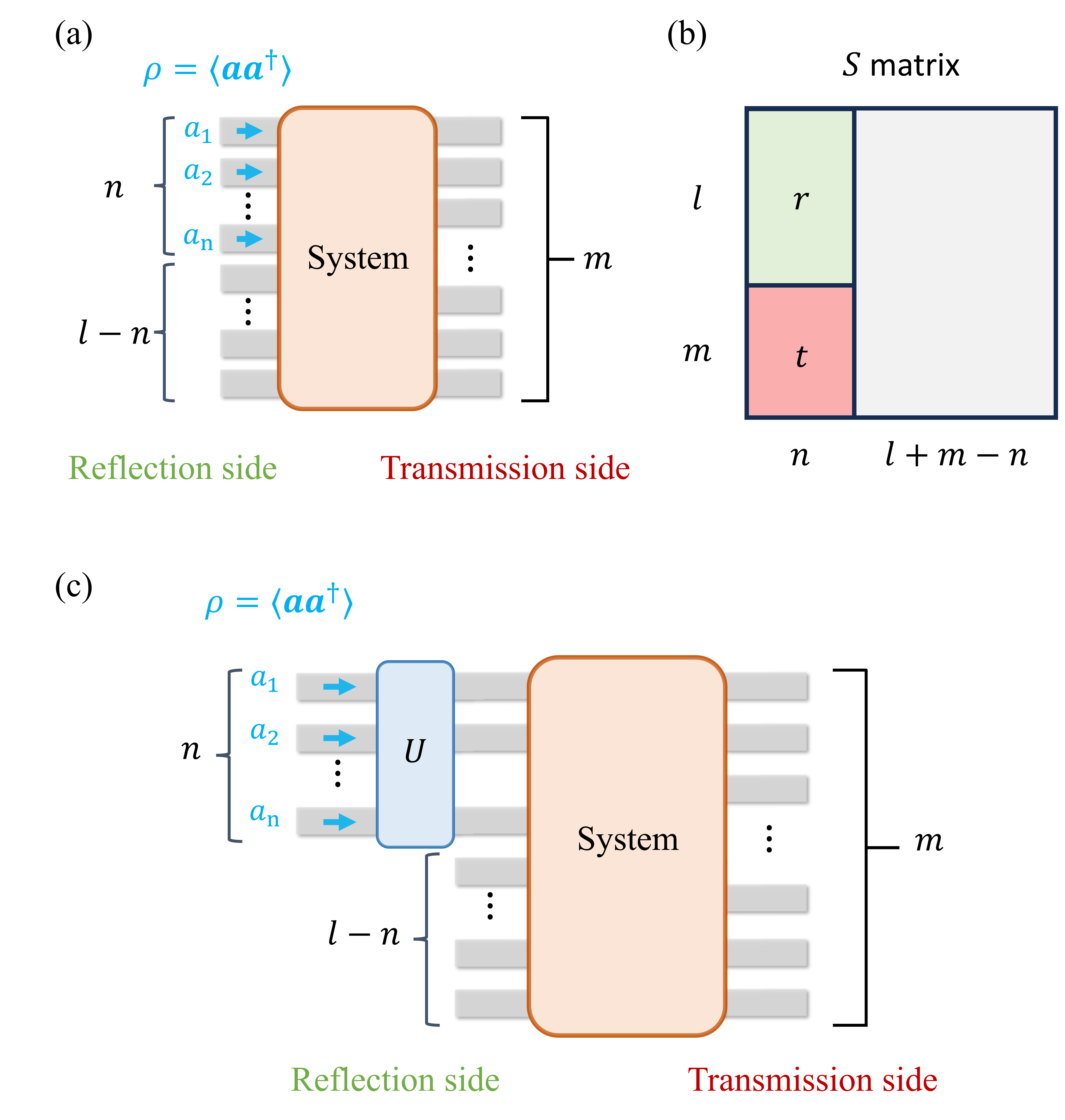}
    \caption{(a) Schematic of a linear time-invariant system with $(l+m)$ ports, where $l$ ports are on the left (reflection) side and $m$ ports are on the right (transmission) side. A partially coherent wave characterized by a density matrix $\rho$ is input into the $n$ input ports on the left side, resulting in transmitted and reflected waves with unnormalized density matrices $\Gamma_t = t \rho t^\dagger$ and $\Gamma_r = r \rho r^\dagger$, respectively. Here, $t$ and $r$ are the transmission and reflection matrices. The total transmittance $T$ or reflectance $R$ is defined as the ratio of the transmitted or reflected power to the input power. (b) Schematic of the $S$-matrix, where $r$ and $t$ are block submatrices of the entire $S$-matrix. (c) Schematic of unitary control of partially coherent wave transmission or reflection. A unitary converter $U$ is applied to the input wave before it interacts with the system, allowing for the manipulation of the total transmittance $T$ and the total reflectance $R$.}
    \label{fig:geometry}
\end{figure}

We study the transmission and reflection of a partially coherent wave in a linear time-invariant system with $(l+m)$ ports, with $l$ ports on the left side and $m$ ports on the right side, as shown in Fig.~\ref{fig:geometry}a. A partially coherent wave characterized by a density matrix $\rho$ is injected into $n \leq l$ input ports on the left side. \red{The cases where $n < l$ represent scenarios in which incident waves are restricted to an accessible $n$-dimensional subspace of the full $l$-dimensional space of waves on the left side.} The output consists of the transmitted wave on the right side and the reflected wave on the left side, characterized by the unnormalized density matrices:
\begin{equation}\label{eq:Gamma}
\Gamma_{t} = t \rho t^{\dagger}, \qquad \Gamma_{r} = r \rho r^{\dagger}.
\end{equation}
Here, $t$ is the $n \times l$ transmission matrix, and $r$ is the $n \times m$ reflection matrix; both are block submatrices of the entire $(l+m) \times (l+m)$ $S$-matrix (Fig.~\ref{fig:geometry}b). The traces of $\Gamma_{t}$ and $\Gamma_{r}$ correspond to the total transmitted and reflected power, respectively. The total transmittance $T$ and reflectance $R$ are defined as the ratios of the transmitted and reflected power to the input power, respectively. Using Eq.~(\ref{eq:rho_normalization}), we obtain:
\begin{align}
T &\coloneqq \operatorname{tr} \Gamma_{t} /\operatorname{tr} \rho = \operatorname{tr} \Gamma_{t} = \operatorname{tr} (\rho t^{\dagger}t),\\
R &\coloneqq \operatorname{tr} \Gamma_{r} /\operatorname{tr} \rho = \operatorname{tr} \Gamma_{r} = \operatorname{tr} (\rho r^{\dagger}r).
\end{align}
Here, $t^\dagger t$ and $r^\dagger r$ are both $n \times n$ positive semidefinite matrices, known as the transmittance matrix and the reflectance matrix, respectively. Their eigenvalues, $\bm{\lambda}(t^\dagger t)$ and $\bm{\lambda}(r^\dagger r)$, are known as the transmission eigenvalues and reflection eigenvalues~\cite{rotter2017,yamilov2016}, respectively.

\subsection{Unitary control of partially coherent wave transmission or reflection}

We briefly review the concept of unitary control. Unitary control involves transforming input waves through a unitary converter, such as spatial light modulators~\cite{vellekoop2007,popoff2014,yu2017e}, Mach-Zehnder interferometers~\cite{reck1994,miller2013c,miller2013a,miller2013b,carolan2015,miller2015,clements2016,ribeiro2016,wilkes2016,annoni2017,miller2017d,perez2017,harris2018,pai2019}, and multiplane light conversion systems~\cite{morizur2010,labroille2014,tanomura2022,kupianskyi2023,taguchi2023,zhang2023b}. Under unitary control, the input wave undergoes modification through unitary similarity~\cite{horn2012}:
\begin{equation}
\rho \to \rho[U] = U \rho U^\dagger.
\end{equation}

As illustrated in Fig.~\ref{fig:geometry}c, we apply unitary control to the input wave within the $n$ input ports. With unitary control, both transmittance and reflectance explicitly depend on $U$:
\begin{align}
\label{eq:T_U}
T \to T [U] &= \operatorname{tr} (U \rho U^\dagger t^\dagger t), \\
R \to R [U] &= \operatorname{tr} (U \rho U^\dagger r^\dagger r). \label{eq:R_U}
\end{align}

\subsection{Major questions}

We pose two fundamental questions: Given a partially coherent incident wave and a system as shown in Fig.~\ref{fig:geometry}c, under unitary control, (1) What total transmittance or reflectance is attainable? (2) How can we achieve a given total transmittance or reflectance?

We now reformulate these key questions mathematically. For Question 1: Given $\rho$ and $t$, what is the set
\begin{equation}
 \{T\} \equiv \Set{T[U] | U\in U(n)}? \label{eq:Question1-1} 
\end{equation}
Or, given $\rho$ and $r$, what is the set
\begin{equation}
\{R\} \equiv \Set{R[U] | U\in U(n)} ? \label{eq:Question1-2} 
\end{equation}
(If $\rho$ needs to be specified, we denote $T[U]$, $R[U]$, $\{T\}$, and $\{R\}$ as $T[U|\rho]$, $R[U|\rho]$, $\{T|\rho\}$, and $\{R|\rho\}$, respectively.)

For Question 2: Given $\rho$, $t$, and $T_0 \in \{T\}$, find a $U_1 \in U(n)$ such that
\begin{equation}
    T[U_1] = T_0.    \label{eq:set_U_T0}
\end{equation}
Or, given $\rho$, $r$, and $R_0 \in \{R\}$, find a $U_2 \in U(n)$ such that
\begin{equation}
    R[U_2] = R_0.    \label{eq:set_U_R0}
\end{equation}

\subsection{Main results}

In this subsection, we provide comprehensive answers to Questions 1 and 2.

\subsubsection{Answer to Question 1}

Let's start with Question 1. The answer is:
\begin{align}
\label{eq:main_result_set_T}
\{T\} &= \left[\bm{\lambda}^\downarrow(\rho)\cdot \bm{\lambda}^\uparrow(t^\dagger t), \bm{\lambda}^\downarrow(\rho)\cdot \bm{\lambda}^\downarrow(t^\dagger t) \right],     \\
\{R\} &= \left[\bm{\lambda}^\downarrow(\rho)\cdot \bm{\lambda}^\uparrow(r^\dagger r), \bm{\lambda}^\downarrow(\rho)\cdot \bm{\lambda}^\downarrow(r^\dagger r) \right], \label{eq:main_result_set_R}
\end{align}
Here, $[\,,\,]$ denotes the closed real interval, and $\cdot$ denotes the usual inner product.
\begin{proof}
The proof is similar to the corresponding proof of Eq.~(21) in Paper~1. 
\end{proof}
Equations~(\ref{eq:main_result_set_T}) and (\ref{eq:main_result_set_R}) are the first main results of our paper. They fully characterize the attainable total transmittance or reflectance via unitary control. These equations show that $\{T\}$ or $\{R\}$ is fully determined by $\bm{\lambda}(\rho)$ and $\bm{\lambda}(t^\dagger t)$ or $\bm{\lambda}(r^\dagger r)$, which are invariant under unitary control.

\begin{figure}[hbtp]
    \centering
    \includegraphics[width=0.48\textwidth]{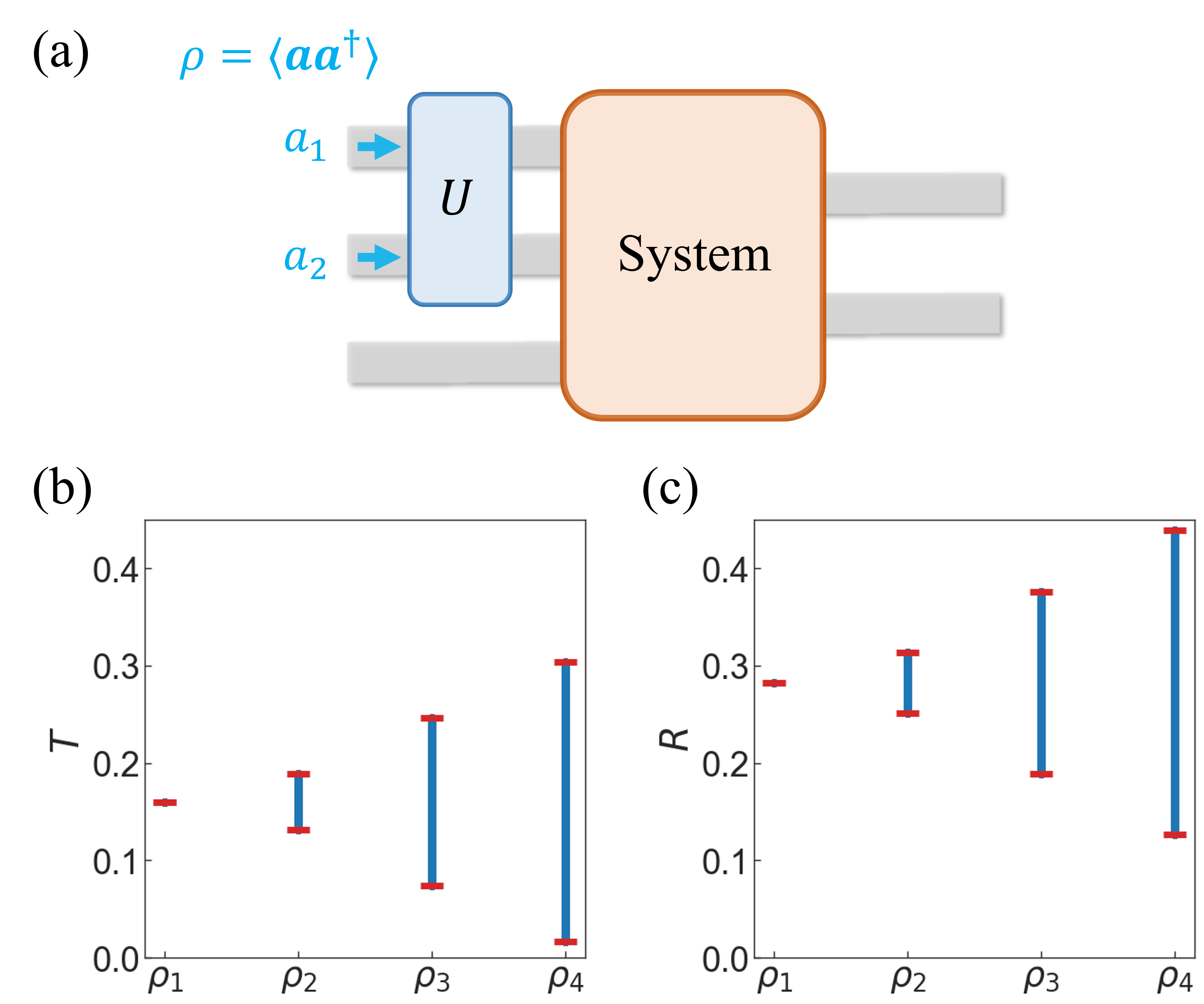}
    \caption{Attainable total transmittance and reflectance under unitary control. (a) Schematic of a $5$-port system. A partially coherent wave characterized by a density matrix $\rho$ is input into $2$ of the $3$ ports on the left side. (b) Blue dots: $T[U_i|\rho_j]$ for $300,000$ random unitary matrices $U_i$ and input density matrices $\rho_j$ with $j=1,2,3,4$. Red lines: calculated interval endpoints by Eq.~(\ref{eq:main_result_set_T}). (c) Blue dots: $R[U_i|\rho_j]$ for $300,000$ random unitary matrices $U_i$ and input density matrices $\rho_j$ with $j=1,2,3,4$. Red lines: calculated interval endpoints by Eq.~(\ref{eq:main_result_set_R}). }
    \label{fig:numerical}
\end{figure}

To illustrate our results, we conduct a numerical experiment. As shown in Fig.~\ref{fig:numerical}a, we consider a $5$-port system with $3$ ports on the left side and $2$ ports on the right side. We input a wave characterized by a random density matrix $\rho$ into $2$ input ports on the left side. \red{The system is characterized by a $5 \times 5$ scattering matrix randomly generated using NumPy~\cite{harris2020}:}
\begin{widetext}
\begin{equation}\label{eq:example1_S_matrix}
S = \left(
\begin{array}{cc|ccc}
0.32 + 0.35 i & -0.19 + 0.07 i & -0.01-0.31 i & -0.10 + 0.07 i & 0.05 - 0.16 i \\
-0.07 - 0.16 i & 0.00-0.44 i & 0.14 + 0.02 i & -0.23 - 0.19i & -0.22 -0.42 i \\
0.13 + 0.03 i & -0.12 + 0.23 i & 0.03 + 0.54 i & -0.14 + 0.08 i & 0.18 - 0.21 i \\
\hline
-0.04 + 0.36 i & 0.20 + 0.03 i & -0.10 - 0.03 i & 0.02 - 0.08 i & -0.32 - 0.27 i \\
0.27 - 0.03i & -0.20 - 0.18i & 0.15 - 0.17 i & 0.09 - 0.22 i & 0.09 - 0.16 i
\end{array}
\right).
\end{equation}    
\end{widetext}
The $r$ and $t$ matrices are the top left and bottom left block matrices of $S$ as indicated by the lines in Eq.~\ref{eq:example1_S_matrix}, respectively, with:
\begin{equation}
\bm{\lambda}^\downarrow(t^\dagger t) = (0.30, 0.02), \quad 
\bm{\lambda}^\downarrow(r^\dagger r) = (0.44, 0.13).     
\end{equation}
We consider four input density matrices $\rho_1$, $\rho_2$, $\rho_3$, and $\rho_4$, with coherence spectra:
\begin{align}
\bm{\lambda}^\downarrow(\rho_1) = (0.50, 0.50),  \quad \bm{\lambda}^\downarrow(\rho_2) = (0.60,0.40), \label{eq:lambda_rho_12}\\
\bm{\lambda}^\downarrow(\rho_3) = (0.80,0.20), \quad 
\bm{\lambda}^\downarrow(\rho_4) = (1.00, 0.00). \label{eq:lambda_rho_34}
\end{align}
Note that $\rho_1$ is completely incoherent, $\rho_2$ and $\rho_3$ are partially coherent, and $\rho_4$ is perfectly coherent. For each input, we generate $300,000$ random unitary matrices $U_i$ from the Circular Unitary Ensemble with Haar measure~\cite{mezzadri2007}, which provides a uniform probability distribution on $U(n)$~\cite{mezzadri2007}. We calculate the transmittance $T[U_i|\rho_j] = \operatorname{tr}(U_i \rho_j U_i^\dagger t^\dagger t)$ and the reflectance $R[U_i|\rho_j] = \operatorname{tr}(U_i \rho_j U_i^\dagger r^\dagger r)$ for each $\rho_j$ using Eqs.~(\ref{eq:T_U}) and (\ref{eq:R_U}). Figures~\ref{fig:numerical}b and~\ref{fig:numerical}c show the scatter plot of $T[U_i|\rho_j]$ and $R[U_i|\rho_j]$, respectively. We verify that the numerical results agree with the theoretical intervals as determined by Eqs.~(\ref{eq:main_result_set_T}) and (\ref{eq:main_result_set_R}):
\begin{align}
&\{T|\rho_1\} = \{0.16\}, \qquad \quad  \{R|\rho_1\} = \{0.28\}; \\
&\{T|\rho_2\} = \left[0.13, 0.19\right], \quad \{R|\rho_2\} = \left[0.25, 0.31\right]; \\
&\{T|\rho_3\} = \left[0.07, 0.25\right], \quad \{R|\rho_3\} = \left[0.19, 0.38\right]; \\
&\{T|\rho_4\} = \left[0.02, 0.30\right], \quad \{R|\rho_4\} = \left[0.13, 0.44\right].
\end{align}

\subsubsection{Answer to Question 2}

Now we turn to Question~2. This problem corresponds to the following physical scenario: Suppose we have a system with a transmission matrix $t$ and a reflection matrix $r$. Consider an incident partially coherent wave characterized by a normalized density matrix $\rho$. Given a total transmittance:
\begin{equation}\label{eq:T0_in_bound}
    \bm{\lambda}^\downarrow(\rho)\cdot \bm{\lambda}^\uparrow(t^\dagger t)
    \leq T_0 \leq \bm{\lambda}^\downarrow(\rho)\cdot \bm{\lambda}^\downarrow(t^\dagger t),
\end{equation}
how can we construct a unitary control scheme described by a unitary matrix $U[T_0]$ that achieves $T_0$? Similarly, given a total reflectance:
\begin{equation}\label{eq:R0_in_bound}
    \bm{\lambda}^\downarrow(\rho)\cdot \bm{\lambda}^\uparrow(r^\dagger r)
    \leq R_0 \leq \bm{\lambda}^\downarrow(\rho)\cdot \bm{\lambda}^\downarrow(r^\dagger r),
\end{equation}
how can we construct a unitary control scheme described by a unitary matrix $U[R_0]$ that achieves $R_0$?

\begin{figure}
    \centering
    \includegraphics[width=0.50\textwidth]{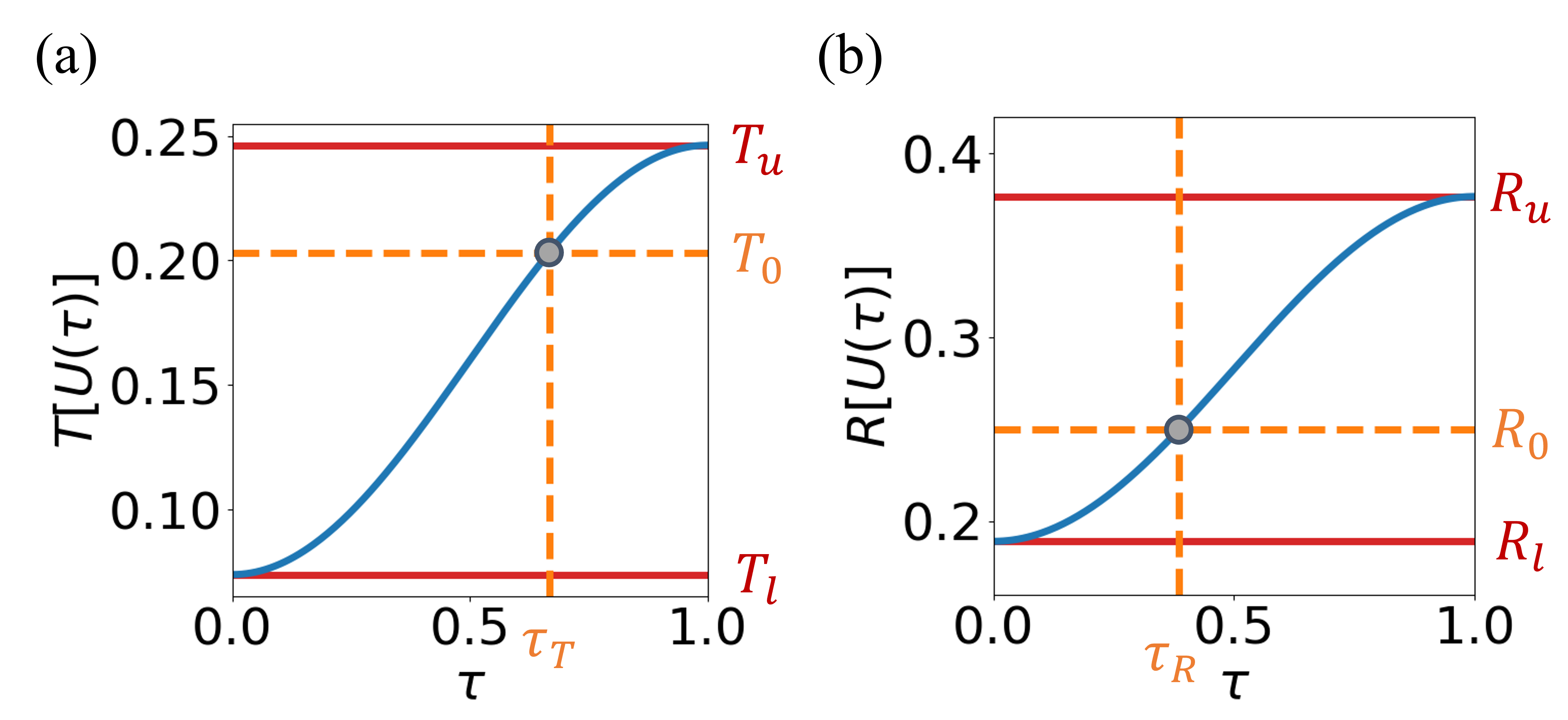}
    \caption{Constructing a unitary matrix for desired transmittance or reflectance (Algorithm~1). (a) Blue curve: $T[U(\tau)]$, where $\tau$ is a parameter that varies from $0$ to $1$, corresponding to a continuous path between the unitary matrices $U_l$ and $U_u$. $U(\tau_T)$ achieves a desired transmittance $T_0 \in [T_l, T_u]$. (b) Corresponding results for constructing $U(\tau_R)$ that achieves a desired reflectance $R_0 \in [R_l, R_u]$. }
    \label{fig:algorithm}
\end{figure}

We solve this problem using Algorithm 1 in Paper~1, modified for $U[T_0]$ or $U[R_0]$ by replacing the absorptivity matrix $A$ with the $t^\dagger t$ or $r^\dagger r$ matrices, respectively. We illustrate this algorithm with a numerical example. We consider the same $S$-matrix in Eq.~(\ref{eq:example1_S_matrix}) and the input density matrix $\rho_3$ as introduced in Eq.~(\ref{eq:lambda_rho_34}). First, we construct a $U[T_0]$ to achieve the desired transmittance:
\begin{equation}
     0.20 = T_0 \in \{T|\rho_3\} = \left[0.07, 0.25\right].
\end{equation}
We use Algorithm~1 modified for $U[T_0]$ and obtain:
\begin{widetext}
\begin{equation}\label{eq:UT_opt}
U[T_0] = 
\begin{pmatrix}
0.49 + 0.70 i & -0.15 + 0.49 i  \\
-0.07 + 0.51 i & 0.74 - 0.43 i 
\end{pmatrix}.
\end{equation}    
\end{widetext}
Second, we construct a $U[R_0]$ to achieve the desired reflectance:
\begin{equation}
     0.25 = R_0 \in \{R|\rho_3\} = \left[0.19, 0.38\right].
\end{equation}
We use Algorithm~1 modified for $U[R_0]$ and obtain:
\begin{widetext}
\begin{equation}\label{eq:UR_opt}
U[R_0] = 
\begin{pmatrix}
0.59 + 0.03 i & -0.68 - 0.44 i  \\
-0.80 + 0.15 i & -0.56 - 0.19 i 
\end{pmatrix}.
\end{equation}    
\end{widetext}

\section{Applications}\label{sec:applications}

Now, we discuss the physical applications of our theory. 

\subsection{Partially coherent perfect transmission or reflection}

First, we examine the conditions for the phenomena of \emph{partially coherent perfect transmission} or \emph{reflection}. Coherent perfect transmission or reflection~\cite{yan2014a,wu2022d} refers to the effect where a coherent wave is perfectly transmitted or reflected by a linear system through unitary control. For a linear system with a transmission matrix $t$ and a reflection matrix $r$, coherent perfect transmission occurs if and only if:
\begin{equation}\label{eq:criterion_CPT}
\operatorname{nullity} (I - t^\dagger t) \ge 1,
\end{equation}
while coherent perfect reflection occurs if and only if:
\begin{equation}\label{eq:criterion_CPR}
\operatorname{nullity} (I - r^\dagger r) \ge 1.
\end{equation}
Similarly, partially coherent perfect transmission or reflection refers to the phenomenon where a partially coherent wave is perfectly transmitted or reflected by a linear system through unitary control. We apply our theory to prove the following criterion: For a linear system with a transmission matrix $t$ and a reflection matrix $r$, and a partially coherent wave characterized by a density matrix $\rho$, partially coherent perfect transmission occurs if and only if:
\begin{equation}\label{eq:null_rank_PCPT}
\operatorname{nullity} (I - t^\dagger t) \ge \operatorname{rank} \rho, 
\end{equation}
while partially coherent perfect reflection occurs if and only if:
\begin{equation}\label{eq:null_rank_PCPR}
\operatorname{nullity} (I - r^\dagger r) \ge \operatorname{rank} \rho.  
\end{equation}
As a sanity check, for a perfectly coherent wave, $\operatorname{rank}\rho = 1$, the criterion (\ref{eq:null_rank_PCPT}) reduces to (\ref{eq:criterion_CPT}), and (\ref{eq:null_rank_PCPR}) reduces to (\ref{eq:criterion_CPR}).
\begin{proof}
The proof is similar to that of the criterion (49) in Paper~1.
\end{proof}

If the criterion (\ref{eq:null_rank_PCPT}) is satisfied, we can unitarily transform the input density matrix $\rho$ such that its support~\footnote{The support of a density matrix $\rho$ is the orthogonal complement of the kernel of $\rho$~\cite{robert2005}.} is a subset of the null space of $(I-t^\dagger t)$, thus achieving partially coherent perfect transmission. We can use Algorithm~1 to obtain such a unitary transformation. A similar analysis applies to the criterion (\ref{eq:null_rank_PCPR}) for partially coherent perfect reflection.

\begin{figure}[hbtp]
    \centering
    \includegraphics[width=0.5\textwidth]{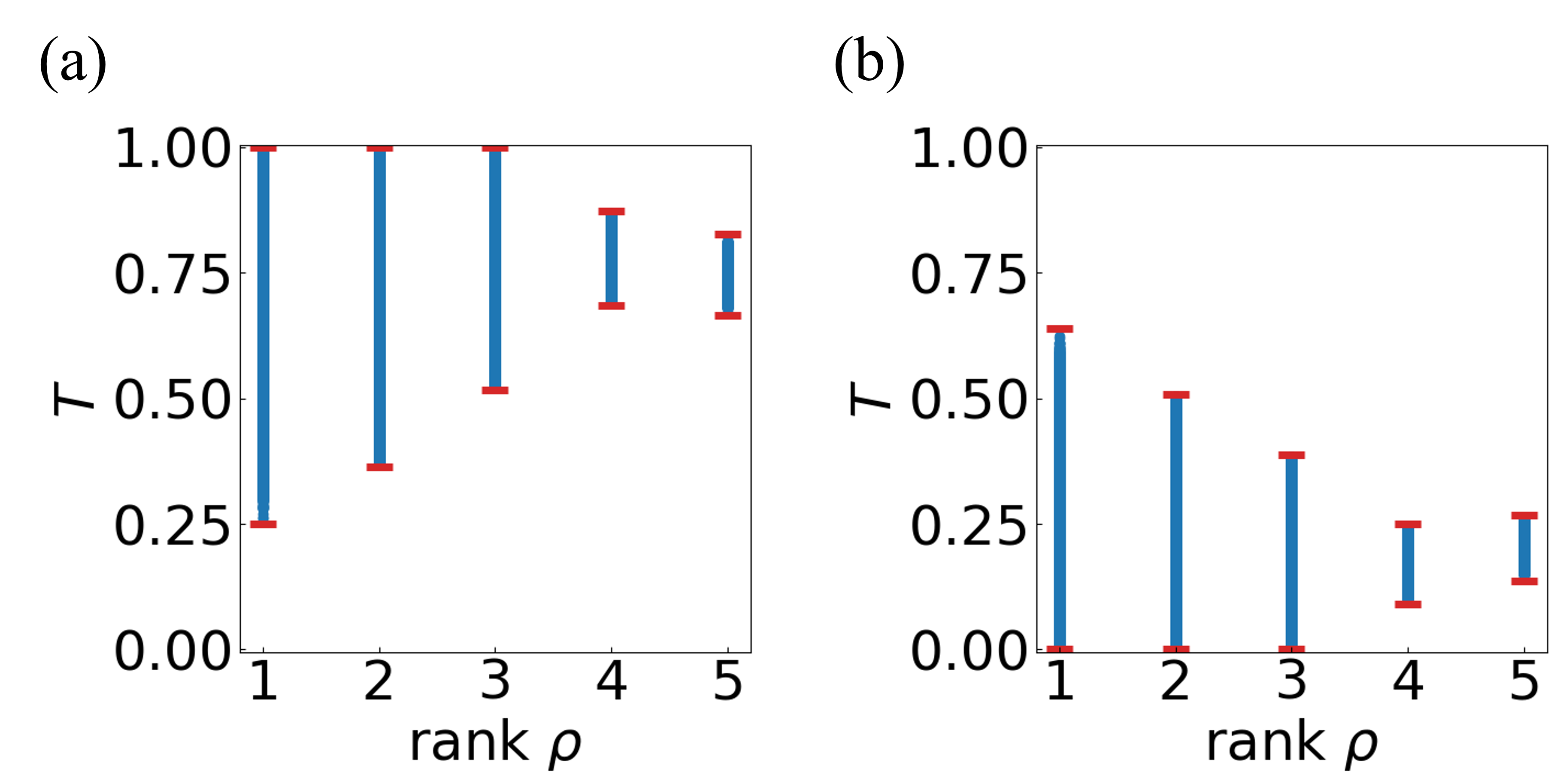}
    \caption{Numerical demonstration of the conditions for (a) partially coherent perfect transmission and (b) partially coherent zero transmission. Blue dots represent $T[U_i|\tilde{\rho}_j]$ for $10,000,000$ random unitary matrices $U_i$ and input density matrices $\tilde{\rho}_j$ with $j=1,2,3,4,5$. Red lines indicate the calculated interval endpoints by Eq.~(\ref{eq:main_result_set_T}).}
    \label{fig:perfect}
\end{figure}

We numerically demonstrate our results on partially coherent perfect transmission using a $5 \times 5$ transmission matrix:
\begin{widetext}
\begin{equation}\label{eq:example2_t_matrix}
{t} = 
\begin{pmatrix}
0.67-0.08i & -0.24+0.07i &  0.11+0.03i &  0.15+0.05i &  0.32+0.35i \\
-0.06-0.04i &  0.18-0.34i &  0.20-0.38i &  0.05-0.35i & -0.35-0.04i \\
0.05+0.05i &  0.23+0.04i & -0.37-0.28i &  0.47+0.63i & -0.06-0.09i \\
-0.40+0.22i &  0.30+0.33i &  0.03-0.21i &  0.11-0.22i &  0.48+0.38i \\
0.00-0.01i &  0.25-0.36i &  0.10+0.46i & -0.13+0.14i &  0.30+0.19i 
\end{pmatrix},
\end{equation}
\end{widetext}
which has:
\begin{equation}
\bm{\lambda}^\downarrow(t^\dagger t) = (1.00, 1.00, 1.00, 0.49, 0.25),      
\end{equation}
thus:
\begin{equation}
\operatorname{nullity} (I - t^\dagger t) = 3. 
\end{equation}
We consider five different incident waves characterized by normalized density matrices $\tilde{\rho}_j$, $1 \le j \le 5$, with coherence spectra:
\begin{align}\label{eq:coherence_spectrum_rho_1}
\bm{\lambda}^\downarrow(\tilde{\rho}_1) &= (1.00, 0.00, 0.00, 0.00, 0.00), \\ \bm{\lambda}^\downarrow(\tilde{\rho}_2) &= (0.53, 0.47, 0.00, 0.00, 0.00), \\
\bm{\lambda}^\downarrow(\tilde{\rho}_3) &= (0.42, 0.33, 0.25, 0.00, 0.00), \\
\bm{\lambda}^\downarrow(\tilde{\rho}_4) &= (0.25, 0.25, 0.25, 0.25, 0.00), \\
\bm{\lambda}^\downarrow(\tilde{\rho}_5) &= (0.29, 0.23, 0.20, 0.15, 0.13),\label{eq:coherence_spectrum_rho_5}
\end{align}
thus, their ranks are different:
\begin{equation}
\operatorname{rank} \tilde{\rho}_j = j.
\end{equation}
For each input, we generate $10,000,000$ random unitary matrices $U_i$ from the Circular Unitary Ensemble. Then, we calculate the transmittance $T[U_i|\tilde{\rho}_j] = \operatorname{tr}(U_i \tilde{\rho}_j U_i^\dagger t^\dagger t)$ for each $\tilde{\rho}_j$ using Eq.~(\ref{eq:T_U}). The results are plotted in Fig.~\ref{fig:perfect}a. We see that partially coherent perfect transmission is achievable when $\operatorname{rank} \tilde{\rho}_j = 1, 2, 3$, but not when $\operatorname{rank} \tilde{\rho}_j = 4, 5$. This verifies the criterion~(\ref{eq:null_rank_PCPT}).

\subsection{Partially coherent zero transmission or reflection}

Second, we examine the conditions for the phenomena of \emph{partially coherent zero transmission} or \emph{reflection}. Coherent zero transmission or reflection refers to the effect where a coherent wave exhibits zero transmission or reflection by a linear system through unitary control. For a linear system with a transmission matrix $t$ and a reflection matrix $r$, coherent zero transmission occurs if and only if:
\begin{equation}\label{eq:criterion_CZT}
\operatorname{nullity} t^\dagger t \ge 1,
\end{equation}
while coherent zero reflection occurs if and only if:
\begin{equation}\label{eq:criterion_CZR}
\operatorname{nullity} r^\dagger r \ge 1.
\end{equation}
Similarly, partially coherent zero transmission or reflection refers to the phenomenon where a partially coherent wave exhibits zero transmission or reflection by a linear system through unitary control. We apply our theory to prove the following criterion: For a linear system with a transmission matrix $t$ and a reflection matrix $r$, and a partially coherent wave characterized by a density matrix $\rho$, partially coherent zero transmission occurs if and only if:
\begin{equation}\label{eq:null_rank_PCZT}
\operatorname{nullity} t^\dagger t \ge \operatorname{rank} \rho,  
\end{equation}
while partially coherent zero reflection occurs if and only if:
\begin{equation}\label{eq:null_rank_PCZR}
\operatorname{nullity} r^\dagger r \ge \operatorname{rank} \rho.  
\end{equation}
As a sanity check, for a perfectly coherent wave, $\operatorname{rank}\rho = 1$, the criterion (\ref{eq:null_rank_PCZT}) reduces to (\ref{eq:criterion_CZT}), and (\ref{eq:null_rank_PCZR}) reduces to (\ref{eq:criterion_CZR}).

\begin{proof}
The proof is similar to that of the criterion~(66) in Paper~1. 
\end{proof}
If the criterion (\ref{eq:null_rank_PCZT}) is satisfied, we can unitarily transform the input density matrix $\rho$ into the null space of the $t^\dagger t$ matrix, thus achieving partially coherent zero transmission. We can use Algorithm~1 to obtain such a unitary transformation. A similar analysis applies to the criterion (\ref{eq:null_rank_PCZR}) for partially coherent zero reflection.

We numerically demonstrate our results on partially coherent zero transmission. We consider a $5 \times 5$ transmission matrix:
\begin{widetext}
\begin{equation}\label{eq:example3_t_matrix}
t = 
\begin{pmatrix}
0.03-0.16i &  0.05+0.02i & -0.21+0.08i & -0.03+0.01i &  0.34-0.12i \\  
-0.24+0.01i & -0.16-0.15i & -0.01+0.03i & -0.07-0.16i & -0.02+0.14i \\   
0.06-0.07i &  0.06-0.10i &  0.04-0.19i &  0.12-0.09i & -0.21+0.30i \\  
-0.11+0.24i & -0.15+0.12i &  0.10+0.15i & -0.16+0.08i &  0.03-0.20i \\  
-0.08+0.15i & -0.10-0.05i &  0.18-0.08i & -0.00-0.06i & -0.30+0.18i
\end{pmatrix},
\end{equation}
\end{widetext}
which has:
\begin{equation}
\bm{\lambda}^\downarrow(t^\dagger t) = (0.64, 0.36, 0.00, 0.00, 0.00),     
\end{equation}
thus:
\begin{equation}
\operatorname{nullity} t^\dagger t = 3. 
\end{equation}
We consider five different incident waves characterized by normalized density matrices $\tilde{\rho}_j$, $1 \le j \le 5$, with coherence spectra as provided in Eqs.~(\ref{eq:coherence_spectrum_rho_1})-(\ref{eq:coherence_spectrum_rho_5}); thus, $\operatorname{rank} \tilde{\rho}_j = j$. For each input, we generate $10,000,000$ random unitary matrices $U_i$ from the Circular Unitary Ensemble. Then, we calculate the transmittance $T[U_i|\tilde{\rho}_j] = \operatorname{tr}(U_i \tilde{\rho}_j U_i^\dagger t^\dagger t)$ for each $\tilde{\rho}_j$ using Eq.~(\ref{eq:T_U}). The results are plotted in Fig.~\ref{fig:perfect}b. We see that partially coherent zero transmission is achievable when $\operatorname{rank} \tilde{\rho}_j = 1, 2, 3$, but not when $\operatorname{rank} \tilde{\rho}_j = 4, 5$. This verifies the criterion~(\ref{eq:null_rank_PCZT}).

\subsection{Majorized coherence implies nested transmission or reflection intervals}

Third, we examine how the degree of coherence affects the attainable transmittance or reflectance intervals. Our main results, Eqs.~(\ref{eq:main_result_set_T}) and (\ref{eq:main_result_set_R}), show that, for a given system, the transmittance interval $\{T\}$ and the reflectance interval $\{R\}$ are controlled by the coherence spectrum $\bm{\lambda}^{\downarrow}(\rho)$. A natural question arises: How will the transmittance or reflectance intervals vary when the degree of coherence changes?

We compare the coherence between waves using the majorization order~\cite{marshall2011,nielsen1999,gour2015,bengtsson2017,gour2018,luis2016}. Consider two waves with density matrices $\rho_1$ and $\rho_2$, respectively. We say that $\rho_1$ is no more coherent than $\rho_2$ if $\bm{\lambda}^{\downarrow}(\rho_1) \prec \bm{\lambda}^{\downarrow}(\rho_2)$. If neither $\bm{\lambda}^{\downarrow}(\rho_1) \prec \bm{\lambda}^{\downarrow}(\rho_2)$ nor $\bm{\lambda}^{\downarrow}(\rho_2) \prec \bm{\lambda}^{\downarrow}(\rho_1)$ holds, we say that $\rho_1$ and $\rho_2$ are incomparable and denote this as $\bm{\lambda}^{\downarrow}(\rho_1) \parallel \bm{\lambda}^{\downarrow}(\rho_2)$. For any $\rho$:
\begin{equation}\label{eq:incoherent_partial_coherent}
(\frac{1}{n},\frac{1}{n},\ldots,\frac{1}{n}) \prec \bm{\lambda}^\downarrow(\rho) \prec (1,0,\ldots,0).
\end{equation}

Now, we state the following theorem: If $\rho_1$ is no more coherent than $\rho_2$, then for any system, the transmittance or reflectance interval of $\rho_1$ is always contained in that of $\rho_2$:
\begin{align}
\label{eq:nested_interval_T}
\bm{\lambda}^\downarrow(\rho_1) \prec \bm{\lambda}^\downarrow(\rho_2) &\implies \{T\}_1 \subseteq \{T\}_2, \\
\bm{\lambda}^\downarrow(\rho_1) \prec \bm{\lambda}^\downarrow(\rho_2) &\implies \{R\}_1 \subseteq \{R\}_2.\label{eq:nested_interval_R}
\end{align}
Using Eqs.~(\ref{eq:main_result_set_T}) and (\ref{eq:main_result_set_R}), we can express the right-hand sides of Eqs.~(\ref{eq:nested_interval_T}) and (\ref{eq:nested_interval_R}) more explicitly as:
\begin{align}
\label{eq:nested_interval_explicit_T}
&\bm{\lambda}^\downarrow(\rho_2)\cdot \bm{\lambda}^\uparrow(t^\dagger t)\leq \bm{\lambda}^\downarrow(\rho_1)\cdot \bm{\lambda}^\uparrow(t^\dagger t) \leq 
\bm{\lambda}^\downarrow(\rho_1)\cdot \bm{\lambda}^\downarrow(t^\dagger t) \leq \bm{\lambda}^\downarrow(\rho_2)\cdot \bm{\lambda}^\downarrow(t^\dagger t), \\
\label{eq:nested_interval_explicit_R}
&\bm{\lambda}^\downarrow(\rho_2)\cdot \bm{\lambda}^\uparrow(r^\dagger r)\leq \bm{\lambda}^\downarrow(\rho_1)\cdot \bm{\lambda}^\uparrow(r^\dagger r) \leq 
\bm{\lambda}^\downarrow(\rho_1)\cdot \bm{\lambda}^\downarrow(r^\dagger r) \leq \bm{\lambda}^\downarrow(\rho_2)\cdot \bm{\lambda}^\downarrow(r^\dagger r).
\end{align}
\begin{proof}
The proof is similar to that of (79) in Paper~1. 
\end{proof}

The statements (\ref{eq:nested_interval_T}) and (\ref{eq:nested_interval_R}) are our main results of this subsection. They can be summarized as: ``Majorized coherence implies nested transmittance and reflectance intervals.'' Now, we examine their implications.

First, we apply Eqs.~(\ref{eq:nested_interval_T}) and (\ref{eq:nested_interval_R}) to Eq.~(\ref{eq:incoherent_partial_coherent}) and deduce that for any density matrix $\rho$ and any transmission matrix $t$ and reflection matrix $r$:
\begin{align}
\label{eq:nested_interval_explicit_coherent_T}
\lambda_{\min}(t^\dagger t) \leq \bm{\lambda}^\downarrow(\rho)\cdot \bm{\lambda}^\uparrow(t^\dagger t) \leq  \frac{1}{n}\sum_i \lambda_i(t^\dagger t) \leq 
\bm{\lambda}^\downarrow(\rho)\cdot \bm{\lambda}^\downarrow(t^\dagger t) \leq \lambda_{\max}(t^\dagger t), \\
\label{eq:nested_interval_explicit_coherent_R}
\lambda_{\min}(r^\dagger r) \leq \bm{\lambda}^\downarrow(\rho)\cdot \bm{\lambda}^\uparrow(r^\dagger r) \leq  \frac{1}{n}\sum_i \lambda_i(r^\dagger r) \leq 
\bm{\lambda}^\downarrow(\rho)\cdot \bm{\lambda}^\downarrow(r^\dagger r) \leq \lambda_{\max}(r^\dagger r).
\end{align}
In particular, the means of $\lambda_i(t^\dagger t)$ and $\lambda_i(r^\dagger r)$ are always contained in the transmittance and reflectance intervals, respectively. Hence, they are attainable via unitary control.

Second, from the contrapositive of Eq.~(\ref{eq:nested_interval_T}), we deduce that if for some system, neither $\{T\}_1 \subseteq \{T\}_2$ nor $\{T\}_2 \subseteq \{T\}_1$ holds (denoted as $\{T\}_1 \parallel \{T\}_2$), then $\rho_1$ and $\rho_2$ are incomparable:
\begin{equation}\label{eq:incomparable_interval_T}
\{T\}_1 \parallel \{T\}_2 \implies   \bm{\lambda}^\downarrow(\rho_1) \parallel \bm{\lambda}^\downarrow(\rho_2). 
\end{equation}
Similarly, from the contrapositive of Eq.~(\ref{eq:nested_interval_R}), we deduce that:
\begin{equation}\label{eq:incomparable_interval_R}
\{R\}_1 \parallel \{R\}_2 \implies   \bm{\lambda}^\downarrow(\rho_1) \parallel \bm{\lambda}^\downarrow(\rho_2). 
\end{equation}

We illustrate these results with previous numerical examples. In Figs.~\ref{fig:numerical}b and~\ref{fig:numerical}c, we observe that:
\begin{align}
 \{T|\rho_1\} \subseteq   \{T|\rho_2\} \subseteq
 \{T|\rho_3\} \subseteq
   \{T|\rho_4\},    \\
 \{R|\rho_1\} \subseteq   \{R|\rho_2\} \subseteq
 \{R|\rho_3\} \subseteq
   \{R|\rho_4\},    
\end{align}
because $\bm{\lambda}^\downarrow(\rho_i)$, as given in Eqs.~(\ref{eq:lambda_rho_12}) and (\ref{eq:lambda_rho_34}), satisfy:
\begin{equation}
\bm{\lambda}^\downarrow(\rho_1) \prec \bm{\lambda}^\downarrow(\rho_2) \prec \bm{\lambda}^\downarrow(\rho_3) \prec \bm{\lambda}^\downarrow(\rho_4).
\end{equation}
In Figs.~\ref{fig:perfect}a and \ref{fig:perfect}b, we observe that:
\begin{align}
 \{T|\tilde{\rho}_1\} \subseteq   \{T|\tilde{\rho}_2\} \subseteq
 \{T|\tilde{\rho}_3\} \subseteq
   \{T|\tilde{\rho}_4\}, \\
 \{R|\tilde{\rho}_1\} \subseteq   \{R|\tilde{\rho}_2\} \subseteq
 \{R|\tilde{\rho}_3\} \subseteq
   \{R|\tilde{\rho}_4\},   
\end{align}
because $\bm{\lambda}^\downarrow(\tilde{\rho}_i)$, as given in Eqs.~(\ref{eq:coherence_spectrum_rho_1})-(\ref{eq:coherence_spectrum_rho_5}), satisfy:
\begin{equation}
\bm{\lambda}^\downarrow(\tilde{\rho}_1) \prec \bm{\lambda}^\downarrow(\tilde{\rho}_2) \prec \bm{\lambda}^\downarrow(\tilde{\rho}_3) \prec \bm{\lambda}^\downarrow(\tilde{\rho}_4).
\end{equation}
We also observe that:
\begin{align}
 \{T|\tilde{\rho}_4\} \parallel  \{T|\tilde{\rho}_5\}, \\
 \{R|\tilde{\rho}_4\} \parallel  \{R|\tilde{\rho}_5\},    
\end{align}
which can occur because:
\begin{equation}
\bm{\lambda}^\downarrow(\tilde{\rho}_4) \parallel \bm{\lambda}^\downarrow(\tilde{\rho}_5).
\end{equation}

\subsection{Symmetry constraints on bilateral transmission and reflection}\label{subsec:symmetry_bilateral}

Fourth, we discuss the constraints imposed by symmetry on the bilateral unitary control of partially coherent transmission and reflection. \red{While symmetry constraints on transmission and reflection eigenvalues are well-established~\cite{beenakker1997b}, their implications for attainable transmission and reflection intervals of partially coherent waves have not been explored.}

\begin{figure}[htbp]
    \centering
\includegraphics[width=0.5\textwidth]{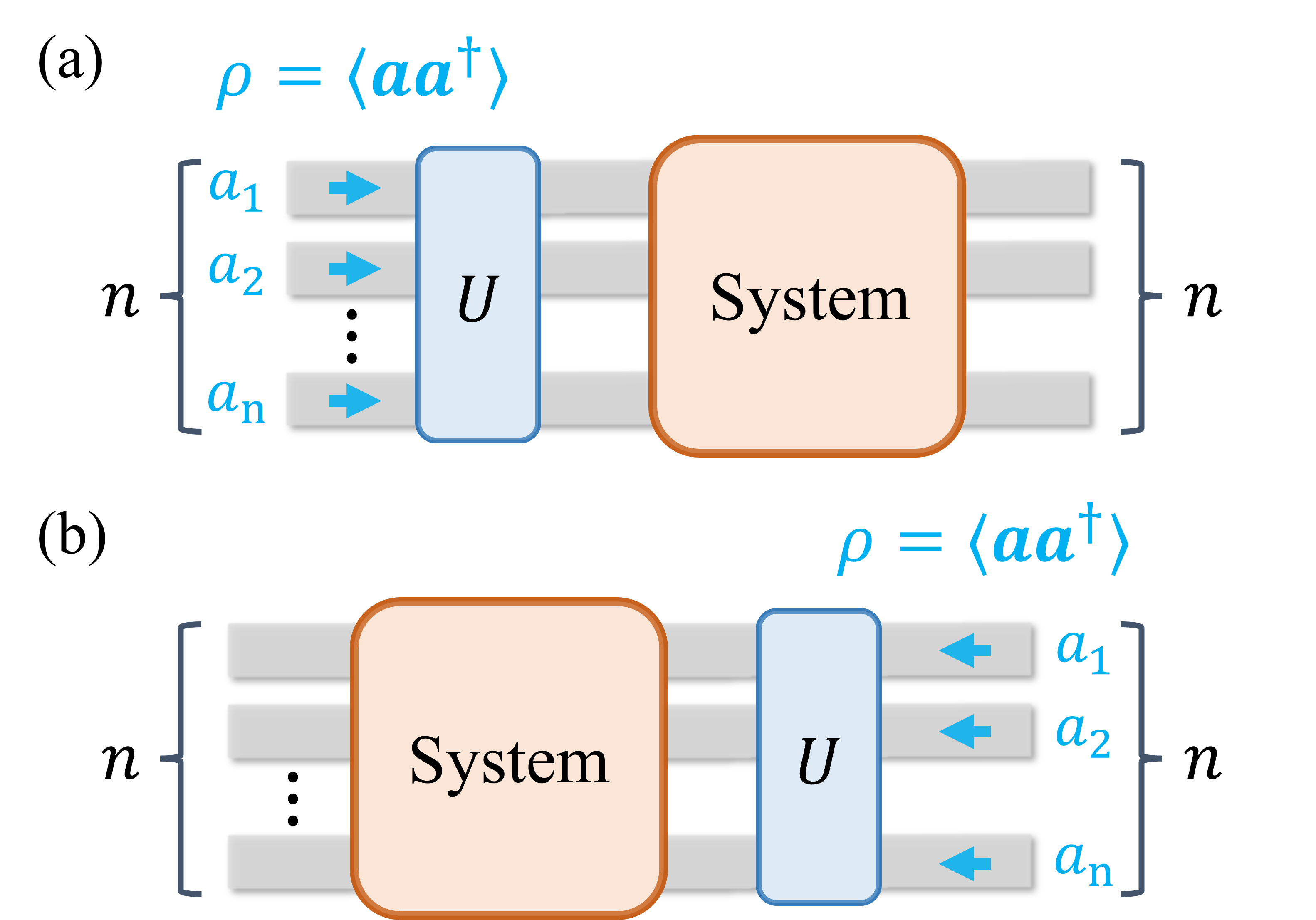}
    \caption{Schematic of unitary control of bilateral transmission or reflection for partially coherent waves in a linear time-invariant system with $2n$ ports, with $n$ ports on either the left or the right side. (a) Unitary control applied to an input wave characterized by a density matrix $\rho$ incident from the left side. The set of attainable total transmittance and reflectance are $\{T\}_l$ and $\{R\}_l$, respectively. (b) Unitary control applied to the same input wave incident from the right side. The set of attainable total transmittance and reflectance are $\{T\}_r$ and $\{R\}_r$, respectively.}
    \label{fig:bilateral}
\end{figure}

For concreteness, we consider a $2n$-port linear time-invariant system with $n$ ports on either the left or right side, as shown in Fig.~\ref{fig:bilateral}. The system is characterized by a $2n \times 2n$ scattering matrix:
\begin{equation}
S = \begin{pmatrix}
r & t' \\
t & r'
\end{pmatrix},    
\end{equation}
where $r$, $t$, $r'$, and $t'$ are $n \times n$ matrices. We input a partially coherent wave characterized by an $n \times n$ density matrix $\rho$ from either side and apply unitary control. We denote the set of attainable total transmittance and reflectance as $\{T\}_{l}$ and $\{R\}_{l}$ ($\{T\}_{r}$ and $\{R\}_{r}$) when the wave is incident from the left (right) side. We study the relationship between these sets imposed by certain symmetries of the system. Here, we examine two important internal symmetries: reciprocity and energy conservation.

If the system is reciprocal,
\begin{equation}\label{eq:reciprocal_S}
S = S^T.    
\end{equation}
It follows that:
\begin{equation}
\bm{\lambda}^{\downarrow}(t^{\dagger}t) = \bm{\lambda}^{\downarrow}(t'^{\dagger}t'),   
\end{equation}
and consequently:
\begin{equation}
\{T\}_{l} = \{T\}_{r}.    
\end{equation}

\begin{proof}
From Eq.~(\ref{eq:reciprocal_S}), we obtain:
\begin{equation}
t' = t^{T},    
\end{equation}
thus:
\begin{equation}
\bm{\lambda}^{\downarrow}(t'^{\dagger}t') = \bm{\lambda}^{\downarrow}(t^{*}t^{T}) = \bm{\lambda}^{\downarrow}(tt^{\dagger}) = \bm{\lambda}^{\downarrow}(t^{\dagger}t).    
\end{equation}
The second equality is because for any square matrix, $\bm{\lambda}(M) = \bm{\lambda}(M^{T})$ (see Ref.~\cite{zhang2011}, p.~102, Theorem~3.14). The last equality is because for any square matrices $M$ and $N$ of the same size, $\bm{\lambda}^{\downarrow}(MN) = \bm{\lambda}^{\downarrow}(NM)$ (see Ref.~\cite{zhang2011}, p.~77, Theorem~2.8). It follows that:
\begin{align}
\{T\}_{l} &= [\bm{\lambda}^{\downarrow}(\rho)\cdot \bm{\lambda}^{\uparrow}(t^{\dagger}t), \bm{\lambda}^{\downarrow}(\rho)\cdot \bm{\lambda}^{\downarrow}(t^{\dagger}t)] \\
&= [\bm{\lambda}^{\downarrow}(\rho)\cdot \bm{\lambda}^{\uparrow}(t'^{\dagger}t'), \bm{\lambda}^{\downarrow}(\rho)\cdot \bm{\lambda}^{\downarrow}(t'^{\dagger}t')] = \{T\}_{r}.     
\end{align}
This completes the proof for the reciprocal case.
\end{proof}

If the system is energy-conserving:
\begin{equation}
S^{\dagger} S = SS^{\dagger}  = I.    
\end{equation}
It follows that:
\begin{align}
\bm{\lambda}^{\downarrow}(t^{\dagger}t) &= \bm{\lambda}^{\downarrow}(t'^{\dagger}t'),  \\ 
\bm{\lambda}^{\downarrow}(r^{\dagger}r) &= \bm{\lambda}^{\downarrow}(r'^{\dagger}r'), \\
\bm{\lambda}^{\downarrow}(t^{\dagger}t) +& \bm{\lambda}^{\uparrow} (r^{\dagger}r) = \bm{1}.
\end{align}
Consequently:
\begin{equation}
\{T\}_{l} = \{T\}_{r},  \quad 
\{R\}_{l} = \{R\}_{r}.    
\end{equation}
Moreover, $\{T\}_{l}$ and $\{R\}_{l}$ are mirror symmetric with respect to $\frac{1}{2}$.

\begin{proof}
From $S^{\dagger}S = I$, we obtain:
\begin{align}
r^{\dagger} r + t^{\dagger}t = I, \label{eq:SdagS-1}\\
r'^{\dagger} r' + t'^{\dagger}t' = I. \label{eq:SdagS-2}
\end{align}
From $SS^{\dagger} = I$, we obtain:
\begin{align}
rr^{\dagger}  + t't'^{\dagger} = I,\label{eq:SSdag-1} \\
r'r'^{\dagger}  + tt^{\dagger} = I. \label{eq:SSdag-2}
\end{align}
Combining Eqs.~(\ref{eq:SdagS-1}) and (\ref{eq:SSdag-1}), we have:
\begin{align}
\bm{\lambda}^{\downarrow}(t^{\dagger}t) &= \bm{\lambda}^{\downarrow}(I - r^{\dagger}r) = \bm{1} - \bm{\lambda}^{\uparrow}(r^{\dagger}r) = \bm{1} - \bm{\lambda}^{\uparrow}(r r^{\dagger}) \label{eq:derivation-middle}
\\ &=  \bm{\lambda}^{\downarrow}(I - r r^{\dagger}) = \bm{\lambda}^{\downarrow}(t't'^{\dagger}) = \bm{\lambda}^{\downarrow}(t'^{\dagger}t').     
\end{align}
Similarly, combining Eqs.~(\ref{eq:SdagS-1}) and (\ref{eq:SSdag-2}), we have:
\begin{align}
\bm{\lambda}^{\downarrow}(r^{\dagger}r) &= \bm{\lambda}^{\downarrow}(I - t^{\dagger}t) = \bm{1} - \bm{\lambda}^{\uparrow}(t^{\dagger}t) = \bm{1} - \bm{\lambda}^{\uparrow}(t t^{\dagger}) \\ &=  \bm{\lambda}^{\downarrow}(I - t t^{\dagger}) = \bm{\lambda}^{\downarrow}(r'r'^{\dagger}) = \bm{\lambda}^{\downarrow}(r'^{\dagger}r').     
\end{align}
Moreover, from Eq.~(\ref{eq:derivation-middle}), we have:
\begin{equation}
\bm{\lambda}^{\downarrow}(t^{\dagger}t) + \bm{\lambda}^{\uparrow} (r^{\dagger}r) = \bm{1}.    
\end{equation}
It follows that:
\begin{align}
\{T\}_{l} = \{T\}_{r} &= [\bm{\lambda}^{\downarrow}(\rho)\cdot \bm{\lambda}^{\uparrow}(t^{\dagger}t), \bm{\lambda}^{\downarrow}(\rho)\cdot \bm{\lambda}^{\downarrow}(t^{\dagger}t)],  \\
\{R\}_{l} = \{R\}_{r} &= [\bm{\lambda}^{\downarrow}(\rho)\cdot \bm{\lambda}^{\uparrow}(r^{\dagger}r), \bm{\lambda}^{\downarrow}(\rho)\cdot \bm{\lambda}^{\downarrow}(r^{\dagger}r)] \\
&= [1-\bm{\lambda}^{\downarrow}(\rho)\cdot \bm{\lambda}^{\downarrow}(t^{\dagger}t), 1-\bm{\lambda}^{\downarrow}(\rho)\cdot \bm{\lambda}^{\uparrow}(t^{\dagger}t)], 
\end{align}
where we have used Eq.~(\ref{eq:rho_normalization}) to obtain:
\begin{equation}
\bm{\lambda}^{\downarrow}(\rho) \cdot \bm{1} = \operatorname{tr} \rho = 1.    
\end{equation}
Hence, $\{T\}_{l}$ and $\{R\}_{l}$ are mirror symmetric with respect to $\frac{1}{2}$. This completes the proof for the energy-conserving case.
\end{proof}

\section{Conclusion}\label{sec:conclusion}

In conclusion, we have developed a comprehensive theory for the unitary control of partially coherent wave transmission and reflection by linear systems. Our key contributions include: (1) analytical expressions [Eqs.~(\ref{eq:main_result_set_T}) and (\ref{eq:main_result_set_R})] that define the ranges of attainable total transmittance and reflectance under arbitrary unitary transformations of the incident field, and (2) an explicit algorithm to construct a unitary control scheme that achieves any desired transmittance or reflectance within the attainable range.

Through this theory, we establish the conditions for four new phenomena: partially coherent perfect transmission, partially coherent perfect reflection, partially coherent zero transmission, and partially coherent zero reflection. We derive precise criteria [Eqs.~(\ref{eq:null_rank_PCPT}), (\ref{eq:null_rank_PCPR}), (\ref{eq:null_rank_PCZR}), and (\ref{eq:null_rank_PCZT})] for their occurrence. Additionally, we prove a fundamental theorem [Eqs.~(\ref{eq:nested_interval_T}) and (\ref{eq:nested_interval_R})] that relates the majorization order of the incident coherence spectra to the nesting order of the resulting transmission or reflection intervals. Furthermore, we reveal the symmetry constraints imposed by reciprocity and energy conservation on the unitary control of bilateral transmission and reflection of partially coherent waves.

The theory established in this work enhances the understanding of partially coherent transmission and reflection control across a diverse range of wave systems. We anticipate that our results will find applications in areas such as imaging, sensing, display, and communication, where partially coherent transmission and reflection play a central role.

\begin{acknowledgments}
This work is funded by the U.~S.~Department of Energy (Grant No.~DE-FG02-07ER46426), and by a Simons Investigator in Physics grant from the Simons Foundation. (Grant No.~827065).

\end{acknowledgments}


\bibliography{main}

\begin{thebibliography}{138}%
\makeatletter
\providecommand \@ifxundefined [1]{%
 \@ifx{#1\undefined}
}%
\providecommand \@ifnum [1]{%
 \ifnum #1\expandafter \@firstoftwo
 \else \expandafter \@secondoftwo
 \fi
}%
\providecommand \@ifx [1]{%
 \ifx #1\expandafter \@firstoftwo
 \else \expandafter \@secondoftwo
 \fi
}%
\providecommand \natexlab [1]{#1}%
\providecommand \enquote  [1]{``#1''}%
\providecommand \bibnamefont  [1]{#1}%
\providecommand \bibfnamefont [1]{#1}%
\providecommand \citenamefont [1]{#1}%
\providecommand \href@noop [0]{\@secondoftwo}%
\providecommand \href [0]{\begingroup \@sanitize@url \@href}%
\providecommand \@href[1]{\@@startlink{#1}\@@href}%
\providecommand \@@href[1]{\endgroup#1\@@endlink}%
\providecommand \@sanitize@url [0]{\catcode `\\12\catcode `\$12\catcode `\&12\catcode `\#12\catcode `\^12\catcode `\_12\catcode `\%12\relax}%
\providecommand \@@startlink[1]{}%
\providecommand \@@endlink[0]{}%
\providecommand \url  [0]{\begingroup\@sanitize@url \@url }%
\providecommand \@url [1]{\endgroup\@href {#1}{\urlprefix }}%
\providecommand \urlprefix  [0]{URL }%
\providecommand \Eprint [0]{\href }%
\providecommand \doibase [0]{https://doi.org/}%
\providecommand \selectlanguage [0]{\@gobble}%
\providecommand \bibinfo  [0]{\@secondoftwo}%
\providecommand \bibfield  [0]{\@secondoftwo}%
\providecommand \translation [1]{[#1]}%
\providecommand \BibitemOpen [0]{}%
\providecommand \bibitemStop [0]{}%
\providecommand \bibitemNoStop [0]{.\EOS\space}%
\providecommand \EOS [0]{\spacefactor3000\relax}%
\providecommand \BibitemShut  [1]{\csname bibitem#1\endcsname}%
\let\auto@bib@innerbib\@empty
\bibitem [{\citenamefont {Planck}(1991)}]{planck1991}%
  \BibitemOpen
  \bibfield  {author} {\bibinfo {author} {\bibfnamefont {M.}~\bibnamefont {Planck}},\ }\href@noop {} {\emph {\bibinfo {title} {The Theory of Heat Radiation}}}\ (\bibinfo  {publisher} {{Dover Publications}},\ \bibinfo {address} {{New York}},\ \bibinfo {year} {1991})\BibitemShut {NoStop}%
\bibitem [{\citenamefont {Chen}(2005)}]{chen2005}%
  \BibitemOpen
  \bibfield  {author} {\bibinfo {author} {\bibfnamefont {G.}~\bibnamefont {Chen}},\ }\href@noop {} {\emph {\bibinfo {title} {Nanoscale Energy Transport and Conversion: A Parallel Treatment of Electrons, Molecules, Phonons, and Photons}}}\ (\bibinfo  {publisher} {{Oxford University Press}},\ \bibinfo {address} {{Oxford}},\ \bibinfo {year} {2005})\BibitemShut {NoStop}%
\bibitem [{\citenamefont {Zhang}(2007)}]{zhang2007}%
  \BibitemOpen
  \bibfield  {author} {\bibinfo {author} {\bibfnamefont {Z.~M.}\ \bibnamefont {Zhang}},\ }\href@noop {} {\emph {\bibinfo {title} {Nano/Microscale Heat Transfer}}}\ (\bibinfo  {publisher} {{McGraw-Hill}},\ \bibinfo {address} {{New York}},\ \bibinfo {year} {2007})\BibitemShut {NoStop}%
\bibitem [{\citenamefont {Howell}\ \emph {et~al.}(2016)\citenamefont {Howell}, \citenamefont {Meng{\"u}{\c c}},\ and\ \citenamefont {Siegel}}]{howell2016}%
  \BibitemOpen
  \bibfield  {author} {\bibinfo {author} {\bibfnamefont {J.~R.}\ \bibnamefont {Howell}}, \bibinfo {author} {\bibfnamefont {M.~P.}\ \bibnamefont {Meng{\"u}{\c c}}},\ and\ \bibinfo {author} {\bibfnamefont {R.}~\bibnamefont {Siegel}},\ }\href@noop {} {\emph {\bibinfo {title} {Thermal Radiation Heat Transfer}}},\ \bibinfo {edition} {sixth}\ ed.\ (\bibinfo  {publisher} {{CRC Press}},\ \bibinfo {address} {{London}},\ \bibinfo {year} {2016})\BibitemShut {NoStop}%
\bibitem [{\citenamefont {Fan}(2017)}]{fan2017}%
  \BibitemOpen
  \bibfield  {author} {\bibinfo {author} {\bibfnamefont {S.}~\bibnamefont {Fan}},\ }\bibfield  {title} {\bibinfo {title} {Thermal {{Photonics}} and {{Energy Applications}}},\ }\href {https://doi.org/10.1016/j.joule.2017.07.012} {\bibfield  {journal} {\bibinfo  {journal} {Joule}\ }\textbf {\bibinfo {volume} {1}},\ \bibinfo {pages} {264} (\bibinfo {year} {2017})}\BibitemShut {NoStop}%
\bibitem [{\citenamefont {Cuevas}\ and\ \citenamefont {{Garc{\'i}a-Vidal}}(2018)}]{cuevas2018b}%
  \BibitemOpen
  \bibfield  {author} {\bibinfo {author} {\bibfnamefont {J.~C.}\ \bibnamefont {Cuevas}}\ and\ \bibinfo {author} {\bibfnamefont {F.~J.}\ \bibnamefont {{Garc{\'i}a-Vidal}}},\ }\bibfield  {title} {\bibinfo {title} {Radiative {{Heat Transfer}}},\ }\href {https://doi.org/10.1021/acsphotonics.8b01031} {\bibfield  {journal} {\bibinfo  {journal} {ACS Photonics}\ }\textbf {\bibinfo {volume} {5}},\ \bibinfo {pages} {3896} (\bibinfo {year} {2018})}\BibitemShut {NoStop}%
\bibitem [{\citenamefont {Li}\ \emph {et~al.}(2021)\citenamefont {Li}, \citenamefont {Li}, \citenamefont {Han}, \citenamefont {Zheng}, \citenamefont {Li}, \citenamefont {Li}, \citenamefont {Fan},\ and\ \citenamefont {Qiu}}]{li2021e}%
  \BibitemOpen
  \bibfield  {author} {\bibinfo {author} {\bibfnamefont {Y.}~\bibnamefont {Li}}, \bibinfo {author} {\bibfnamefont {W.}~\bibnamefont {Li}}, \bibinfo {author} {\bibfnamefont {T.}~\bibnamefont {Han}}, \bibinfo {author} {\bibfnamefont {X.}~\bibnamefont {Zheng}}, \bibinfo {author} {\bibfnamefont {J.}~\bibnamefont {Li}}, \bibinfo {author} {\bibfnamefont {B.}~\bibnamefont {Li}}, \bibinfo {author} {\bibfnamefont {S.}~\bibnamefont {Fan}},\ and\ \bibinfo {author} {\bibfnamefont {C.-W.}\ \bibnamefont {Qiu}},\ }\bibfield  {title} {\bibinfo {title} {Transforming heat transfer with thermal metamaterials and devices},\ }\href {https://doi.org/10.1038/s41578-021-00283-2} {\bibfield  {journal} {\bibinfo  {journal} {Nature Reviews Materials}\ }\textbf {\bibinfo {volume} {6}},\ \bibinfo {pages} {488} (\bibinfo {year} {2021})}\BibitemShut {NoStop}%
\bibitem [{\citenamefont {Miller}(2013{\natexlab{a}})}]{miller2013c}%
  \BibitemOpen
  \bibfield  {author} {\bibinfo {author} {\bibfnamefont {D.~A.~B.}\ \bibnamefont {Miller}},\ }\bibfield  {title} {\bibinfo {title} {Establishing {{Optimal Wave Communication Channels Automatically}}},\ }\href {https://doi.org/10.1109/JLT.2013.2278809} {\bibfield  {journal} {\bibinfo  {journal} {Journal of Lightwave Technology}\ }\textbf {\bibinfo {volume} {31}},\ \bibinfo {pages} {3987} (\bibinfo {year} {2013}{\natexlab{a}})}\BibitemShut {NoStop}%
\bibitem [{\citenamefont {Miller}(2019)}]{miller2019}%
  \BibitemOpen
  \bibfield  {author} {\bibinfo {author} {\bibfnamefont {D.~A.~B.}\ \bibnamefont {Miller}},\ }\bibfield  {title} {\bibinfo {title} {Waves, modes, communications, and optics: A tutorial},\ }\href {https://doi.org/10.1364/AOP.11.000679} {\bibfield  {journal} {\bibinfo  {journal} {Advances in Optics and Photonics}\ }\textbf {\bibinfo {volume} {11}},\ \bibinfo {pages} {679} (\bibinfo {year} {2019})}\BibitemShut {NoStop}%
\bibitem [{\citenamefont {SeyedinNavadeh}\ \emph {et~al.}(2024)\citenamefont {SeyedinNavadeh}, \citenamefont {Milanizadeh}, \citenamefont {Zanetto}, \citenamefont {Ferrari}, \citenamefont {Sampietro}, \citenamefont {Sorel}, \citenamefont {Miller}, \citenamefont {Melloni},\ and\ \citenamefont {Morichetti}}]{seyedinnavadeh2023}%
  \BibitemOpen
  \bibfield  {author} {\bibinfo {author} {\bibfnamefont {S.}~\bibnamefont {SeyedinNavadeh}}, \bibinfo {author} {\bibfnamefont {M.}~\bibnamefont {Milanizadeh}}, \bibinfo {author} {\bibfnamefont {F.}~\bibnamefont {Zanetto}}, \bibinfo {author} {\bibfnamefont {G.}~\bibnamefont {Ferrari}}, \bibinfo {author} {\bibfnamefont {M.}~\bibnamefont {Sampietro}}, \bibinfo {author} {\bibfnamefont {M.}~\bibnamefont {Sorel}}, \bibinfo {author} {\bibfnamefont {D.~A.~B.}\ \bibnamefont {Miller}}, \bibinfo {author} {\bibfnamefont {A.}~\bibnamefont {Melloni}},\ and\ \bibinfo {author} {\bibfnamefont {F.}~\bibnamefont {Morichetti}},\ }\bibfield  {title} {\bibinfo {title} {Determining the optimal communication channels of arbitrary optical systems using integrated photonic processors},\ }\href {https://doi.org/10.1038/s41566-023-01330-w} {\bibfield  {journal} {\bibinfo  {journal} {Nature Photonics}\ }\textbf {\bibinfo {volume} {18}},\ \bibinfo {pages} {149} (\bibinfo {year} {2024})}\BibitemShut {NoStop}%
\bibitem [{\citenamefont {Sebbah}(2001)}]{sebbah2001a}%
  \BibitemOpen
  \bibinfo {editor} {\bibfnamefont {P.}~\bibnamefont {Sebbah}},\ ed.,\ \href@noop {} {\emph {\bibinfo {title} {Waves and Imaging through Complex Media}}}\ (\bibinfo  {publisher} {{Kluwer Academic Publishers}},\ \bibinfo {address} {{Dordrecht ; Boston}},\ \bibinfo {year} {2001})\BibitemShut {NoStop}%
\bibitem [{\citenamefont {Kittel}\ \emph {et~al.}(2005)\citenamefont {Kittel}, \citenamefont {{M{\"u}ller-Hirsch}}, \citenamefont {Parisi}, \citenamefont {Biehs}, \citenamefont {Reddig},\ and\ \citenamefont {Holthaus}}]{kittel2005}%
  \BibitemOpen
  \bibfield  {author} {\bibinfo {author} {\bibfnamefont {A.}~\bibnamefont {Kittel}}, \bibinfo {author} {\bibfnamefont {W.}~\bibnamefont {{M{\"u}ller-Hirsch}}}, \bibinfo {author} {\bibfnamefont {J.}~\bibnamefont {Parisi}}, \bibinfo {author} {\bibfnamefont {S.-A.}\ \bibnamefont {Biehs}}, \bibinfo {author} {\bibfnamefont {D.}~\bibnamefont {Reddig}},\ and\ \bibinfo {author} {\bibfnamefont {M.}~\bibnamefont {Holthaus}},\ }\bibfield  {title} {\bibinfo {title} {Near-{{Field Heat Transfer}} in a {{Scanning Thermal Microscope}}},\ }\href@noop {} {\bibfield  {journal} {\bibinfo  {journal} {Physical Review Letters}\ }\textbf {\bibinfo {volume} {95}},\ \bibinfo {pages} {224301} (\bibinfo {year} {2005})}\BibitemShut {NoStop}%
\bibitem [{\citenamefont {Ntziachristos}(2010)}]{ntziachristos2010a}%
  \BibitemOpen
  \bibfield  {author} {\bibinfo {author} {\bibfnamefont {V.}~\bibnamefont {Ntziachristos}},\ }\bibfield  {title} {\bibinfo {title} {Going deeper than microscopy: The optical imaging frontier in biology},\ }\href {https://doi.org/10.1038/nmeth.1483} {\bibfield  {journal} {\bibinfo  {journal} {Nature Methods}\ }\textbf {\bibinfo {volume} {7}},\ \bibinfo {pages} {603} (\bibinfo {year} {2010})}\BibitemShut {NoStop}%
\bibitem [{\citenamefont {{\v C}i{\v z}m{\'a}r}\ and\ \citenamefont {Dholakia}(2012)}]{cizmar2012}%
  \BibitemOpen
  \bibfield  {author} {\bibinfo {author} {\bibfnamefont {T.}~\bibnamefont {{\v C}i{\v z}m{\'a}r}}\ and\ \bibinfo {author} {\bibfnamefont {K.}~\bibnamefont {Dholakia}},\ }\bibfield  {title} {\bibinfo {title} {Exploiting multimode waveguides for pure fibre-based imaging},\ }\href {https://doi.org/10.1038/ncomms2024} {\bibfield  {journal} {\bibinfo  {journal} {Nature Communications}\ }\textbf {\bibinfo {volume} {3}},\ \bibinfo {pages} {1027} (\bibinfo {year} {2012})}\BibitemShut {NoStop}%
\bibitem [{\citenamefont {Kang}\ \emph {et~al.}(2015)\citenamefont {Kang}, \citenamefont {Jeong}, \citenamefont {Choi}, \citenamefont {Ko}, \citenamefont {Yang}, \citenamefont {Joo}, \citenamefont {Lee}, \citenamefont {Lim}, \citenamefont {Park},\ and\ \citenamefont {Choi}}]{kang2015}%
  \BibitemOpen
  \bibfield  {author} {\bibinfo {author} {\bibfnamefont {S.}~\bibnamefont {Kang}}, \bibinfo {author} {\bibfnamefont {S.}~\bibnamefont {Jeong}}, \bibinfo {author} {\bibfnamefont {W.}~\bibnamefont {Choi}}, \bibinfo {author} {\bibfnamefont {H.}~\bibnamefont {Ko}}, \bibinfo {author} {\bibfnamefont {T.~D.}\ \bibnamefont {Yang}}, \bibinfo {author} {\bibfnamefont {J.~H.}\ \bibnamefont {Joo}}, \bibinfo {author} {\bibfnamefont {J.-S.}\ \bibnamefont {Lee}}, \bibinfo {author} {\bibfnamefont {Y.-S.}\ \bibnamefont {Lim}}, \bibinfo {author} {\bibfnamefont {Q.-H.}\ \bibnamefont {Park}},\ and\ \bibinfo {author} {\bibfnamefont {W.}~\bibnamefont {Choi}},\ }\bibfield  {title} {\bibinfo {title} {Imaging deep within a scattering medium using collective accumulation of single-scattered waves},\ }\href {https://doi.org/10.1038/nphoton.2015.24} {\bibfield  {journal} {\bibinfo  {journal} {Nature Photonics}\ }\textbf {\bibinfo {volume} {9}},\ \bibinfo {pages} {253} (\bibinfo {year} {2015})}\BibitemShut {NoStop}%
\bibitem [{\citenamefont {Guo}\ \emph {et~al.}(2018{\natexlab{a}})\citenamefont {Guo}, \citenamefont {Xiao}, \citenamefont {Minkov}, \citenamefont {Shi},\ and\ \citenamefont {Fan}}]{guo2018}%
  \BibitemOpen
  \bibfield  {author} {\bibinfo {author} {\bibfnamefont {C.}~\bibnamefont {Guo}}, \bibinfo {author} {\bibfnamefont {M.}~\bibnamefont {Xiao}}, \bibinfo {author} {\bibfnamefont {M.}~\bibnamefont {Minkov}}, \bibinfo {author} {\bibfnamefont {Y.}~\bibnamefont {Shi}},\ and\ \bibinfo {author} {\bibfnamefont {S.}~\bibnamefont {Fan}},\ }\bibfield  {title} {\bibinfo {title} {Photonic crystal slab {{Laplace}} operator for image differentiation},\ }\href {https://doi.org/10.1364/OPTICA.5.000251} {\bibfield  {journal} {\bibinfo  {journal} {Optica}\ }\textbf {\bibinfo {volume} {5}},\ \bibinfo {pages} {251} (\bibinfo {year} {2018}{\natexlab{a}})}\BibitemShut {NoStop}%
\bibitem [{\citenamefont {Guo}\ \emph {et~al.}(2018{\natexlab{b}})\citenamefont {Guo}, \citenamefont {Xiao}, \citenamefont {Minkov}, \citenamefont {Shi},\ and\ \citenamefont {Fan}}]{guo2018a}%
  \BibitemOpen
  \bibfield  {author} {\bibinfo {author} {\bibfnamefont {C.}~\bibnamefont {Guo}}, \bibinfo {author} {\bibfnamefont {M.}~\bibnamefont {Xiao}}, \bibinfo {author} {\bibfnamefont {M.}~\bibnamefont {Minkov}}, \bibinfo {author} {\bibfnamefont {Y.}~\bibnamefont {Shi}},\ and\ \bibinfo {author} {\bibfnamefont {S.}~\bibnamefont {Fan}},\ }\bibfield  {title} {\bibinfo {title} {Isotropic wavevector domain image filters by a photonic crystal slab device},\ }\href {https://doi.org/10.1364/JOSAA.35.001685} {\bibfield  {journal} {\bibinfo  {journal} {JOSA A}\ }\textbf {\bibinfo {volume} {35}},\ \bibinfo {pages} {1685} (\bibinfo {year} {2018}{\natexlab{b}})}\BibitemShut {NoStop}%
\bibitem [{\citenamefont {Yoon}\ \emph {et~al.}(2020)\citenamefont {Yoon}, \citenamefont {Kim}, \citenamefont {Jang}, \citenamefont {Choi}, \citenamefont {Choi}, \citenamefont {Kang},\ and\ \citenamefont {Choi}}]{yoon2020b}%
  \BibitemOpen
  \bibfield  {author} {\bibinfo {author} {\bibfnamefont {S.}~\bibnamefont {Yoon}}, \bibinfo {author} {\bibfnamefont {M.}~\bibnamefont {Kim}}, \bibinfo {author} {\bibfnamefont {M.}~\bibnamefont {Jang}}, \bibinfo {author} {\bibfnamefont {Y.}~\bibnamefont {Choi}}, \bibinfo {author} {\bibfnamefont {W.}~\bibnamefont {Choi}}, \bibinfo {author} {\bibfnamefont {S.}~\bibnamefont {Kang}},\ and\ \bibinfo {author} {\bibfnamefont {W.}~\bibnamefont {Choi}},\ }\bibfield  {title} {\bibinfo {title} {Deep optical imaging within complex scattering media},\ }\href {https://doi.org/10.1038/s42254-019-0143-2} {\bibfield  {journal} {\bibinfo  {journal} {Nature Reviews Physics}\ }\textbf {\bibinfo {volume} {2}},\ \bibinfo {pages} {141} (\bibinfo {year} {2020})}\BibitemShut {NoStop}%
\bibitem [{\citenamefont {Wang}\ \emph {et~al.}(2020{\natexlab{a}})\citenamefont {Wang}, \citenamefont {Guo}, \citenamefont {Zhao},\ and\ \citenamefont {Fan}}]{wang2020p}%
  \BibitemOpen
  \bibfield  {author} {\bibinfo {author} {\bibfnamefont {H.}~\bibnamefont {Wang}}, \bibinfo {author} {\bibfnamefont {C.}~\bibnamefont {Guo}}, \bibinfo {author} {\bibfnamefont {Z.}~\bibnamefont {Zhao}},\ and\ \bibinfo {author} {\bibfnamefont {S.}~\bibnamefont {Fan}},\ }\bibfield  {title} {\bibinfo {title} {Compact {{Incoherent Image Differentiation}} with {{Nanophotonic Structures}}},\ }\href {https://doi.org/10.1021/acsphotonics.9b01465} {\bibfield  {journal} {\bibinfo  {journal} {ACS Photonics}\ }\textbf {\bibinfo {volume} {7}},\ \bibinfo {pages} {338} (\bibinfo {year} {2020}{\natexlab{a}})}\BibitemShut {NoStop}%
\bibitem [{\citenamefont {Long}\ \emph {et~al.}(2021)\citenamefont {Long}, \citenamefont {Guo}, \citenamefont {Wang},\ and\ \citenamefont {Fan}}]{long2021}%
  \BibitemOpen
  \bibfield  {author} {\bibinfo {author} {\bibfnamefont {O.~Y.}\ \bibnamefont {Long}}, \bibinfo {author} {\bibfnamefont {C.}~\bibnamefont {Guo}}, \bibinfo {author} {\bibfnamefont {H.}~\bibnamefont {Wang}},\ and\ \bibinfo {author} {\bibfnamefont {S.}~\bibnamefont {Fan}},\ }\bibfield  {title} {\bibinfo {title} {Isotropic topological second-order spatial differentiator operating in transmission mode},\ }\href {https://doi.org/10.1364/OL.430699} {\bibfield  {journal} {\bibinfo  {journal} {Optics Letters}\ }\textbf {\bibinfo {volume} {46}},\ \bibinfo {pages} {3247} (\bibinfo {year} {2021})}\BibitemShut {NoStop}%
\bibitem [{\citenamefont {Bertolotti}\ and\ \citenamefont {Katz}(2022)}]{bertolotti2022}%
  \BibitemOpen
  \bibfield  {author} {\bibinfo {author} {\bibfnamefont {J.}~\bibnamefont {Bertolotti}}\ and\ \bibinfo {author} {\bibfnamefont {O.}~\bibnamefont {Katz}},\ }\bibfield  {title} {\bibinfo {title} {Imaging in complex media},\ }\href {https://doi.org/10.1038/s41567-022-01723-8} {\bibfield  {journal} {\bibinfo  {journal} {Nature Physics}\ }\textbf {\bibinfo {volume} {18}},\ \bibinfo {pages} {1008} (\bibinfo {year} {2022})}\BibitemShut {NoStop}%
\bibitem [{\citenamefont {Wang}\ \emph {et~al.}(2022)\citenamefont {Wang}, \citenamefont {Jin}, \citenamefont {Guo}, \citenamefont {Zhao}, \citenamefont {Rodrigues},\ and\ \citenamefont {Fan}}]{wang2022}%
  \BibitemOpen
  \bibfield  {author} {\bibinfo {author} {\bibfnamefont {H.}~\bibnamefont {Wang}}, \bibinfo {author} {\bibfnamefont {W.}~\bibnamefont {Jin}}, \bibinfo {author} {\bibfnamefont {C.}~\bibnamefont {Guo}}, \bibinfo {author} {\bibfnamefont {N.}~\bibnamefont {Zhao}}, \bibinfo {author} {\bibfnamefont {S.~P.}\ \bibnamefont {Rodrigues}},\ and\ \bibinfo {author} {\bibfnamefont {S.}~\bibnamefont {Fan}},\ }\bibfield  {title} {\bibinfo {title} {Design of {{Compact Meta-Crystal Slab}} for {{General Optical Convolution}}},\ }\href {https://doi.org/10.1021/acsphotonics.1c02005} {\bibfield  {journal} {\bibinfo  {journal} {ACS Photonics}\ }\textbf {\bibinfo {volume} {9}},\ \bibinfo {pages} {1358} (\bibinfo {year} {2022})}\BibitemShut {NoStop}%
\bibitem [{\citenamefont {Long}\ \emph {et~al.}(2022)\citenamefont {Long}, \citenamefont {Guo}, \citenamefont {Jin},\ and\ \citenamefont {Fan}}]{long2022}%
  \BibitemOpen
  \bibfield  {author} {\bibinfo {author} {\bibfnamefont {O.~Y.}\ \bibnamefont {Long}}, \bibinfo {author} {\bibfnamefont {C.}~\bibnamefont {Guo}}, \bibinfo {author} {\bibfnamefont {W.}~\bibnamefont {Jin}},\ and\ \bibinfo {author} {\bibfnamefont {S.}~\bibnamefont {Fan}},\ }\bibfield  {title} {\bibinfo {title} {Polarization-{{Independent Isotropic Nonlocal Metasurfaces}} with {{Wavelength-Controlled Functionality}}},\ }\href {https://doi.org/10.1103/PhysRevApplied.17.024029} {\bibfield  {journal} {\bibinfo  {journal} {Physical Review Applied}\ }\textbf {\bibinfo {volume} {17}},\ \bibinfo {pages} {024029} (\bibinfo {year} {2022})}\BibitemShut {NoStop}%
\bibitem [{\citenamefont {Aulbach}\ \emph {et~al.}(2011)\citenamefont {Aulbach}, \citenamefont {Gjonaj}, \citenamefont {Johnson}, \citenamefont {Mosk},\ and\ \citenamefont {Lagendijk}}]{aulbach2011}%
  \BibitemOpen
  \bibfield  {author} {\bibinfo {author} {\bibfnamefont {J.}~\bibnamefont {Aulbach}}, \bibinfo {author} {\bibfnamefont {B.}~\bibnamefont {Gjonaj}}, \bibinfo {author} {\bibfnamefont {P.~M.}\ \bibnamefont {Johnson}}, \bibinfo {author} {\bibfnamefont {A.~P.}\ \bibnamefont {Mosk}},\ and\ \bibinfo {author} {\bibfnamefont {A.}~\bibnamefont {Lagendijk}},\ }\bibfield  {title} {\bibinfo {title} {Control of {{Light Transmission}} through {{Opaque Scattering Media}} in {{Space}} and {{Time}}},\ }\href {https://doi.org/10.1103/physrevlett.106.103901} {\bibfield  {journal} {\bibinfo  {journal} {Physical Review Letters}\ }\textbf {\bibinfo {volume} {106}},\ \bibinfo {pages} {103901} (\bibinfo {year} {2011})}\BibitemShut {NoStop}%
\bibitem [{\citenamefont {Sarma}\ \emph {et~al.}(2015)\citenamefont {Sarma}, \citenamefont {Yamilov}, \citenamefont {Liew}, \citenamefont {Guy},\ and\ \citenamefont {Cao}}]{sarma2015}%
  \BibitemOpen
  \bibfield  {author} {\bibinfo {author} {\bibfnamefont {R.}~\bibnamefont {Sarma}}, \bibinfo {author} {\bibfnamefont {A.}~\bibnamefont {Yamilov}}, \bibinfo {author} {\bibfnamefont {S.~F.}\ \bibnamefont {Liew}}, \bibinfo {author} {\bibfnamefont {M.}~\bibnamefont {Guy}},\ and\ \bibinfo {author} {\bibfnamefont {H.}~\bibnamefont {Cao}},\ }\bibfield  {title} {\bibinfo {title} {Control of mesoscopic transport by modifying transmission channels in opaque media},\ }\href {https://doi.org/10.1103/PhysRevB.92.214206} {\bibfield  {journal} {\bibinfo  {journal} {Physical Review B}\ }\textbf {\bibinfo {volume} {92}},\ \bibinfo {pages} {214206} (\bibinfo {year} {2015})}\BibitemShut {NoStop}%
\bibitem [{\citenamefont {Mounaix}\ \emph {et~al.}(2016)\citenamefont {Mounaix}, \citenamefont {Andreoli}, \citenamefont {Defienne}, \citenamefont {Volpe}, \citenamefont {Katz}, \citenamefont {Gr{\'e}sillon},\ and\ \citenamefont {Gigan}}]{mounaix2016}%
  \BibitemOpen
  \bibfield  {author} {\bibinfo {author} {\bibfnamefont {M.}~\bibnamefont {Mounaix}}, \bibinfo {author} {\bibfnamefont {D.}~\bibnamefont {Andreoli}}, \bibinfo {author} {\bibfnamefont {H.}~\bibnamefont {Defienne}}, \bibinfo {author} {\bibfnamefont {G.}~\bibnamefont {Volpe}}, \bibinfo {author} {\bibfnamefont {O.}~\bibnamefont {Katz}}, \bibinfo {author} {\bibfnamefont {S.}~\bibnamefont {Gr{\'e}sillon}},\ and\ \bibinfo {author} {\bibfnamefont {S.}~\bibnamefont {Gigan}},\ }\bibfield  {title} {\bibinfo {title} {Spatiotemporal {{Coherent Control}} of {{Light}} through a {{Multiple Scattering Medium}} with the {{Multispectral Transmission Matrix}}},\ }\href {https://doi.org/10.1103/physrevlett.116.253901} {\bibfield  {journal} {\bibinfo  {journal} {Physical Review Letters}\ }\textbf {\bibinfo {volume} {116}},\ \bibinfo {pages} {253901} (\bibinfo {year} {2016})}\BibitemShut {NoStop}%
\bibitem [{\citenamefont {Jeong}\ \emph {et~al.}(2018)\citenamefont {Jeong}, \citenamefont {Lee}, \citenamefont {Choi}, \citenamefont {Kang}, \citenamefont {Hong}, \citenamefont {Park}, \citenamefont {Lim}, \citenamefont {Park},\ and\ \citenamefont {Choi}}]{jeong2018a}%
  \BibitemOpen
  \bibfield  {author} {\bibinfo {author} {\bibfnamefont {S.}~\bibnamefont {Jeong}}, \bibinfo {author} {\bibfnamefont {Y.-R.}\ \bibnamefont {Lee}}, \bibinfo {author} {\bibfnamefont {W.}~\bibnamefont {Choi}}, \bibinfo {author} {\bibfnamefont {S.}~\bibnamefont {Kang}}, \bibinfo {author} {\bibfnamefont {J.~H.}\ \bibnamefont {Hong}}, \bibinfo {author} {\bibfnamefont {J.-S.}\ \bibnamefont {Park}}, \bibinfo {author} {\bibfnamefont {Y.-S.}\ \bibnamefont {Lim}}, \bibinfo {author} {\bibfnamefont {H.-G.}\ \bibnamefont {Park}},\ and\ \bibinfo {author} {\bibfnamefont {W.}~\bibnamefont {Choi}},\ }\bibfield  {title} {\bibinfo {title} {Focusing of light energy inside a scattering medium by controlling the time-gated multiple light scattering},\ }\href {https://doi.org/10.1038/s41566-018-0120-9} {\bibfield  {journal} {\bibinfo  {journal} {Nature Photonics}\ }\textbf {\bibinfo {volume} {12}},\ \bibinfo {pages} {277} (\bibinfo {year} {2018})}\BibitemShut {NoStop}%
\bibitem [{\citenamefont {Muraviev}\ \emph {et~al.}(2018)\citenamefont {Muraviev}, \citenamefont {Smolski}, \citenamefont {Loparo},\ and\ \citenamefont {Vodopyanov}}]{muraviev2018}%
  \BibitemOpen
  \bibfield  {author} {\bibinfo {author} {\bibfnamefont {A.~V.}\ \bibnamefont {Muraviev}}, \bibinfo {author} {\bibfnamefont {V.~O.}\ \bibnamefont {Smolski}}, \bibinfo {author} {\bibfnamefont {Z.~E.}\ \bibnamefont {Loparo}},\ and\ \bibinfo {author} {\bibfnamefont {K.~L.}\ \bibnamefont {Vodopyanov}},\ }\bibfield  {title} {\bibinfo {title} {Massively parallel sensing of trace molecules and their isotopologues with broadband subharmonic mid-infrared frequency combs},\ }\href {https://doi.org/10.1038/s41566-018-0135-2} {\bibfield  {journal} {\bibinfo  {journal} {Nature Photonics}\ }\textbf {\bibinfo {volume} {12}},\ \bibinfo {pages} {209} (\bibinfo {year} {2018})}\BibitemShut {NoStop}%
\bibitem [{\citenamefont {Tan}\ \emph {et~al.}(2020)\citenamefont {Tan}, \citenamefont {Zhang}, \citenamefont {Li}, \citenamefont {Wan}, \citenamefont {Guo}, \citenamefont {Zhu}, \citenamefont {Liu},\ and\ \citenamefont {Yi}}]{tan2020}%
  \BibitemOpen
  \bibfield  {author} {\bibinfo {author} {\bibfnamefont {X.}~\bibnamefont {Tan}}, \bibinfo {author} {\bibfnamefont {H.}~\bibnamefont {Zhang}}, \bibinfo {author} {\bibfnamefont {J.}~\bibnamefont {Li}}, \bibinfo {author} {\bibfnamefont {H.}~\bibnamefont {Wan}}, \bibinfo {author} {\bibfnamefont {Q.}~\bibnamefont {Guo}}, \bibinfo {author} {\bibfnamefont {H.}~\bibnamefont {Zhu}}, \bibinfo {author} {\bibfnamefont {H.}~\bibnamefont {Liu}},\ and\ \bibinfo {author} {\bibfnamefont {F.}~\bibnamefont {Yi}},\ }\bibfield  {title} {\bibinfo {title} {Non-dispersive infrared multi-gas sensing via nanoantenna integrated narrowband detectors},\ }\href {https://doi.org/10.1038/s41467-020-19085-1} {\bibfield  {journal} {\bibinfo  {journal} {Nature Communications}\ }\textbf {\bibinfo {volume} {11}},\ \bibinfo {pages} {5245} (\bibinfo {year} {2020})}\BibitemShut {NoStop}%
\bibitem [{\citenamefont {Liu}\ and\ \citenamefont {Fiore}(2020)}]{liu2020s}%
  \BibitemOpen
  \bibfield  {author} {\bibinfo {author} {\bibfnamefont {T.}~\bibnamefont {Liu}}\ and\ \bibinfo {author} {\bibfnamefont {A.}~\bibnamefont {Fiore}},\ }\bibfield  {title} {\bibinfo {title} {Designing open channels in random scattering media for on-chip spectrometers},\ }\href {https://doi.org/10.1364/optica.391612} {\bibfield  {journal} {\bibinfo  {journal} {Optica}\ }\textbf {\bibinfo {volume} {7}},\ \bibinfo {pages} {934} (\bibinfo {year} {2020})}\BibitemShut {NoStop}%
\bibitem [{\citenamefont {Greffet}\ \emph {et~al.}(2002)\citenamefont {Greffet}, \citenamefont {Carminati}, \citenamefont {Joulain}, \citenamefont {Mulet}, \citenamefont {Mainguy},\ and\ \citenamefont {Chen}}]{greffet2002a}%
  \BibitemOpen
  \bibfield  {author} {\bibinfo {author} {\bibfnamefont {J.-J.}\ \bibnamefont {Greffet}}, \bibinfo {author} {\bibfnamefont {R.}~\bibnamefont {Carminati}}, \bibinfo {author} {\bibfnamefont {K.}~\bibnamefont {Joulain}}, \bibinfo {author} {\bibfnamefont {J.-P.}\ \bibnamefont {Mulet}}, \bibinfo {author} {\bibfnamefont {S.}~\bibnamefont {Mainguy}},\ and\ \bibinfo {author} {\bibfnamefont {Y.}~\bibnamefont {Chen}},\ }\bibfield  {title} {\bibinfo {title} {Coherent emission of light by thermal sources},\ }\href@noop {} {\bibfield  {journal} {\bibinfo  {journal} {Nature}\ }\textbf {\bibinfo {volume} {416}},\ \bibinfo {pages} {61} (\bibinfo {year} {2002})}\BibitemShut {NoStop}%
\bibitem [{\citenamefont {Guo}\ \emph {et~al.}(2012)\citenamefont {Guo}, \citenamefont {Cortes}, \citenamefont {Molesky},\ and\ \citenamefont {Jacob}}]{guo2012b}%
  \BibitemOpen
  \bibfield  {author} {\bibinfo {author} {\bibfnamefont {Y.}~\bibnamefont {Guo}}, \bibinfo {author} {\bibfnamefont {C.~L.}\ \bibnamefont {Cortes}}, \bibinfo {author} {\bibfnamefont {S.}~\bibnamefont {Molesky}},\ and\ \bibinfo {author} {\bibfnamefont {Z.}~\bibnamefont {Jacob}},\ }\bibfield  {title} {\bibinfo {title} {Broadband super-{{Planckian}} thermal emission from hyperbolic metamaterials},\ }\href@noop {} {\bibfield  {journal} {\bibinfo  {journal} {Applied Physics Letters}\ }\textbf {\bibinfo {volume} {101}},\ \bibinfo {pages} {131106} (\bibinfo {year} {2012})}\BibitemShut {NoStop}%
\bibitem [{\citenamefont {De~Zoysa}\ \emph {et~al.}(2012)\citenamefont {De~Zoysa}, \citenamefont {Asano}, \citenamefont {Mochizuki}, \citenamefont {Oskooi}, \citenamefont {Inoue},\ and\ \citenamefont {Noda}}]{dezoysa2012}%
  \BibitemOpen
  \bibfield  {author} {\bibinfo {author} {\bibfnamefont {M.}~\bibnamefont {De~Zoysa}}, \bibinfo {author} {\bibfnamefont {T.}~\bibnamefont {Asano}}, \bibinfo {author} {\bibfnamefont {K.}~\bibnamefont {Mochizuki}}, \bibinfo {author} {\bibfnamefont {A.}~\bibnamefont {Oskooi}}, \bibinfo {author} {\bibfnamefont {T.}~\bibnamefont {Inoue}},\ and\ \bibinfo {author} {\bibfnamefont {S.}~\bibnamefont {Noda}},\ }\bibfield  {title} {\bibinfo {title} {Conversion of broadband to narrowband thermal emission through energy recycling},\ }\href@noop {} {\bibfield  {journal} {\bibinfo  {journal} {Nature Photonics}\ }\textbf {\bibinfo {volume} {6}},\ \bibinfo {pages} {535} (\bibinfo {year} {2012})}\BibitemShut {NoStop}%
\bibitem [{\citenamefont {Yu}\ \emph {et~al.}(2013{\natexlab{a}})\citenamefont {Yu}, \citenamefont {Sergeant}, \citenamefont {Skauli}, \citenamefont {Zhang}, \citenamefont {Wang},\ and\ \citenamefont {Fan}}]{yu2013}%
  \BibitemOpen
  \bibfield  {author} {\bibinfo {author} {\bibfnamefont {Z.}~\bibnamefont {Yu}}, \bibinfo {author} {\bibfnamefont {N.~P.}\ \bibnamefont {Sergeant}}, \bibinfo {author} {\bibfnamefont {T.}~\bibnamefont {Skauli}}, \bibinfo {author} {\bibfnamefont {G.}~\bibnamefont {Zhang}}, \bibinfo {author} {\bibfnamefont {H.}~\bibnamefont {Wang}},\ and\ \bibinfo {author} {\bibfnamefont {S.}~\bibnamefont {Fan}},\ }\bibfield  {title} {\bibinfo {title} {Enhancing far-field thermal emission with thermal extraction},\ }\href@noop {} {\bibfield  {journal} {\bibinfo  {journal} {Nature Communications}\ }\textbf {\bibinfo {volume} {4}},\ \bibinfo {pages} {1730} (\bibinfo {year} {2013}{\natexlab{a}})}\BibitemShut {NoStop}%
\bibitem [{\citenamefont {Pelton}(2015)}]{pelton2015}%
  \BibitemOpen
  \bibfield  {author} {\bibinfo {author} {\bibfnamefont {M.}~\bibnamefont {Pelton}},\ }\bibfield  {title} {\bibinfo {title} {Modified spontaneous emission in nanophotonic structures},\ }\href@noop {} {\bibfield  {journal} {\bibinfo  {journal} {Nature Photonics}\ }\textbf {\bibinfo {volume} {9}},\ \bibinfo {pages} {427} (\bibinfo {year} {2015})}\BibitemShut {NoStop}%
\bibitem [{\citenamefont {Thompson}\ \emph {et~al.}(2018)\citenamefont {Thompson}, \citenamefont {Zhu}, \citenamefont {Mittapally}, \citenamefont {Sadat}, \citenamefont {Xing}, \citenamefont {McArdle}, \citenamefont {Qazilbash}, \citenamefont {Reddy},\ and\ \citenamefont {Meyhofer}}]{thompson2018}%
  \BibitemOpen
  \bibfield  {author} {\bibinfo {author} {\bibfnamefont {D.}~\bibnamefont {Thompson}}, \bibinfo {author} {\bibfnamefont {L.}~\bibnamefont {Zhu}}, \bibinfo {author} {\bibfnamefont {R.}~\bibnamefont {Mittapally}}, \bibinfo {author} {\bibfnamefont {S.}~\bibnamefont {Sadat}}, \bibinfo {author} {\bibfnamefont {Z.}~\bibnamefont {Xing}}, \bibinfo {author} {\bibfnamefont {P.}~\bibnamefont {McArdle}}, \bibinfo {author} {\bibfnamefont {M.~M.}\ \bibnamefont {Qazilbash}}, \bibinfo {author} {\bibfnamefont {P.}~\bibnamefont {Reddy}},\ and\ \bibinfo {author} {\bibfnamefont {E.}~\bibnamefont {Meyhofer}},\ }\bibfield  {title} {\bibinfo {title} {Hundred-fold enhancement in far-field radiative heat transfer over the blackbody limit},\ }\href@noop {} {\bibfield  {journal} {\bibinfo  {journal} {Nature}\ }\textbf {\bibinfo {volume} {561}},\ \bibinfo {pages} {216} (\bibinfo {year} {2018})}\BibitemShut {NoStop}%
\bibitem [{\citenamefont {Baranov}\ \emph {et~al.}(2019)\citenamefont {Baranov}, \citenamefont {Xiao}, \citenamefont {Nechepurenko}, \citenamefont {Krasnok}, \citenamefont {Al{\`u}},\ and\ \citenamefont {Kats}}]{baranov2019a}%
  \BibitemOpen
  \bibfield  {author} {\bibinfo {author} {\bibfnamefont {D.~G.}\ \bibnamefont {Baranov}}, \bibinfo {author} {\bibfnamefont {Y.}~\bibnamefont {Xiao}}, \bibinfo {author} {\bibfnamefont {I.~A.}\ \bibnamefont {Nechepurenko}}, \bibinfo {author} {\bibfnamefont {A.}~\bibnamefont {Krasnok}}, \bibinfo {author} {\bibfnamefont {A.}~\bibnamefont {Al{\`u}}},\ and\ \bibinfo {author} {\bibfnamefont {M.~A.}\ \bibnamefont {Kats}},\ }\bibfield  {title} {\bibinfo {title} {Nanophotonic engineering of far-field thermal emitters},\ }\href@noop {} {\bibfield  {journal} {\bibinfo  {journal} {Nature Materials}\ }\textbf {\bibinfo {volume} {18}},\ \bibinfo {pages} {920} (\bibinfo {year} {2019})}\BibitemShut {NoStop}%
\bibitem [{\citenamefont {Guo}\ \emph {et~al.}(2021{\natexlab{a}})\citenamefont {Guo}, \citenamefont {Guo}, \citenamefont {Lou},\ and\ \citenamefont {Fan}}]{guo2021a}%
  \BibitemOpen
  \bibfield  {author} {\bibinfo {author} {\bibfnamefont {C.}~\bibnamefont {Guo}}, \bibinfo {author} {\bibfnamefont {Y.}~\bibnamefont {Guo}}, \bibinfo {author} {\bibfnamefont {B.}~\bibnamefont {Lou}},\ and\ \bibinfo {author} {\bibfnamefont {S.}~\bibnamefont {Fan}},\ }\bibfield  {title} {\bibinfo {title} {Wide wavelength-tunable narrow-band thermal radiation from moir\'e patterns},\ }\href@noop {} {\bibfield  {journal} {\bibinfo  {journal} {Applied Physics Letters}\ }\textbf {\bibinfo {volume} {118}},\ \bibinfo {pages} {131111} (\bibinfo {year} {2021}{\natexlab{a}})}\BibitemShut {NoStop}%
\bibitem [{\citenamefont {Guo}\ \emph {et~al.}(2023)\citenamefont {Guo}, \citenamefont {Li}, \citenamefont {Xiao},\ and\ \citenamefont {Fan}}]{guo2023c}%
  \BibitemOpen
  \bibfield  {author} {\bibinfo {author} {\bibfnamefont {C.}~\bibnamefont {Guo}}, \bibinfo {author} {\bibfnamefont {J.}~\bibnamefont {Li}}, \bibinfo {author} {\bibfnamefont {M.}~\bibnamefont {Xiao}},\ and\ \bibinfo {author} {\bibfnamefont {S.}~\bibnamefont {Fan}},\ }\bibfield  {title} {\bibinfo {title} {Singular topology of scattering matrices},\ }\href {https://doi.org/10.1103/PhysRevB.108.155418} {\bibfield  {journal} {\bibinfo  {journal} {Physical Review B}\ }\textbf {\bibinfo {volume} {108}},\ \bibinfo {pages} {155418} (\bibinfo {year} {2023})}\BibitemShut {NoStop}%
\bibitem [{\citenamefont {Veselago}(1968)}]{veselago1968}%
  \BibitemOpen
  \bibfield  {author} {\bibinfo {author} {\bibfnamefont {V.~G.}\ \bibnamefont {Veselago}},\ }\bibfield  {title} {\bibinfo {title} {{{THE ELECTRODYNAMICS OF SUBSTANCES WITH SIMULTANEOUSLY NEGATIVE VALUES OF AND}} {$\mu$}},\ }\href {https://doi.org/10.1070/PU1968v010n04ABEH003699} {\bibfield  {journal} {\bibinfo  {journal} {Soviet Physics Uspekhi}\ }\textbf {\bibinfo {volume} {10}},\ \bibinfo {pages} {509} (\bibinfo {year} {1968})}\BibitemShut {NoStop}%
\bibitem [{\citenamefont {Pendry}(2000)}]{pendry2000}%
  \BibitemOpen
  \bibfield  {author} {\bibinfo {author} {\bibfnamefont {J.~B.}\ \bibnamefont {Pendry}},\ }\bibfield  {title} {\bibinfo {title} {Negative refraction makes a perfect lens},\ }\href {https://doi.org/10.1103/PhysRevLett.85.3966} {\bibfield  {journal} {\bibinfo  {journal} {Physical Review Letters}\ }\textbf {\bibinfo {volume} {85}},\ \bibinfo {pages} {3966} (\bibinfo {year} {2000})}\BibitemShut {NoStop}%
\bibitem [{\citenamefont {Smith}\ \emph {et~al.}(2000)\citenamefont {Smith}, \citenamefont {Padilla}, \citenamefont {Vier}, \citenamefont {{Nemat-Nasser}},\ and\ \citenamefont {Schultz}}]{smith2000}%
  \BibitemOpen
  \bibfield  {author} {\bibinfo {author} {\bibfnamefont {D.~R.}\ \bibnamefont {Smith}}, \bibinfo {author} {\bibfnamefont {W.~J.}\ \bibnamefont {Padilla}}, \bibinfo {author} {\bibfnamefont {D.~C.}\ \bibnamefont {Vier}}, \bibinfo {author} {\bibfnamefont {S.~C.}\ \bibnamefont {{Nemat-Nasser}}},\ and\ \bibinfo {author} {\bibfnamefont {S.}~\bibnamefont {Schultz}},\ }\bibfield  {title} {\bibinfo {title} {Composite {{Medium}} with {{Simultaneously Negative Permeability}} and {{Permittivity}}},\ }\href {https://doi.org/10.1103/PhysRevLett.84.4184} {\bibfield  {journal} {\bibinfo  {journal} {Physical Review Letters}\ }\textbf {\bibinfo {volume} {84}},\ \bibinfo {pages} {4184} (\bibinfo {year} {2000})}\BibitemShut {NoStop}%
\bibitem [{\citenamefont {Shelby}\ \emph {et~al.}(2001)\citenamefont {Shelby}, \citenamefont {Smith},\ and\ \citenamefont {Schultz}}]{shelby2001}%
  \BibitemOpen
  \bibfield  {author} {\bibinfo {author} {\bibfnamefont {R.~A.}\ \bibnamefont {Shelby}}, \bibinfo {author} {\bibfnamefont {D.~R.}\ \bibnamefont {Smith}},\ and\ \bibinfo {author} {\bibfnamefont {S.}~\bibnamefont {Schultz}},\ }\bibfield  {title} {\bibinfo {title} {Experimental {{Verification}} of a {{Negative Index}} of {{Refraction}}},\ }\href {https://doi.org/10.1126/science.1058847} {\bibfield  {journal} {\bibinfo  {journal} {Science}\ }\textbf {\bibinfo {volume} {292}},\ \bibinfo {pages} {77} (\bibinfo {year} {2001})}\BibitemShut {NoStop}%
\bibitem [{\citenamefont {Fan}\ and\ \citenamefont {Kahn}(2005)}]{fan2005}%
  \BibitemOpen
  \bibfield  {author} {\bibinfo {author} {\bibfnamefont {S.}~\bibnamefont {Fan}}\ and\ \bibinfo {author} {\bibfnamefont {J.~M.}\ \bibnamefont {Kahn}},\ }\bibfield  {title} {\bibinfo {title} {Principal modes in multimode waveguides},\ }\href {https://doi.org/10.1364/OL.30.000135} {\bibfield  {journal} {\bibinfo  {journal} {Optics Letters}\ }\textbf {\bibinfo {volume} {30}},\ \bibinfo {pages} {135} (\bibinfo {year} {2005})}\BibitemShut {NoStop}%
\bibitem [{\citenamefont {Valentine}\ \emph {et~al.}(2008)\citenamefont {Valentine}, \citenamefont {Zhang}, \citenamefont {Zentgraf}, \citenamefont {{Ulin-Avila}}, \citenamefont {Genov}, \citenamefont {Bartal},\ and\ \citenamefont {Zhang}}]{valentine2008}%
  \BibitemOpen
  \bibfield  {author} {\bibinfo {author} {\bibfnamefont {J.}~\bibnamefont {Valentine}}, \bibinfo {author} {\bibfnamefont {S.}~\bibnamefont {Zhang}}, \bibinfo {author} {\bibfnamefont {T.}~\bibnamefont {Zentgraf}}, \bibinfo {author} {\bibfnamefont {E.}~\bibnamefont {{Ulin-Avila}}}, \bibinfo {author} {\bibfnamefont {D.~A.}\ \bibnamefont {Genov}}, \bibinfo {author} {\bibfnamefont {G.}~\bibnamefont {Bartal}},\ and\ \bibinfo {author} {\bibfnamefont {X.}~\bibnamefont {Zhang}},\ }\bibfield  {title} {\bibinfo {title} {Three-dimensional optical metamaterial with a negative refractive index},\ }\href {https://doi.org/10.1038/nature07247} {\bibfield  {journal} {\bibinfo  {journal} {Nature}\ }\textbf {\bibinfo {volume} {455}},\ \bibinfo {pages} {376} (\bibinfo {year} {2008})}\BibitemShut {NoStop}%
\bibitem [{\citenamefont {Zhu}\ \emph {et~al.}(2021)\citenamefont {Zhu}, \citenamefont {Guo}, \citenamefont {Huang}, \citenamefont {Wang}, \citenamefont {Orenstein}, \citenamefont {Ruan},\ and\ \citenamefont {Fan}}]{zhu2021}%
  \BibitemOpen
  \bibfield  {author} {\bibinfo {author} {\bibfnamefont {T.}~\bibnamefont {Zhu}}, \bibinfo {author} {\bibfnamefont {C.}~\bibnamefont {Guo}}, \bibinfo {author} {\bibfnamefont {J.}~\bibnamefont {Huang}}, \bibinfo {author} {\bibfnamefont {H.}~\bibnamefont {Wang}}, \bibinfo {author} {\bibfnamefont {M.}~\bibnamefont {Orenstein}}, \bibinfo {author} {\bibfnamefont {Z.}~\bibnamefont {Ruan}},\ and\ \bibinfo {author} {\bibfnamefont {S.}~\bibnamefont {Fan}},\ }\bibfield  {title} {\bibinfo {title} {Topological optical differentiator},\ }\href {https://doi.org/10.1038/s41467-021-20972-4} {\bibfield  {journal} {\bibinfo  {journal} {Nature Communications}\ }\textbf {\bibinfo {volume} {12}},\ \bibinfo {pages} {680} (\bibinfo {year} {2021})}\BibitemShut {NoStop}%
\bibitem [{\citenamefont {Guo}\ \emph {et~al.}(2020)\citenamefont {Guo}, \citenamefont {Wang},\ and\ \citenamefont {Fan}}]{guo2020a}%
  \BibitemOpen
  \bibfield  {author} {\bibinfo {author} {\bibfnamefont {C.}~\bibnamefont {Guo}}, \bibinfo {author} {\bibfnamefont {H.}~\bibnamefont {Wang}},\ and\ \bibinfo {author} {\bibfnamefont {S.}~\bibnamefont {Fan}},\ }\bibfield  {title} {\bibinfo {title} {Squeeze free space with nonlocal flat optics},\ }\href {https://doi.org/10.1364/OPTICA.392978} {\bibfield  {journal} {\bibinfo  {journal} {Optica}\ }\textbf {\bibinfo {volume} {7}},\ \bibinfo {pages} {1133} (\bibinfo {year} {2020})},\ \Eprint {https://arxiv.org/abs/2003.06918} {arxiv:2003.06918} \BibitemShut {NoStop}%
\bibitem [{\citenamefont {Reshef}\ \emph {et~al.}(2021)\citenamefont {Reshef}, \citenamefont {DelMastro}, \citenamefont {Bearne}, \citenamefont {Alhulaymi}, \citenamefont {Giner}, \citenamefont {Boyd},\ and\ \citenamefont {Lundeen}}]{reshef2021}%
  \BibitemOpen
  \bibfield  {author} {\bibinfo {author} {\bibfnamefont {O.}~\bibnamefont {Reshef}}, \bibinfo {author} {\bibfnamefont {M.~P.}\ \bibnamefont {DelMastro}}, \bibinfo {author} {\bibfnamefont {K.~K.~M.}\ \bibnamefont {Bearne}}, \bibinfo {author} {\bibfnamefont {A.~H.}\ \bibnamefont {Alhulaymi}}, \bibinfo {author} {\bibfnamefont {L.}~\bibnamefont {Giner}}, \bibinfo {author} {\bibfnamefont {R.~W.}\ \bibnamefont {Boyd}},\ and\ \bibinfo {author} {\bibfnamefont {J.~S.}\ \bibnamefont {Lundeen}},\ }\bibfield  {title} {\bibinfo {title} {An optic to replace space and its application towards ultra-thin imaging systems},\ }\href {https://doi.org/10.1038/s41467-021-23358-8} {\bibfield  {journal} {\bibinfo  {journal} {Nature Communications}\ }\textbf {\bibinfo {volume} {12}},\ \bibinfo {pages} {3512} (\bibinfo {year} {2021})}\BibitemShut {NoStop}%
\bibitem [{\citenamefont {Long}\ \emph {et~al.}(2023)\citenamefont {Long}, \citenamefont {Guo},\ and\ \citenamefont {Fan}}]{long2023}%
  \BibitemOpen
  \bibfield  {author} {\bibinfo {author} {\bibfnamefont {O.~Y.}\ \bibnamefont {Long}}, \bibinfo {author} {\bibfnamefont {C.}~\bibnamefont {Guo}},\ and\ \bibinfo {author} {\bibfnamefont {S.}~\bibnamefont {Fan}},\ }\bibfield  {title} {\bibinfo {title} {Topological nature of non-{{Hermitian}} degenerate bands in structural parameter space},\ }\href {https://doi.org/10.1103/PhysRevApplied.20.L051001} {\bibfield  {journal} {\bibinfo  {journal} {Physical Review Applied}\ }\textbf {\bibinfo {volume} {20}},\ \bibinfo {pages} {L051001} (\bibinfo {year} {2023})}\BibitemShut {NoStop}%
\bibitem [{\citenamefont {Guo}\ \emph {et~al.}(2021{\natexlab{b}})\citenamefont {Guo}, \citenamefont {Xiao}, \citenamefont {Orenstein},\ and\ \citenamefont {Fan}}]{guo2021c}%
  \BibitemOpen
  \bibfield  {author} {\bibinfo {author} {\bibfnamefont {C.}~\bibnamefont {Guo}}, \bibinfo {author} {\bibfnamefont {M.}~\bibnamefont {Xiao}}, \bibinfo {author} {\bibfnamefont {M.}~\bibnamefont {Orenstein}},\ and\ \bibinfo {author} {\bibfnamefont {S.}~\bibnamefont {Fan}},\ }\bibfield  {title} {\bibinfo {title} {Structured {{3D}} linear space\textendash time light bullets by nonlocal nanophotonics},\ }\href {https://doi.org/10.1038/s41377-021-00595-6} {\bibfield  {journal} {\bibinfo  {journal} {Light: Science \& Applications}\ }\textbf {\bibinfo {volume} {10}},\ \bibinfo {pages} {160} (\bibinfo {year} {2021}{\natexlab{b}})}\BibitemShut {NoStop}%
\bibitem [{\citenamefont {Popoff}\ \emph {et~al.}(2014)\citenamefont {Popoff}, \citenamefont {Goetschy}, \citenamefont {Liew}, \citenamefont {Stone},\ and\ \citenamefont {Cao}}]{popoff2014}%
  \BibitemOpen
  \bibfield  {author} {\bibinfo {author} {\bibfnamefont {S.~M.}\ \bibnamefont {Popoff}}, \bibinfo {author} {\bibfnamefont {A.}~\bibnamefont {Goetschy}}, \bibinfo {author} {\bibfnamefont {S.~F.}\ \bibnamefont {Liew}}, \bibinfo {author} {\bibfnamefont {A.~D.}\ \bibnamefont {Stone}},\ and\ \bibinfo {author} {\bibfnamefont {H.}~\bibnamefont {Cao}},\ }\bibfield  {title} {\bibinfo {title} {Coherent {{Control}} of {{Total Transmission}} of {{Light}} through {{Disordered Media}}},\ }\href {https://doi.org/10.1103/physrevlett.112.133903} {\bibfield  {journal} {\bibinfo  {journal} {Physical Review Letters}\ }\textbf {\bibinfo {volume} {112}},\ \bibinfo {pages} {133903} (\bibinfo {year} {2014})}\BibitemShut {NoStop}%
\bibitem [{\citenamefont {Liew}\ \emph {et~al.}(2016)\citenamefont {Liew}, \citenamefont {Popoff}, \citenamefont {Sheehan}, \citenamefont {Goetschy}, \citenamefont {Schmuttenmaer}, \citenamefont {Stone},\ and\ \citenamefont {Cao}}]{liew2016}%
  \BibitemOpen
  \bibfield  {author} {\bibinfo {author} {\bibfnamefont {S.~F.}\ \bibnamefont {Liew}}, \bibinfo {author} {\bibfnamefont {S.~M.}\ \bibnamefont {Popoff}}, \bibinfo {author} {\bibfnamefont {S.~W.}\ \bibnamefont {Sheehan}}, \bibinfo {author} {\bibfnamefont {A.}~\bibnamefont {Goetschy}}, \bibinfo {author} {\bibfnamefont {C.~A.}\ \bibnamefont {Schmuttenmaer}}, \bibinfo {author} {\bibfnamefont {A.~D.}\ \bibnamefont {Stone}},\ and\ \bibinfo {author} {\bibfnamefont {H.}~\bibnamefont {Cao}},\ }\bibfield  {title} {\bibinfo {title} {Coherent {{Control}} of {{Photocurrent}} in a {{Strongly Scattering Photoelectrochemical System}}},\ }\href {https://doi.org/10.1021/acsphotonics.5b00642} {\bibfield  {journal} {\bibinfo  {journal} {ACS Photonics}\ }\textbf {\bibinfo {volume} {3}},\ \bibinfo {pages} {449} (\bibinfo {year} {2016})}\BibitemShut {NoStop}%
\bibitem [{\citenamefont {Vellekoop}\ and\ \citenamefont {Mosk}(2007)}]{vellekoop2007}%
  \BibitemOpen
  \bibfield  {author} {\bibinfo {author} {\bibfnamefont {I.~M.}\ \bibnamefont {Vellekoop}}\ and\ \bibinfo {author} {\bibfnamefont {A.~P.}\ \bibnamefont {Mosk}},\ }\bibfield  {title} {\bibinfo {title} {Focusing coherent light through opaque strongly scattering media},\ }\href {https://doi.org/10.1364/OL.32.002309} {\bibfield  {journal} {\bibinfo  {journal} {Optics Letters}\ }\textbf {\bibinfo {volume} {32}},\ \bibinfo {pages} {2309} (\bibinfo {year} {2007})}\BibitemShut {NoStop}%
\bibitem [{\citenamefont {Yu}\ \emph {et~al.}(2017)\citenamefont {Yu}, \citenamefont {Lee},\ and\ \citenamefont {Park}}]{yu2017e}%
  \BibitemOpen
  \bibfield  {author} {\bibinfo {author} {\bibfnamefont {H.}~\bibnamefont {Yu}}, \bibinfo {author} {\bibfnamefont {K.}~\bibnamefont {Lee}},\ and\ \bibinfo {author} {\bibfnamefont {Y.}~\bibnamefont {Park}},\ }\bibfield  {title} {\bibinfo {title} {Ultrahigh enhancement of light focusing through disordered media controlled by mega-pixel modes},\ }\href {https://doi.org/10.1364/oe.25.008036} {\bibfield  {journal} {\bibinfo  {journal} {Optics Express}\ }\textbf {\bibinfo {volume} {25}},\ \bibinfo {pages} {8036} (\bibinfo {year} {2017})}\BibitemShut {NoStop}%
\bibitem [{\citenamefont {Sweeney}\ \emph {et~al.}(2020)\citenamefont {Sweeney}, \citenamefont {Hsu},\ and\ \citenamefont {Stone}}]{sweeney2020a}%
  \BibitemOpen
  \bibfield  {author} {\bibinfo {author} {\bibfnamefont {W.~R.}\ \bibnamefont {Sweeney}}, \bibinfo {author} {\bibfnamefont {C.~W.}\ \bibnamefont {Hsu}},\ and\ \bibinfo {author} {\bibfnamefont {A.~D.}\ \bibnamefont {Stone}},\ }\bibfield  {title} {\bibinfo {title} {Theory of reflectionless scattering modes},\ }\href {https://doi.org/10.1103/PhysRevA.102.063511} {\bibfield  {journal} {\bibinfo  {journal} {Physical Review A}\ }\textbf {\bibinfo {volume} {102}},\ \bibinfo {pages} {063511} (\bibinfo {year} {2020})}\BibitemShut {NoStop}%
\bibitem [{\citenamefont {Stone}\ \emph {et~al.}(2021)\citenamefont {Stone}, \citenamefont {Sweeney}, \citenamefont {Hsu}, \citenamefont {Wisal},\ and\ \citenamefont {Wang}}]{stone2021}%
  \BibitemOpen
  \bibfield  {author} {\bibinfo {author} {\bibfnamefont {A.~D.}\ \bibnamefont {Stone}}, \bibinfo {author} {\bibfnamefont {W.~R.}\ \bibnamefont {Sweeney}}, \bibinfo {author} {\bibfnamefont {C.~W.}\ \bibnamefont {Hsu}}, \bibinfo {author} {\bibfnamefont {K.}~\bibnamefont {Wisal}},\ and\ \bibinfo {author} {\bibfnamefont {Z.}~\bibnamefont {Wang}},\ }\bibfield  {title} {\bibinfo {title} {Reflectionless excitation of arbitrary photonic structures: A general theory},\ }\href {https://doi.org/10.1515/nanoph-2020-0403} {\bibfield  {journal} {\bibinfo  {journal} {Nanophotonics}\ }\textbf {\bibinfo {volume} {10}},\ \bibinfo {pages} {343} (\bibinfo {year} {2021})}\BibitemShut {NoStop}%
\bibitem [{\citenamefont {Horodynski}\ \emph {et~al.}(2022)\citenamefont {Horodynski}, \citenamefont {K{\"u}hmayer}, \citenamefont {Ferise}, \citenamefont {Rotter},\ and\ \citenamefont {Davy}}]{horodynski2022}%
  \BibitemOpen
  \bibfield  {author} {\bibinfo {author} {\bibfnamefont {M.}~\bibnamefont {Horodynski}}, \bibinfo {author} {\bibfnamefont {M.}~\bibnamefont {K{\"u}hmayer}}, \bibinfo {author} {\bibfnamefont {C.}~\bibnamefont {Ferise}}, \bibinfo {author} {\bibfnamefont {S.}~\bibnamefont {Rotter}},\ and\ \bibinfo {author} {\bibfnamefont {M.}~\bibnamefont {Davy}},\ }\bibfield  {title} {\bibinfo {title} {Anti-reflection structure for perfect transmission through complex media},\ }\href {https://doi.org/10.1038/s41586-022-04843-6} {\bibfield  {journal} {\bibinfo  {journal} {Nature}\ }\textbf {\bibinfo {volume} {607}},\ \bibinfo {pages} {281} (\bibinfo {year} {2022})}\BibitemShut {NoStop}%
\bibitem [{\citenamefont {Berdagu{\'e}}\ and\ \citenamefont {Facq}(1982)}]{berdague1982}%
  \BibitemOpen
  \bibfield  {author} {\bibinfo {author} {\bibfnamefont {S.}~\bibnamefont {Berdagu{\'e}}}\ and\ \bibinfo {author} {\bibfnamefont {P.}~\bibnamefont {Facq}},\ }\bibfield  {title} {\bibinfo {title} {Mode division multiplexing in optical fibers},\ }\href {https://doi.org/10.1364/AO.21.001950} {\bibfield  {journal} {\bibinfo  {journal} {Applied Optics}\ }\textbf {\bibinfo {volume} {21}},\ \bibinfo {pages} {1950} (\bibinfo {year} {1982})}\BibitemShut {NoStop}%
\bibitem [{\citenamefont {Luo}\ \emph {et~al.}(2014)\citenamefont {Luo}, \citenamefont {Ophir}, \citenamefont {Chen}, \citenamefont {Gabrielli}, \citenamefont {Poitras}, \citenamefont {Bergmen},\ and\ \citenamefont {Lipson}}]{luo2014}%
  \BibitemOpen
  \bibfield  {author} {\bibinfo {author} {\bibfnamefont {L.-W.}\ \bibnamefont {Luo}}, \bibinfo {author} {\bibfnamefont {N.}~\bibnamefont {Ophir}}, \bibinfo {author} {\bibfnamefont {C.~P.}\ \bibnamefont {Chen}}, \bibinfo {author} {\bibfnamefont {L.~H.}\ \bibnamefont {Gabrielli}}, \bibinfo {author} {\bibfnamefont {C.~B.}\ \bibnamefont {Poitras}}, \bibinfo {author} {\bibfnamefont {K.}~\bibnamefont {Bergmen}},\ and\ \bibinfo {author} {\bibfnamefont {M.}~\bibnamefont {Lipson}},\ }\bibfield  {title} {\bibinfo {title} {{{WDM-compatible}} mode-division multiplexing on a silicon chip},\ }\href {https://doi.org/10.1038/ncomms4069} {\bibfield  {journal} {\bibinfo  {journal} {Nature Communications}\ }\textbf {\bibinfo {volume} {5}},\ \bibinfo {pages} {3069} (\bibinfo {year} {2014})}\BibitemShut {NoStop}%
\bibitem [{\citenamefont {Su}\ \emph {et~al.}(2021)\citenamefont {Su}, \citenamefont {He}, \citenamefont {Chen}, \citenamefont {Li},\ and\ \citenamefont {Li}}]{su2021}%
  \BibitemOpen
  \bibfield  {author} {\bibinfo {author} {\bibfnamefont {Y.}~\bibnamefont {Su}}, \bibinfo {author} {\bibfnamefont {Y.}~\bibnamefont {He}}, \bibinfo {author} {\bibfnamefont {H.}~\bibnamefont {Chen}}, \bibinfo {author} {\bibfnamefont {X.}~\bibnamefont {Li}},\ and\ \bibinfo {author} {\bibfnamefont {G.}~\bibnamefont {Li}},\ }\bibfield  {title} {\bibinfo {title} {Perspective on mode-division multiplexing},\ }\href {https://doi.org/10.1063/5.0046071} {\bibfield  {journal} {\bibinfo  {journal} {Applied Physics Letters}\ }\textbf {\bibinfo {volume} {118}},\ \bibinfo {pages} {200502} (\bibinfo {year} {2021})}\BibitemShut {NoStop}%
\bibitem [{\citenamefont {Reck}\ \emph {et~al.}(1994)\citenamefont {Reck}, \citenamefont {Zeilinger}, \citenamefont {Bernstein},\ and\ \citenamefont {Bertani}}]{reck1994}%
  \BibitemOpen
  \bibfield  {author} {\bibinfo {author} {\bibfnamefont {M.}~\bibnamefont {Reck}}, \bibinfo {author} {\bibfnamefont {A.}~\bibnamefont {Zeilinger}}, \bibinfo {author} {\bibfnamefont {H.~J.}\ \bibnamefont {Bernstein}},\ and\ \bibinfo {author} {\bibfnamefont {P.}~\bibnamefont {Bertani}},\ }\bibfield  {title} {\bibinfo {title} {Experimental realization of any discrete unitary operator},\ }\href {https://doi.org/10.1103/PhysRevLett.73.58} {\bibfield  {journal} {\bibinfo  {journal} {Physical Review Letters}\ }\textbf {\bibinfo {volume} {73}},\ \bibinfo {pages} {58} (\bibinfo {year} {1994})}\BibitemShut {NoStop}%
\bibitem [{\citenamefont {Miller}(2013{\natexlab{b}})}]{miller2013a}%
  \BibitemOpen
  \bibfield  {author} {\bibinfo {author} {\bibfnamefont {D.~A.~B.}\ \bibnamefont {Miller}},\ }\bibfield  {title} {\bibinfo {title} {Self-aligning universal beam coupler},\ }\href {https://doi.org/10.1364/OE.21.006360} {\bibfield  {journal} {\bibinfo  {journal} {Optics Express}\ }\textbf {\bibinfo {volume} {21}},\ \bibinfo {pages} {6360} (\bibinfo {year} {2013}{\natexlab{b}})}\BibitemShut {NoStop}%
\bibitem [{\citenamefont {Miller}(2013{\natexlab{c}})}]{miller2013b}%
  \BibitemOpen
  \bibfield  {author} {\bibinfo {author} {\bibfnamefont {D.~A.~B.}\ \bibnamefont {Miller}},\ }\bibfield  {title} {\bibinfo {title} {Self-configuring universal linear optical component [{{Invited}}]},\ }\href {https://doi.org/10.1364/PRJ.1.000001} {\bibfield  {journal} {\bibinfo  {journal} {Photonics Research}\ }\textbf {\bibinfo {volume} {1}},\ \bibinfo {pages} {1} (\bibinfo {year} {2013}{\natexlab{c}})},\ \Eprint {https://arxiv.org/abs/1303.4602} {arxiv:1303.4602} \BibitemShut {NoStop}%
\bibitem [{\citenamefont {Carolan}\ \emph {et~al.}(2015)\citenamefont {Carolan}, \citenamefont {Harrold}, \citenamefont {Sparrow}, \citenamefont {{Mart{\'i}n-L{\'o}pez}}, \citenamefont {Russell}, \citenamefont {Silverstone}, \citenamefont {Shadbolt}, \citenamefont {Matsuda}, \citenamefont {Oguma}, \citenamefont {Itoh}, \citenamefont {Marshall}, \citenamefont {Thompson}, \citenamefont {Matthews}, \citenamefont {Hashimoto}, \citenamefont {O'Brien},\ and\ \citenamefont {Laing}}]{carolan2015}%
  \BibitemOpen
  \bibfield  {author} {\bibinfo {author} {\bibfnamefont {J.}~\bibnamefont {Carolan}}, \bibinfo {author} {\bibfnamefont {C.}~\bibnamefont {Harrold}}, \bibinfo {author} {\bibfnamefont {C.}~\bibnamefont {Sparrow}}, \bibinfo {author} {\bibfnamefont {E.}~\bibnamefont {{Mart{\'i}n-L{\'o}pez}}}, \bibinfo {author} {\bibfnamefont {N.~J.}\ \bibnamefont {Russell}}, \bibinfo {author} {\bibfnamefont {J.~W.}\ \bibnamefont {Silverstone}}, \bibinfo {author} {\bibfnamefont {P.~J.}\ \bibnamefont {Shadbolt}}, \bibinfo {author} {\bibfnamefont {N.}~\bibnamefont {Matsuda}}, \bibinfo {author} {\bibfnamefont {M.}~\bibnamefont {Oguma}}, \bibinfo {author} {\bibfnamefont {M.}~\bibnamefont {Itoh}}, \bibinfo {author} {\bibfnamefont {G.~D.}\ \bibnamefont {Marshall}}, \bibinfo {author} {\bibfnamefont {M.~G.}\ \bibnamefont {Thompson}}, \bibinfo {author} {\bibfnamefont {J.~C.~F.}\ \bibnamefont {Matthews}}, \bibinfo {author} {\bibfnamefont {T.}~\bibnamefont {Hashimoto}}, \bibinfo {author} {\bibfnamefont {J.~L.}\ \bibnamefont {O'Brien}},\ and\
  \bibinfo {author} {\bibfnamefont {A.}~\bibnamefont {Laing}},\ }\bibfield  {title} {\bibinfo {title} {Universal linear optics},\ }\href {https://doi.org/10.1126/science.aab3642} {\bibfield  {journal} {\bibinfo  {journal} {Science}\ }\textbf {\bibinfo {volume} {349}},\ \bibinfo {pages} {711} (\bibinfo {year} {2015})}\BibitemShut {NoStop}%
\bibitem [{\citenamefont {Miller}(2015)}]{miller2015}%
  \BibitemOpen
  \bibfield  {author} {\bibinfo {author} {\bibfnamefont {D.~A.~B.}\ \bibnamefont {Miller}},\ }\bibfield  {title} {\bibinfo {title} {Perfect optics with imperfect components},\ }\href {https://doi.org/10.1364/OPTICA.2.000747} {\bibfield  {journal} {\bibinfo  {journal} {Optica}\ }\textbf {\bibinfo {volume} {2}},\ \bibinfo {pages} {747} (\bibinfo {year} {2015})}\BibitemShut {NoStop}%
\bibitem [{\citenamefont {Clements}\ \emph {et~al.}(2016)\citenamefont {Clements}, \citenamefont {Humphreys}, \citenamefont {Metcalf}, \citenamefont {Kolthammer},\ and\ \citenamefont {Walmsley}}]{clements2016}%
  \BibitemOpen
  \bibfield  {author} {\bibinfo {author} {\bibfnamefont {W.~R.}\ \bibnamefont {Clements}}, \bibinfo {author} {\bibfnamefont {P.~C.}\ \bibnamefont {Humphreys}}, \bibinfo {author} {\bibfnamefont {B.~J.}\ \bibnamefont {Metcalf}}, \bibinfo {author} {\bibfnamefont {W.~S.}\ \bibnamefont {Kolthammer}},\ and\ \bibinfo {author} {\bibfnamefont {I.~A.}\ \bibnamefont {Walmsley}},\ }\bibfield  {title} {\bibinfo {title} {Optimal design for universal multiport interferometers},\ }\href {https://doi.org/10.1364/OPTICA.3.001460} {\bibfield  {journal} {\bibinfo  {journal} {Optica}\ }\textbf {\bibinfo {volume} {3}},\ \bibinfo {pages} {1460} (\bibinfo {year} {2016})}\BibitemShut {NoStop}%
\bibitem [{\citenamefont {Ribeiro}\ \emph {et~al.}(2016)\citenamefont {Ribeiro}, \citenamefont {Ruocco}, \citenamefont {Vanacker},\ and\ \citenamefont {Bogaerts}}]{ribeiro2016}%
  \BibitemOpen
  \bibfield  {author} {\bibinfo {author} {\bibfnamefont {A.}~\bibnamefont {Ribeiro}}, \bibinfo {author} {\bibfnamefont {A.}~\bibnamefont {Ruocco}}, \bibinfo {author} {\bibfnamefont {L.}~\bibnamefont {Vanacker}},\ and\ \bibinfo {author} {\bibfnamefont {W.}~\bibnamefont {Bogaerts}},\ }\bibfield  {title} {\bibinfo {title} {Demonstration of a 4\,{\texttimes}\,4-port universal linear circuit},\ }\href {https://doi.org/10.1364/OPTICA.3.001348} {\bibfield  {journal} {\bibinfo  {journal} {Optica}\ }\textbf {\bibinfo {volume} {3}},\ \bibinfo {pages} {1348} (\bibinfo {year} {2016})}\BibitemShut {NoStop}%
\bibitem [{\citenamefont {Wilkes}\ \emph {et~al.}(2016)\citenamefont {Wilkes}, \citenamefont {Qiang}, \citenamefont {Wang}, \citenamefont {Santagati}, \citenamefont {Paesani}, \citenamefont {Zhou}, \citenamefont {Miller}, \citenamefont {Marshall}, \citenamefont {Thompson},\ and\ \citenamefont {O'Brien}}]{wilkes2016}%
  \BibitemOpen
  \bibfield  {author} {\bibinfo {author} {\bibfnamefont {C.~M.}\ \bibnamefont {Wilkes}}, \bibinfo {author} {\bibfnamefont {X.}~\bibnamefont {Qiang}}, \bibinfo {author} {\bibfnamefont {J.}~\bibnamefont {Wang}}, \bibinfo {author} {\bibfnamefont {R.}~\bibnamefont {Santagati}}, \bibinfo {author} {\bibfnamefont {S.}~\bibnamefont {Paesani}}, \bibinfo {author} {\bibfnamefont {X.}~\bibnamefont {Zhou}}, \bibinfo {author} {\bibfnamefont {D.~a.~B.}\ \bibnamefont {Miller}}, \bibinfo {author} {\bibfnamefont {G.~D.}\ \bibnamefont {Marshall}}, \bibinfo {author} {\bibfnamefont {M.~G.}\ \bibnamefont {Thompson}},\ and\ \bibinfo {author} {\bibfnamefont {J.~L.}\ \bibnamefont {O'Brien}},\ }\bibfield  {title} {\bibinfo {title} {60 {{dB}} high-extinction auto-configured {{Mach}}{\textendash}{{Zehnder}} interferometer},\ }\href {https://doi.org/10.1364/OL.41.005318} {\bibfield  {journal} {\bibinfo  {journal} {Optics Letters}\ }\textbf {\bibinfo {volume} {41}},\ \bibinfo {pages} {5318} (\bibinfo {year} {2016})}\BibitemShut {NoStop}%
\bibitem [{\citenamefont {Annoni}\ \emph {et~al.}(2017)\citenamefont {Annoni}, \citenamefont {Guglielmi}, \citenamefont {Carminati}, \citenamefont {Ferrari}, \citenamefont {Sampietro}, \citenamefont {Miller}, \citenamefont {Melloni},\ and\ \citenamefont {Morichetti}}]{annoni2017}%
  \BibitemOpen
  \bibfield  {author} {\bibinfo {author} {\bibfnamefont {A.}~\bibnamefont {Annoni}}, \bibinfo {author} {\bibfnamefont {E.}~\bibnamefont {Guglielmi}}, \bibinfo {author} {\bibfnamefont {M.}~\bibnamefont {Carminati}}, \bibinfo {author} {\bibfnamefont {G.}~\bibnamefont {Ferrari}}, \bibinfo {author} {\bibfnamefont {M.}~\bibnamefont {Sampietro}}, \bibinfo {author} {\bibfnamefont {D.~A.}\ \bibnamefont {Miller}}, \bibinfo {author} {\bibfnamefont {A.}~\bibnamefont {Melloni}},\ and\ \bibinfo {author} {\bibfnamefont {F.}~\bibnamefont {Morichetti}},\ }\bibfield  {title} {\bibinfo {title} {Unscrambling light{\textemdash}automatically undoing strong mixing between modes},\ }\href {https://doi.org/10.1038/lsa.2017.110} {\bibfield  {journal} {\bibinfo  {journal} {Light: Science \& Applications}\ }\textbf {\bibinfo {volume} {6}},\ \bibinfo {pages} {e17110} (\bibinfo {year} {2017})}\BibitemShut {NoStop}%
\bibitem [{\citenamefont {Miller}(2017)}]{miller2017d}%
  \BibitemOpen
  \bibfield  {author} {\bibinfo {author} {\bibfnamefont {D.~A.~B.}\ \bibnamefont {Miller}},\ }\bibfield  {title} {\bibinfo {title} {Setting up meshes of interferometers {\textendash} reversed local light interference method},\ }\href {https://doi.org/10.1364/OE.25.029233} {\bibfield  {journal} {\bibinfo  {journal} {Optics Express}\ }\textbf {\bibinfo {volume} {25}},\ \bibinfo {pages} {29233} (\bibinfo {year} {2017})}\BibitemShut {NoStop}%
\bibitem [{\citenamefont {Perez}\ \emph {et~al.}(2017)\citenamefont {Perez}, \citenamefont {Gasulla}, \citenamefont {Fraile}, \citenamefont {Crudgington}, \citenamefont {Thomson}, \citenamefont {Khokhar}, \citenamefont {Li}, \citenamefont {Cao}, \citenamefont {Mashanovich},\ and\ \citenamefont {Capmany}}]{perez2017}%
  \BibitemOpen
  \bibfield  {author} {\bibinfo {author} {\bibfnamefont {D.}~\bibnamefont {Perez}}, \bibinfo {author} {\bibfnamefont {I.}~\bibnamefont {Gasulla}}, \bibinfo {author} {\bibfnamefont {F.~J.}\ \bibnamefont {Fraile}}, \bibinfo {author} {\bibfnamefont {L.}~\bibnamefont {Crudgington}}, \bibinfo {author} {\bibfnamefont {D.~J.}\ \bibnamefont {Thomson}}, \bibinfo {author} {\bibfnamefont {A.~Z.}\ \bibnamefont {Khokhar}}, \bibinfo {author} {\bibfnamefont {K.}~\bibnamefont {Li}}, \bibinfo {author} {\bibfnamefont {W.}~\bibnamefont {Cao}}, \bibinfo {author} {\bibfnamefont {G.~Z.}\ \bibnamefont {Mashanovich}},\ and\ \bibinfo {author} {\bibfnamefont {J.}~\bibnamefont {Capmany}},\ }\bibfield  {title} {\bibinfo {title} {Silicon {{Photonics Rectangular Universal Interferometer}}},\ }\href {https://doi.org/10.1002/lpor.201700219} {\bibfield  {journal} {\bibinfo  {journal} {Laser \& Photonics Reviews}\ }\textbf {\bibinfo {volume} {11}},\ \bibinfo {pages} {1700219} (\bibinfo {year} {2017})}\BibitemShut {NoStop}%
\bibitem [{\citenamefont {Harris}\ \emph {et~al.}(2018)\citenamefont {Harris}, \citenamefont {Carolan}, \citenamefont {Bunandar}, \citenamefont {Prabhu}, \citenamefont {Hochberg}, \citenamefont {{Baehr-Jones}}, \citenamefont {Fanto}, \citenamefont {Smith}, \citenamefont {Tison}, \citenamefont {Alsing},\ and\ \citenamefont {Englund}}]{harris2018}%
  \BibitemOpen
  \bibfield  {author} {\bibinfo {author} {\bibfnamefont {N.~C.}\ \bibnamefont {Harris}}, \bibinfo {author} {\bibfnamefont {J.}~\bibnamefont {Carolan}}, \bibinfo {author} {\bibfnamefont {D.}~\bibnamefont {Bunandar}}, \bibinfo {author} {\bibfnamefont {M.}~\bibnamefont {Prabhu}}, \bibinfo {author} {\bibfnamefont {M.}~\bibnamefont {Hochberg}}, \bibinfo {author} {\bibfnamefont {T.}~\bibnamefont {{Baehr-Jones}}}, \bibinfo {author} {\bibfnamefont {M.~L.}\ \bibnamefont {Fanto}}, \bibinfo {author} {\bibfnamefont {A.~M.}\ \bibnamefont {Smith}}, \bibinfo {author} {\bibfnamefont {C.~C.}\ \bibnamefont {Tison}}, \bibinfo {author} {\bibfnamefont {P.~M.}\ \bibnamefont {Alsing}},\ and\ \bibinfo {author} {\bibfnamefont {D.}~\bibnamefont {Englund}},\ }\bibfield  {title} {\bibinfo {title} {Linear programmable nanophotonic processors},\ }\href {https://doi.org/10.1364/OPTICA.5.001623} {\bibfield  {journal} {\bibinfo  {journal} {Optica}\ }\textbf {\bibinfo {volume} {5}},\ \bibinfo {pages} {1623} (\bibinfo {year}
  {2018})}\BibitemShut {NoStop}%
\bibitem [{\citenamefont {Pai}\ \emph {et~al.}(2019)\citenamefont {Pai}, \citenamefont {Bartlett}, \citenamefont {Solgaard},\ and\ \citenamefont {Miller}}]{pai2019}%
  \BibitemOpen
  \bibfield  {author} {\bibinfo {author} {\bibfnamefont {S.}~\bibnamefont {Pai}}, \bibinfo {author} {\bibfnamefont {B.}~\bibnamefont {Bartlett}}, \bibinfo {author} {\bibfnamefont {O.}~\bibnamefont {Solgaard}},\ and\ \bibinfo {author} {\bibfnamefont {D.~A.~B.}\ \bibnamefont {Miller}},\ }\bibfield  {title} {\bibinfo {title} {Matrix {{Optimization}} on {{Universal Unitary Photonic Devices}}},\ }\href {https://doi.org/10.1103/physrevapplied.11.064044} {\bibfield  {journal} {\bibinfo  {journal} {Physical Review Applied}\ }\textbf {\bibinfo {volume} {11}},\ \bibinfo {pages} {064044} (\bibinfo {year} {2019})}\BibitemShut {NoStop}%
\bibitem [{\citenamefont {Morizur}\ \emph {et~al.}(2010)\citenamefont {Morizur}, \citenamefont {Nicholls}, \citenamefont {Jian}, \citenamefont {Armstrong}, \citenamefont {Treps}, \citenamefont {Hage}, \citenamefont {Hsu}, \citenamefont {Bowen}, \citenamefont {Janousek},\ and\ \citenamefont {Bachor}}]{morizur2010}%
  \BibitemOpen
  \bibfield  {author} {\bibinfo {author} {\bibfnamefont {J.-F.}\ \bibnamefont {Morizur}}, \bibinfo {author} {\bibfnamefont {L.}~\bibnamefont {Nicholls}}, \bibinfo {author} {\bibfnamefont {P.}~\bibnamefont {Jian}}, \bibinfo {author} {\bibfnamefont {S.}~\bibnamefont {Armstrong}}, \bibinfo {author} {\bibfnamefont {N.}~\bibnamefont {Treps}}, \bibinfo {author} {\bibfnamefont {B.}~\bibnamefont {Hage}}, \bibinfo {author} {\bibfnamefont {M.}~\bibnamefont {Hsu}}, \bibinfo {author} {\bibfnamefont {W.}~\bibnamefont {Bowen}}, \bibinfo {author} {\bibfnamefont {J.}~\bibnamefont {Janousek}},\ and\ \bibinfo {author} {\bibfnamefont {H.-A.}\ \bibnamefont {Bachor}},\ }\bibfield  {title} {\bibinfo {title} {Programmable unitary spatial mode manipulation},\ }\href {https://doi.org/10.1364/JOSAA.27.002524} {\bibfield  {journal} {\bibinfo  {journal} {JOSA A}\ }\textbf {\bibinfo {volume} {27}},\ \bibinfo {pages} {2524} (\bibinfo {year} {2010})}\BibitemShut {NoStop}%
\bibitem [{\citenamefont {Labroille}\ \emph {et~al.}(2014)\citenamefont {Labroille}, \citenamefont {Denolle}, \citenamefont {Jian}, \citenamefont {Genevaux}, \citenamefont {Treps},\ and\ \citenamefont {Morizur}}]{labroille2014}%
  \BibitemOpen
  \bibfield  {author} {\bibinfo {author} {\bibfnamefont {G.}~\bibnamefont {Labroille}}, \bibinfo {author} {\bibfnamefont {B.}~\bibnamefont {Denolle}}, \bibinfo {author} {\bibfnamefont {P.}~\bibnamefont {Jian}}, \bibinfo {author} {\bibfnamefont {P.}~\bibnamefont {Genevaux}}, \bibinfo {author} {\bibfnamefont {N.}~\bibnamefont {Treps}},\ and\ \bibinfo {author} {\bibfnamefont {J.-F.}\ \bibnamefont {Morizur}},\ }\bibfield  {title} {\bibinfo {title} {Efficient and mode selective spatial mode multiplexer based on multi-plane light conversion},\ }\href {https://doi.org/10.1364/OE.22.015599} {\bibfield  {journal} {\bibinfo  {journal} {Optics Express}\ }\textbf {\bibinfo {volume} {22}},\ \bibinfo {pages} {15599} (\bibinfo {year} {2014})}\BibitemShut {NoStop}%
\bibitem [{\citenamefont {Tanomura}\ \emph {et~al.}(2022)\citenamefont {Tanomura}, \citenamefont {Taguchi}, \citenamefont {Tang}, \citenamefont {Tanemura},\ and\ \citenamefont {Nakano}}]{tanomura2022}%
  \BibitemOpen
  \bibfield  {author} {\bibinfo {author} {\bibfnamefont {R.}~\bibnamefont {Tanomura}}, \bibinfo {author} {\bibfnamefont {Y.}~\bibnamefont {Taguchi}}, \bibinfo {author} {\bibfnamefont {R.}~\bibnamefont {Tang}}, \bibinfo {author} {\bibfnamefont {T.}~\bibnamefont {Tanemura}},\ and\ \bibinfo {author} {\bibfnamefont {Y.}~\bibnamefont {Nakano}},\ }\bibfield  {title} {\bibinfo {title} {Entropy of {{Mode Mixers}} for {{Optical Unitary Converter}} based on {{Multi-Plane Light Conversion}}},\ }in\ \href {https://doi.org/10.1364/CLEOPR.2022.CWP13A_02} {\emph {\bibinfo {booktitle} {Proceedings of the 2022 {{Conference}} on {{Lasers}} and {{Electro-Optics Pacific Rim}} (2022), Paper {{CWP13A}}\_02}}}\ (\bibinfo  {publisher} {{Optica Publishing Group}},\ \bibinfo {year} {2022})\ p.\ \bibinfo {pages} {CWP13A\_02}\BibitemShut {NoStop}%
\bibitem [{\citenamefont {Kupianskyi}\ \emph {et~al.}(2023)\citenamefont {Kupianskyi}, \citenamefont {Horsley},\ and\ \citenamefont {Phillips}}]{kupianskyi2023}%
  \BibitemOpen
  \bibfield  {author} {\bibinfo {author} {\bibfnamefont {H.}~\bibnamefont {Kupianskyi}}, \bibinfo {author} {\bibfnamefont {S.~A.~R.}\ \bibnamefont {Horsley}},\ and\ \bibinfo {author} {\bibfnamefont {D.~B.}\ \bibnamefont {Phillips}},\ }\bibfield  {title} {\bibinfo {title} {High-dimensional spatial mode sorting and optical circuit design using multi-plane light conversion},\ }\href {https://doi.org/10.1063/5.0128431} {\bibfield  {journal} {\bibinfo  {journal} {APL Photonics}\ }\textbf {\bibinfo {volume} {8}},\ \bibinfo {pages} {026101} (\bibinfo {year} {2023})}\BibitemShut {NoStop}%
\bibitem [{\citenamefont {Taguchi}\ \emph {et~al.}(2023)\citenamefont {Taguchi}, \citenamefont {Wang}, \citenamefont {Tanomura}, \citenamefont {Tanemura},\ and\ \citenamefont {Ozeki}}]{taguchi2023}%
  \BibitemOpen
  \bibfield  {author} {\bibinfo {author} {\bibfnamefont {Y.}~\bibnamefont {Taguchi}}, \bibinfo {author} {\bibfnamefont {Y.}~\bibnamefont {Wang}}, \bibinfo {author} {\bibfnamefont {R.}~\bibnamefont {Tanomura}}, \bibinfo {author} {\bibfnamefont {T.}~\bibnamefont {Tanemura}},\ and\ \bibinfo {author} {\bibfnamefont {Y.}~\bibnamefont {Ozeki}},\ }\bibfield  {title} {\bibinfo {title} {Iterative {{Configuration}} of {{Programmable Unitary Converter Based}} on {{Few-Layer Redundant Multiplane Light Conversion}}},\ }\href {https://doi.org/10.1103/PhysRevApplied.19.054002} {\bibfield  {journal} {\bibinfo  {journal} {Physical Review Applied}\ }\textbf {\bibinfo {volume} {19}},\ \bibinfo {pages} {054002} (\bibinfo {year} {2023})}\BibitemShut {NoStop}%
\bibitem [{\citenamefont {Zhang}\ and\ \citenamefont {Fontaine}(2023)}]{zhang2023b}%
  \BibitemOpen
  \bibfield  {author} {\bibinfo {author} {\bibfnamefont {Y.}~\bibnamefont {Zhang}}\ and\ \bibinfo {author} {\bibfnamefont {N.~K.}\ \bibnamefont {Fontaine}},\ }\href {https://doi.org/10.48550/arXiv.2304.11323} {\bibinfo {title} {Multi-{{Plane Light Conversion}}: {{A Practical Tutorial}}}} (\bibinfo {year} {2023}),\ \Eprint {https://arxiv.org/abs/2304.11323} {arXiv:2304.11323 [physics]} \BibitemShut {NoStop}%
\bibitem [{\citenamefont {Carolan}\ \emph {et~al.}(2020)\citenamefont {Carolan}, \citenamefont {Mohseni}, \citenamefont {Olson}, \citenamefont {Prabhu}, \citenamefont {Chen}, \citenamefont {Bunandar}, \citenamefont {Niu}, \citenamefont {Harris}, \citenamefont {Wong}, \citenamefont {Hochberg}, \citenamefont {Lloyd},\ and\ \citenamefont {Englund}}]{carolan2020}%
  \BibitemOpen
  \bibfield  {author} {\bibinfo {author} {\bibfnamefont {J.}~\bibnamefont {Carolan}}, \bibinfo {author} {\bibfnamefont {M.}~\bibnamefont {Mohseni}}, \bibinfo {author} {\bibfnamefont {J.~P.}\ \bibnamefont {Olson}}, \bibinfo {author} {\bibfnamefont {M.}~\bibnamefont {Prabhu}}, \bibinfo {author} {\bibfnamefont {C.}~\bibnamefont {Chen}}, \bibinfo {author} {\bibfnamefont {D.}~\bibnamefont {Bunandar}}, \bibinfo {author} {\bibfnamefont {M.~Y.}\ \bibnamefont {Niu}}, \bibinfo {author} {\bibfnamefont {N.~C.}\ \bibnamefont {Harris}}, \bibinfo {author} {\bibfnamefont {F.~N.~C.}\ \bibnamefont {Wong}}, \bibinfo {author} {\bibfnamefont {M.}~\bibnamefont {Hochberg}}, \bibinfo {author} {\bibfnamefont {S.}~\bibnamefont {Lloyd}},\ and\ \bibinfo {author} {\bibfnamefont {D.}~\bibnamefont {Englund}},\ }\bibfield  {title} {\bibinfo {title} {Variational quantum unsampling on a quantum photonic processor},\ }\href {https://doi.org/10.1038/s41567-019-0747-6} {\bibfield  {journal} {\bibinfo  {journal} {Nature Physics}\ }\textbf
  {\bibinfo {volume} {16}},\ \bibinfo {pages} {322} (\bibinfo {year} {2020})}\BibitemShut {NoStop}%
\bibitem [{\citenamefont {Elshaari}\ \emph {et~al.}(2020)\citenamefont {Elshaari}, \citenamefont {Pernice}, \citenamefont {Srinivasan}, \citenamefont {Benson},\ and\ \citenamefont {Zwiller}}]{elshaari2020}%
  \BibitemOpen
  \bibfield  {author} {\bibinfo {author} {\bibfnamefont {A.~W.}\ \bibnamefont {Elshaari}}, \bibinfo {author} {\bibfnamefont {W.}~\bibnamefont {Pernice}}, \bibinfo {author} {\bibfnamefont {K.}~\bibnamefont {Srinivasan}}, \bibinfo {author} {\bibfnamefont {O.}~\bibnamefont {Benson}},\ and\ \bibinfo {author} {\bibfnamefont {V.}~\bibnamefont {Zwiller}},\ }\bibfield  {title} {\bibinfo {title} {Hybrid integrated quantum photonic circuits},\ }\href {https://doi.org/10.1038/s41566-020-0609-x} {\bibfield  {journal} {\bibinfo  {journal} {Nature Photonics}\ }\textbf {\bibinfo {volume} {14}},\ \bibinfo {pages} {285} (\bibinfo {year} {2020})}\BibitemShut {NoStop}%
\bibitem [{\citenamefont {Wang}\ \emph {et~al.}(2020{\natexlab{b}})\citenamefont {Wang}, \citenamefont {Sciarrino}, \citenamefont {Laing},\ and\ \citenamefont {Thompson}}]{wang2020aw}%
  \BibitemOpen
  \bibfield  {author} {\bibinfo {author} {\bibfnamefont {J.}~\bibnamefont {Wang}}, \bibinfo {author} {\bibfnamefont {F.}~\bibnamefont {Sciarrino}}, \bibinfo {author} {\bibfnamefont {A.}~\bibnamefont {Laing}},\ and\ \bibinfo {author} {\bibfnamefont {M.~G.}\ \bibnamefont {Thompson}},\ }\bibfield  {title} {\bibinfo {title} {Integrated photonic quantum technologies},\ }\href {https://doi.org/10.1038/s41566-019-0532-1} {\bibfield  {journal} {\bibinfo  {journal} {Nature Photonics}\ }\textbf {\bibinfo {volume} {14}},\ \bibinfo {pages} {273} (\bibinfo {year} {2020}{\natexlab{b}})}\BibitemShut {NoStop}%
\bibitem [{\citenamefont {Chi}\ \emph {et~al.}(2022)\citenamefont {Chi}, \citenamefont {Huang}, \citenamefont {Zhang}, \citenamefont {Mao}, \citenamefont {Zhou}, \citenamefont {Chen}, \citenamefont {Zhai}, \citenamefont {Bao}, \citenamefont {Dai}, \citenamefont {Yuan}, \citenamefont {Zhang}, \citenamefont {Dai}, \citenamefont {Tang}, \citenamefont {Yang}, \citenamefont {Li}, \citenamefont {Ding}, \citenamefont {Oxenl{\o}we}, \citenamefont {Thompson}, \citenamefont {O'Brien}, \citenamefont {Li}, \citenamefont {Gong},\ and\ \citenamefont {Wang}}]{chi2022}%
  \BibitemOpen
  \bibfield  {author} {\bibinfo {author} {\bibfnamefont {Y.}~\bibnamefont {Chi}}, \bibinfo {author} {\bibfnamefont {J.}~\bibnamefont {Huang}}, \bibinfo {author} {\bibfnamefont {Z.}~\bibnamefont {Zhang}}, \bibinfo {author} {\bibfnamefont {J.}~\bibnamefont {Mao}}, \bibinfo {author} {\bibfnamefont {Z.}~\bibnamefont {Zhou}}, \bibinfo {author} {\bibfnamefont {X.}~\bibnamefont {Chen}}, \bibinfo {author} {\bibfnamefont {C.}~\bibnamefont {Zhai}}, \bibinfo {author} {\bibfnamefont {J.}~\bibnamefont {Bao}}, \bibinfo {author} {\bibfnamefont {T.}~\bibnamefont {Dai}}, \bibinfo {author} {\bibfnamefont {H.}~\bibnamefont {Yuan}}, \bibinfo {author} {\bibfnamefont {M.}~\bibnamefont {Zhang}}, \bibinfo {author} {\bibfnamefont {D.}~\bibnamefont {Dai}}, \bibinfo {author} {\bibfnamefont {B.}~\bibnamefont {Tang}}, \bibinfo {author} {\bibfnamefont {Y.}~\bibnamefont {Yang}}, \bibinfo {author} {\bibfnamefont {Z.}~\bibnamefont {Li}}, \bibinfo {author} {\bibfnamefont {Y.}~\bibnamefont {Ding}}, \bibinfo {author} {\bibfnamefont {L.~K.}\
  \bibnamefont {Oxenl{\o}we}}, \bibinfo {author} {\bibfnamefont {M.~G.}\ \bibnamefont {Thompson}}, \bibinfo {author} {\bibfnamefont {J.~L.}\ \bibnamefont {O'Brien}}, \bibinfo {author} {\bibfnamefont {Y.}~\bibnamefont {Li}}, \bibinfo {author} {\bibfnamefont {Q.}~\bibnamefont {Gong}},\ and\ \bibinfo {author} {\bibfnamefont {J.}~\bibnamefont {Wang}},\ }\bibfield  {title} {\bibinfo {title} {A programmable qudit-based quantum processor},\ }\href {https://doi.org/10.1038/s41467-022-28767-x} {\bibfield  {journal} {\bibinfo  {journal} {Nature Communications}\ }\textbf {\bibinfo {volume} {13}},\ \bibinfo {pages} {1166} (\bibinfo {year} {2022})}\BibitemShut {NoStop}%
\bibitem [{\citenamefont {Madsen}\ \emph {et~al.}(2022)\citenamefont {Madsen}, \citenamefont {Laudenbach}, \citenamefont {Askarani}, \citenamefont {Rortais}, \citenamefont {Vincent}, \citenamefont {Bulmer}, \citenamefont {Miatto}, \citenamefont {Neuhaus}, \citenamefont {Helt}, \citenamefont {Collins}, \citenamefont {Lita}, \citenamefont {Gerrits}, \citenamefont {Nam}, \citenamefont {Vaidya}, \citenamefont {Menotti}, \citenamefont {Dhand}, \citenamefont {Vernon}, \citenamefont {Quesada},\ and\ \citenamefont {Lavoie}}]{madsen2022}%
  \BibitemOpen
  \bibfield  {author} {\bibinfo {author} {\bibfnamefont {L.~S.}\ \bibnamefont {Madsen}}, \bibinfo {author} {\bibfnamefont {F.}~\bibnamefont {Laudenbach}}, \bibinfo {author} {\bibfnamefont {M.~F.}\ \bibnamefont {Askarani}}, \bibinfo {author} {\bibfnamefont {F.}~\bibnamefont {Rortais}}, \bibinfo {author} {\bibfnamefont {T.}~\bibnamefont {Vincent}}, \bibinfo {author} {\bibfnamefont {J.~F.~F.}\ \bibnamefont {Bulmer}}, \bibinfo {author} {\bibfnamefont {F.~M.}\ \bibnamefont {Miatto}}, \bibinfo {author} {\bibfnamefont {L.}~\bibnamefont {Neuhaus}}, \bibinfo {author} {\bibfnamefont {L.~G.}\ \bibnamefont {Helt}}, \bibinfo {author} {\bibfnamefont {M.~J.}\ \bibnamefont {Collins}}, \bibinfo {author} {\bibfnamefont {A.~E.}\ \bibnamefont {Lita}}, \bibinfo {author} {\bibfnamefont {T.}~\bibnamefont {Gerrits}}, \bibinfo {author} {\bibfnamefont {S.~W.}\ \bibnamefont {Nam}}, \bibinfo {author} {\bibfnamefont {V.~D.}\ \bibnamefont {Vaidya}}, \bibinfo {author} {\bibfnamefont {M.}~\bibnamefont {Menotti}}, \bibinfo {author}
  {\bibfnamefont {I.}~\bibnamefont {Dhand}}, \bibinfo {author} {\bibfnamefont {Z.}~\bibnamefont {Vernon}}, \bibinfo {author} {\bibfnamefont {N.}~\bibnamefont {Quesada}},\ and\ \bibinfo {author} {\bibfnamefont {J.}~\bibnamefont {Lavoie}},\ }\bibfield  {title} {\bibinfo {title} {Quantum computational advantage with a programmable photonic processor},\ }\href {https://doi.org/10.1038/s41586-022-04725-x} {\bibfield  {journal} {\bibinfo  {journal} {Nature}\ }\textbf {\bibinfo {volume} {606}},\ \bibinfo {pages} {75} (\bibinfo {year} {2022})}\BibitemShut {NoStop}%
\bibitem [{\citenamefont {Pelucchi}\ \emph {et~al.}(2022)\citenamefont {Pelucchi}, \citenamefont {Fagas}, \citenamefont {Aharonovich}, \citenamefont {Englund}, \citenamefont {Figueroa}, \citenamefont {Gong}, \citenamefont {Hannes}, \citenamefont {Liu}, \citenamefont {Lu}, \citenamefont {Matsuda}, \citenamefont {Pan}, \citenamefont {Schreck}, \citenamefont {Sciarrino}, \citenamefont {Silberhorn}, \citenamefont {Wang},\ and\ \citenamefont {J{\"o}ns}}]{pelucchi2022}%
  \BibitemOpen
  \bibfield  {author} {\bibinfo {author} {\bibfnamefont {E.}~\bibnamefont {Pelucchi}}, \bibinfo {author} {\bibfnamefont {G.}~\bibnamefont {Fagas}}, \bibinfo {author} {\bibfnamefont {I.}~\bibnamefont {Aharonovich}}, \bibinfo {author} {\bibfnamefont {D.}~\bibnamefont {Englund}}, \bibinfo {author} {\bibfnamefont {E.}~\bibnamefont {Figueroa}}, \bibinfo {author} {\bibfnamefont {Q.}~\bibnamefont {Gong}}, \bibinfo {author} {\bibfnamefont {H.}~\bibnamefont {Hannes}}, \bibinfo {author} {\bibfnamefont {J.}~\bibnamefont {Liu}}, \bibinfo {author} {\bibfnamefont {C.-Y.}\ \bibnamefont {Lu}}, \bibinfo {author} {\bibfnamefont {N.}~\bibnamefont {Matsuda}}, \bibinfo {author} {\bibfnamefont {J.-W.}\ \bibnamefont {Pan}}, \bibinfo {author} {\bibfnamefont {F.}~\bibnamefont {Schreck}}, \bibinfo {author} {\bibfnamefont {F.}~\bibnamefont {Sciarrino}}, \bibinfo {author} {\bibfnamefont {C.}~\bibnamefont {Silberhorn}}, \bibinfo {author} {\bibfnamefont {J.}~\bibnamefont {Wang}},\ and\ \bibinfo {author} {\bibfnamefont {K.~D.}\
  \bibnamefont {J{\"o}ns}},\ }\bibfield  {title} {\bibinfo {title} {The potential and global outlook of integrated photonics for quantum technologies},\ }\href {https://doi.org/10.1038/s42254-021-00398-z} {\bibfield  {journal} {\bibinfo  {journal} {Nature Reviews Physics}\ }\textbf {\bibinfo {volume} {4}},\ \bibinfo {pages} {194} (\bibinfo {year} {2022})}\BibitemShut {NoStop}%
\bibitem [{\citenamefont {Shen}\ \emph {et~al.}(2017)\citenamefont {Shen}, \citenamefont {Harris}, \citenamefont {Skirlo}, \citenamefont {Prabhu}, \citenamefont {{Baehr-Jones}}, \citenamefont {Hochberg}, \citenamefont {Sun}, \citenamefont {Zhao}, \citenamefont {Larochelle}, \citenamefont {Englund},\ and\ \citenamefont {Solja{\v c}i{\'c}}}]{shen2017}%
  \BibitemOpen
  \bibfield  {author} {\bibinfo {author} {\bibfnamefont {Y.}~\bibnamefont {Shen}}, \bibinfo {author} {\bibfnamefont {N.~C.}\ \bibnamefont {Harris}}, \bibinfo {author} {\bibfnamefont {S.}~\bibnamefont {Skirlo}}, \bibinfo {author} {\bibfnamefont {M.}~\bibnamefont {Prabhu}}, \bibinfo {author} {\bibfnamefont {T.}~\bibnamefont {{Baehr-Jones}}}, \bibinfo {author} {\bibfnamefont {M.}~\bibnamefont {Hochberg}}, \bibinfo {author} {\bibfnamefont {X.}~\bibnamefont {Sun}}, \bibinfo {author} {\bibfnamefont {S.}~\bibnamefont {Zhao}}, \bibinfo {author} {\bibfnamefont {H.}~\bibnamefont {Larochelle}}, \bibinfo {author} {\bibfnamefont {D.}~\bibnamefont {Englund}},\ and\ \bibinfo {author} {\bibfnamefont {M.}~\bibnamefont {Solja{\v c}i{\'c}}},\ }\bibfield  {title} {\bibinfo {title} {Deep learning with coherent nanophotonic circuits},\ }\href {https://doi.org/10.1038/nphoton.2017.93} {\bibfield  {journal} {\bibinfo  {journal} {Nature Photonics}\ }\textbf {\bibinfo {volume} {11}},\ \bibinfo {pages} {441} (\bibinfo {year}
  {2017})}\BibitemShut {NoStop}%
\bibitem [{\citenamefont {Prabhu}\ \emph {et~al.}(2020)\citenamefont {Prabhu}, \citenamefont {{Roques-Carmes}}, \citenamefont {Shen}, \citenamefont {Harris}, \citenamefont {Jing}, \citenamefont {Carolan}, \citenamefont {Hamerly}, \citenamefont {{Baehr-Jones}}, \citenamefont {Hochberg}, \citenamefont {{\v C}eperi{\'c}}, \citenamefont {Joannopoulos}, \citenamefont {Englund},\ and\ \citenamefont {Solja{\v c}i{\'c}}}]{prabhu2020}%
  \BibitemOpen
  \bibfield  {author} {\bibinfo {author} {\bibfnamefont {M.}~\bibnamefont {Prabhu}}, \bibinfo {author} {\bibfnamefont {C.}~\bibnamefont {{Roques-Carmes}}}, \bibinfo {author} {\bibfnamefont {Y.}~\bibnamefont {Shen}}, \bibinfo {author} {\bibfnamefont {N.}~\bibnamefont {Harris}}, \bibinfo {author} {\bibfnamefont {L.}~\bibnamefont {Jing}}, \bibinfo {author} {\bibfnamefont {J.}~\bibnamefont {Carolan}}, \bibinfo {author} {\bibfnamefont {R.}~\bibnamefont {Hamerly}}, \bibinfo {author} {\bibfnamefont {T.}~\bibnamefont {{Baehr-Jones}}}, \bibinfo {author} {\bibfnamefont {M.}~\bibnamefont {Hochberg}}, \bibinfo {author} {\bibfnamefont {V.}~\bibnamefont {{\v C}eperi{\'c}}}, \bibinfo {author} {\bibfnamefont {J.~D.}\ \bibnamefont {Joannopoulos}}, \bibinfo {author} {\bibfnamefont {D.~R.}\ \bibnamefont {Englund}},\ and\ \bibinfo {author} {\bibfnamefont {M.}~\bibnamefont {Solja{\v c}i{\'c}}},\ }\bibfield  {title} {\bibinfo {title} {Accelerating recurrent {{Ising}} machines in photonic integrated circuits},\ }\href
  {https://doi.org/10.1364/OPTICA.386613} {\bibfield  {journal} {\bibinfo  {journal} {Optica}\ }\textbf {\bibinfo {volume} {7}},\ \bibinfo {pages} {551} (\bibinfo {year} {2020})}\BibitemShut {NoStop}%
\bibitem [{\citenamefont {Zhang}\ \emph {et~al.}(2021)\citenamefont {Zhang}, \citenamefont {Gu}, \citenamefont {Jiang}, \citenamefont {Thompson}, \citenamefont {Cai}, \citenamefont {Paesani}, \citenamefont {Santagati}, \citenamefont {Laing}, \citenamefont {Zhang}, \citenamefont {Yung}, \citenamefont {Shi}, \citenamefont {Muhammad}, \citenamefont {Lo}, \citenamefont {Luo}, \citenamefont {Dong}, \citenamefont {Kwong}, \citenamefont {Kwek},\ and\ \citenamefont {Liu}}]{zhang2021f}%
  \BibitemOpen
  \bibfield  {author} {\bibinfo {author} {\bibfnamefont {H.}~\bibnamefont {Zhang}}, \bibinfo {author} {\bibfnamefont {M.}~\bibnamefont {Gu}}, \bibinfo {author} {\bibfnamefont {X.~D.}\ \bibnamefont {Jiang}}, \bibinfo {author} {\bibfnamefont {J.}~\bibnamefont {Thompson}}, \bibinfo {author} {\bibfnamefont {H.}~\bibnamefont {Cai}}, \bibinfo {author} {\bibfnamefont {S.}~\bibnamefont {Paesani}}, \bibinfo {author} {\bibfnamefont {R.}~\bibnamefont {Santagati}}, \bibinfo {author} {\bibfnamefont {A.}~\bibnamefont {Laing}}, \bibinfo {author} {\bibfnamefont {Y.}~\bibnamefont {Zhang}}, \bibinfo {author} {\bibfnamefont {M.~H.}\ \bibnamefont {Yung}}, \bibinfo {author} {\bibfnamefont {Y.~Z.}\ \bibnamefont {Shi}}, \bibinfo {author} {\bibfnamefont {F.~K.}\ \bibnamefont {Muhammad}}, \bibinfo {author} {\bibfnamefont {G.~Q.}\ \bibnamefont {Lo}}, \bibinfo {author} {\bibfnamefont {X.~S.}\ \bibnamefont {Luo}}, \bibinfo {author} {\bibfnamefont {B.}~\bibnamefont {Dong}}, \bibinfo {author} {\bibfnamefont {D.~L.}\ \bibnamefont {Kwong}},
  \bibinfo {author} {\bibfnamefont {L.~C.}\ \bibnamefont {Kwek}},\ and\ \bibinfo {author} {\bibfnamefont {A.~Q.}\ \bibnamefont {Liu}},\ }\bibfield  {title} {\bibinfo {title} {An optical neural chip for implementing complex-valued neural network},\ }\href {https://doi.org/10.1038/s41467-020-20719-7} {\bibfield  {journal} {\bibinfo  {journal} {Nature Communications}\ }\textbf {\bibinfo {volume} {12}},\ \bibinfo {pages} {457} (\bibinfo {year} {2021})}\BibitemShut {NoStop}%
\bibitem [{\citenamefont {Ashtiani}\ \emph {et~al.}(2022)\citenamefont {Ashtiani}, \citenamefont {Geers},\ and\ \citenamefont {Aflatouni}}]{ashtiani2022}%
  \BibitemOpen
  \bibfield  {author} {\bibinfo {author} {\bibfnamefont {F.}~\bibnamefont {Ashtiani}}, \bibinfo {author} {\bibfnamefont {A.~J.}\ \bibnamefont {Geers}},\ and\ \bibinfo {author} {\bibfnamefont {F.}~\bibnamefont {Aflatouni}},\ }\bibfield  {title} {\bibinfo {title} {An on-chip photonic deep neural network for image classification},\ }\href {https://doi.org/10.1038/s41586-022-04714-0} {\bibfield  {journal} {\bibinfo  {journal} {Nature}\ }\textbf {\bibinfo {volume} {606}},\ \bibinfo {pages} {501} (\bibinfo {year} {2022})}\BibitemShut {NoStop}%
\bibitem [{\citenamefont {Bandyopadhyay}\ \emph {et~al.}(2022)\citenamefont {Bandyopadhyay}, \citenamefont {Sludds}, \citenamefont {Krastanov}, \citenamefont {Hamerly}, \citenamefont {Harris}, \citenamefont {Bunandar}, \citenamefont {Streshinsky}, \citenamefont {Hochberg},\ and\ \citenamefont {Englund}}]{bandyopadhyay2022}%
  \BibitemOpen
  \bibfield  {author} {\bibinfo {author} {\bibfnamefont {S.}~\bibnamefont {Bandyopadhyay}}, \bibinfo {author} {\bibfnamefont {A.}~\bibnamefont {Sludds}}, \bibinfo {author} {\bibfnamefont {S.}~\bibnamefont {Krastanov}}, \bibinfo {author} {\bibfnamefont {R.}~\bibnamefont {Hamerly}}, \bibinfo {author} {\bibfnamefont {N.}~\bibnamefont {Harris}}, \bibinfo {author} {\bibfnamefont {D.}~\bibnamefont {Bunandar}}, \bibinfo {author} {\bibfnamefont {M.}~\bibnamefont {Streshinsky}}, \bibinfo {author} {\bibfnamefont {M.}~\bibnamefont {Hochberg}},\ and\ \bibinfo {author} {\bibfnamefont {D.}~\bibnamefont {Englund}},\ }\href@noop {} {\bibinfo {title} {Single chip photonic deep neural network with accelerated training}} (\bibinfo {year} {2022})\BibitemShut {NoStop}%
\bibitem [{\citenamefont {Ohno}\ \emph {et~al.}(2022)\citenamefont {Ohno}, \citenamefont {Tang}, \citenamefont {Toprasertpong}, \citenamefont {Takagi},\ and\ \citenamefont {Takenaka}}]{ohno2022}%
  \BibitemOpen
  \bibfield  {author} {\bibinfo {author} {\bibfnamefont {S.}~\bibnamefont {Ohno}}, \bibinfo {author} {\bibfnamefont {R.}~\bibnamefont {Tang}}, \bibinfo {author} {\bibfnamefont {K.}~\bibnamefont {Toprasertpong}}, \bibinfo {author} {\bibfnamefont {S.}~\bibnamefont {Takagi}},\ and\ \bibinfo {author} {\bibfnamefont {M.}~\bibnamefont {Takenaka}},\ }\bibfield  {title} {\bibinfo {title} {Si {{Microring Resonator Crossbar Array}} for {{On-Chip Inference}} and {{Training}} of the {{Optical Neural Network}}},\ }\href {https://doi.org/10.1021/acsphotonics.1c01777} {\bibfield  {journal} {\bibinfo  {journal} {ACS Photonics}\ }\textbf {\bibinfo {volume} {9}},\ \bibinfo {pages} {2614} (\bibinfo {year} {2022})}\BibitemShut {NoStop}%
\bibitem [{\citenamefont {Chen}\ \emph {et~al.}(2023)\citenamefont {Chen}, \citenamefont {Sludds}, \citenamefont {Davis}, \citenamefont {Christen}, \citenamefont {Bernstein}, \citenamefont {Ateshian}, \citenamefont {Heuser}, \citenamefont {Heermeier}, \citenamefont {Lott}, \citenamefont {Reitzenstein}, \citenamefont {Hamerly},\ and\ \citenamefont {Englund}}]{chen2023}%
  \BibitemOpen
  \bibfield  {author} {\bibinfo {author} {\bibfnamefont {Z.}~\bibnamefont {Chen}}, \bibinfo {author} {\bibfnamefont {A.}~\bibnamefont {Sludds}}, \bibinfo {author} {\bibfnamefont {R.}~\bibnamefont {Davis}}, \bibinfo {author} {\bibfnamefont {I.}~\bibnamefont {Christen}}, \bibinfo {author} {\bibfnamefont {L.}~\bibnamefont {Bernstein}}, \bibinfo {author} {\bibfnamefont {L.}~\bibnamefont {Ateshian}}, \bibinfo {author} {\bibfnamefont {T.}~\bibnamefont {Heuser}}, \bibinfo {author} {\bibfnamefont {N.}~\bibnamefont {Heermeier}}, \bibinfo {author} {\bibfnamefont {J.~A.}\ \bibnamefont {Lott}}, \bibinfo {author} {\bibfnamefont {S.}~\bibnamefont {Reitzenstein}}, \bibinfo {author} {\bibfnamefont {R.}~\bibnamefont {Hamerly}},\ and\ \bibinfo {author} {\bibfnamefont {D.}~\bibnamefont {Englund}},\ }\bibfield  {title} {\bibinfo {title} {Deep learning with coherent {{VCSEL}} neural networks},\ }\href {https://doi.org/10.1038/s41566-023-01233-w} {\bibfield  {journal} {\bibinfo  {journal} {Nature Photonics}\ }\textbf {\bibinfo
  {volume} {17}},\ \bibinfo {pages} {723} (\bibinfo {year} {2023})}\BibitemShut {NoStop}%
\bibitem [{\citenamefont {Pai}\ \emph {et~al.}(2023)\citenamefont {Pai}, \citenamefont {Sun}, \citenamefont {Hughes}, \citenamefont {Park}, \citenamefont {Bartlett}, \citenamefont {Williamson}, \citenamefont {Minkov}, \citenamefont {Milanizadeh}, \citenamefont {Abebe}, \citenamefont {Morichetti}, \citenamefont {Melloni}, \citenamefont {Fan}, \citenamefont {Solgaard},\ and\ \citenamefont {Miller}}]{pai2023}%
  \BibitemOpen
  \bibfield  {author} {\bibinfo {author} {\bibfnamefont {S.}~\bibnamefont {Pai}}, \bibinfo {author} {\bibfnamefont {Z.}~\bibnamefont {Sun}}, \bibinfo {author} {\bibfnamefont {T.~W.}\ \bibnamefont {Hughes}}, \bibinfo {author} {\bibfnamefont {T.}~\bibnamefont {Park}}, \bibinfo {author} {\bibfnamefont {B.}~\bibnamefont {Bartlett}}, \bibinfo {author} {\bibfnamefont {I.~A.~D.}\ \bibnamefont {Williamson}}, \bibinfo {author} {\bibfnamefont {M.}~\bibnamefont {Minkov}}, \bibinfo {author} {\bibfnamefont {M.}~\bibnamefont {Milanizadeh}}, \bibinfo {author} {\bibfnamefont {N.}~\bibnamefont {Abebe}}, \bibinfo {author} {\bibfnamefont {F.}~\bibnamefont {Morichetti}}, \bibinfo {author} {\bibfnamefont {A.}~\bibnamefont {Melloni}}, \bibinfo {author} {\bibfnamefont {S.}~\bibnamefont {Fan}}, \bibinfo {author} {\bibfnamefont {O.}~\bibnamefont {Solgaard}},\ and\ \bibinfo {author} {\bibfnamefont {D.~A.~B.}\ \bibnamefont {Miller}},\ }\bibfield  {title} {\bibinfo {title} {Experimentally realized in situ backpropagation for deep
  learning in photonic neural networks},\ }\href {https://doi.org/10.1126/science.ade8450} {\bibfield  {journal} {\bibinfo  {journal} {Science}\ }\textbf {\bibinfo {volume} {380}},\ \bibinfo {pages} {398} (\bibinfo {year} {2023})}\BibitemShut {NoStop}%
\bibitem [{\citenamefont {Burgwal}\ \emph {et~al.}(2017)\citenamefont {Burgwal}, \citenamefont {Clements}, \citenamefont {Smith}, \citenamefont {Gates}, \citenamefont {Kolthammer}, \citenamefont {Renema},\ and\ \citenamefont {Walmsley}}]{burgwal2017}%
  \BibitemOpen
  \bibfield  {author} {\bibinfo {author} {\bibfnamefont {R.}~\bibnamefont {Burgwal}}, \bibinfo {author} {\bibfnamefont {W.~R.}\ \bibnamefont {Clements}}, \bibinfo {author} {\bibfnamefont {D.~H.}\ \bibnamefont {Smith}}, \bibinfo {author} {\bibfnamefont {J.~C.}\ \bibnamefont {Gates}}, \bibinfo {author} {\bibfnamefont {W.~S.}\ \bibnamefont {Kolthammer}}, \bibinfo {author} {\bibfnamefont {J.~J.}\ \bibnamefont {Renema}},\ and\ \bibinfo {author} {\bibfnamefont {I.~A.}\ \bibnamefont {Walmsley}},\ }\bibfield  {title} {\bibinfo {title} {Using an imperfect photonic network to implement random unitaries},\ }\href {https://doi.org/10.1364/OE.25.028236} {\bibfield  {journal} {\bibinfo  {journal} {Optics Express}\ }\textbf {\bibinfo {volume} {25}},\ \bibinfo {pages} {28236} (\bibinfo {year} {2017})}\BibitemShut {NoStop}%
\bibitem [{\citenamefont {Melati}\ \emph {et~al.}(2017)\citenamefont {Melati}, \citenamefont {Alippi}, \citenamefont {Annoni}, \citenamefont {Peserico},\ and\ \citenamefont {Melloni}}]{melati2017}%
  \BibitemOpen
  \bibfield  {author} {\bibinfo {author} {\bibfnamefont {D.}~\bibnamefont {Melati}}, \bibinfo {author} {\bibfnamefont {A.}~\bibnamefont {Alippi}}, \bibinfo {author} {\bibfnamefont {A.}~\bibnamefont {Annoni}}, \bibinfo {author} {\bibfnamefont {N.}~\bibnamefont {Peserico}},\ and\ \bibinfo {author} {\bibfnamefont {A.}~\bibnamefont {Melloni}},\ }\bibfield  {title} {\bibinfo {title} {Integrated all-optical {{MIMO}} demultiplexer for mode- and wavelength-division-multiplexed transmission},\ }\href {https://doi.org/10.1364/OL.42.000342} {\bibfield  {journal} {\bibinfo  {journal} {Optics Letters}\ }\textbf {\bibinfo {volume} {42}},\ \bibinfo {pages} {342} (\bibinfo {year} {2017})}\BibitemShut {NoStop}%
\bibitem [{\citenamefont {Choutagunta}\ \emph {et~al.}(2020)\citenamefont {Choutagunta}, \citenamefont {Roberts}, \citenamefont {Miller},\ and\ \citenamefont {Kahn}}]{choutagunta2020}%
  \BibitemOpen
  \bibfield  {author} {\bibinfo {author} {\bibfnamefont {K.}~\bibnamefont {Choutagunta}}, \bibinfo {author} {\bibfnamefont {I.}~\bibnamefont {Roberts}}, \bibinfo {author} {\bibfnamefont {D.~A.~B.}\ \bibnamefont {Miller}},\ and\ \bibinfo {author} {\bibfnamefont {J.~M.}\ \bibnamefont {Kahn}},\ }\bibfield  {title} {\bibinfo {title} {Adapting {{Mach}}{\textendash}{{Zehnder Mesh Equalizers}} in {{Direct-Detection Mode-Division-Multiplexed Links}}},\ }\href {https://doi.org/10.1109/JLT.2019.2952060} {\bibfield  {journal} {\bibinfo  {journal} {Journal of Lightwave Technology}\ }\textbf {\bibinfo {volume} {38}},\ \bibinfo {pages} {723} (\bibinfo {year} {2020})}\BibitemShut {NoStop}%
\bibitem [{\citenamefont {Buddhiraju}\ \emph {et~al.}(2021)\citenamefont {Buddhiraju}, \citenamefont {Dutt}, \citenamefont {Minkov}, \citenamefont {Williamson},\ and\ \citenamefont {Fan}}]{buddhiraju2021}%
  \BibitemOpen
  \bibfield  {author} {\bibinfo {author} {\bibfnamefont {S.}~\bibnamefont {Buddhiraju}}, \bibinfo {author} {\bibfnamefont {A.}~\bibnamefont {Dutt}}, \bibinfo {author} {\bibfnamefont {M.}~\bibnamefont {Minkov}}, \bibinfo {author} {\bibfnamefont {I.~A.~D.}\ \bibnamefont {Williamson}},\ and\ \bibinfo {author} {\bibfnamefont {S.}~\bibnamefont {Fan}},\ }\bibfield  {title} {\bibinfo {title} {Arbitrary linear transformations for photons in the frequency synthetic dimension},\ }\href {https://doi.org/10.1038/s41467-021-22670-7} {\bibfield  {journal} {\bibinfo  {journal} {Nature Communications}\ }\textbf {\bibinfo {volume} {12}},\ \bibinfo {pages} {2401} (\bibinfo {year} {2021})}\BibitemShut {NoStop}%
\bibitem [{\citenamefont {Guo}\ and\ \citenamefont {Fan}(2023)}]{guo2023b}%
  \BibitemOpen
  \bibfield  {author} {\bibinfo {author} {\bibfnamefont {C.}~\bibnamefont {Guo}}\ and\ \bibinfo {author} {\bibfnamefont {S.}~\bibnamefont {Fan}},\ }\bibfield  {title} {\bibinfo {title} {Majorization {{Theory}} for {{Unitary Control}} of {{Optical Absorption}} and {{Emission}}},\ }\href {https://doi.org/10.1103/PhysRevLett.130.146202} {\bibfield  {journal} {\bibinfo  {journal} {Physical Review Letters}\ }\textbf {\bibinfo {volume} {130}},\ \bibinfo {pages} {146202} (\bibinfo {year} {2023})}\BibitemShut {NoStop}%
\bibitem [{\citenamefont {Vellekoop}\ and\ \citenamefont {Mosk}(2008)}]{vellekoop2008}%
  \BibitemOpen
  \bibfield  {author} {\bibinfo {author} {\bibfnamefont {I.~M.}\ \bibnamefont {Vellekoop}}\ and\ \bibinfo {author} {\bibfnamefont {A.~P.}\ \bibnamefont {Mosk}},\ }\bibfield  {title} {\bibinfo {title} {Universal {{Optimal Transmission}} of {{Light Through Disordered Materials}}},\ }\href {https://doi.org/10.1103/physrevlett.101.120601} {\bibfield  {journal} {\bibinfo  {journal} {Physical Review Letters}\ }\textbf {\bibinfo {volume} {101}},\ \bibinfo {pages} {120601} (\bibinfo {year} {2008})}\BibitemShut {NoStop}%
\bibitem [{\citenamefont {Popoff}\ \emph {et~al.}(2010)\citenamefont {Popoff}, \citenamefont {Lerosey}, \citenamefont {Carminati}, \citenamefont {Fink}, \citenamefont {Boccara},\ and\ \citenamefont {Gigan}}]{popoff2010}%
  \BibitemOpen
  \bibfield  {author} {\bibinfo {author} {\bibfnamefont {S.~M.}\ \bibnamefont {Popoff}}, \bibinfo {author} {\bibfnamefont {G.}~\bibnamefont {Lerosey}}, \bibinfo {author} {\bibfnamefont {R.}~\bibnamefont {Carminati}}, \bibinfo {author} {\bibfnamefont {M.}~\bibnamefont {Fink}}, \bibinfo {author} {\bibfnamefont {A.~C.}\ \bibnamefont {Boccara}},\ and\ \bibinfo {author} {\bibfnamefont {S.}~\bibnamefont {Gigan}},\ }\bibfield  {title} {\bibinfo {title} {Measuring the {{Transmission Matrix}} in {{Optics}}: {{An Approach}} to the {{Study}} and {{Control}} of {{Light Propagation}} in {{Disordered Media}}},\ }\href {https://doi.org/10.1103/physrevlett.104.100601} {\bibfield  {journal} {\bibinfo  {journal} {Physical Review Letters}\ }\textbf {\bibinfo {volume} {104}},\ \bibinfo {pages} {100601} (\bibinfo {year} {2010})}\BibitemShut {NoStop}%
\bibitem [{\citenamefont {Kim}\ \emph {et~al.}(2012)\citenamefont {Kim}, \citenamefont {Choi}, \citenamefont {Yoon}, \citenamefont {Choi}, \citenamefont {Kim}, \citenamefont {Park},\ and\ \citenamefont {Choi}}]{kim2012a}%
  \BibitemOpen
  \bibfield  {author} {\bibinfo {author} {\bibfnamefont {M.}~\bibnamefont {Kim}}, \bibinfo {author} {\bibfnamefont {Y.}~\bibnamefont {Choi}}, \bibinfo {author} {\bibfnamefont {C.}~\bibnamefont {Yoon}}, \bibinfo {author} {\bibfnamefont {W.}~\bibnamefont {Choi}}, \bibinfo {author} {\bibfnamefont {J.}~\bibnamefont {Kim}}, \bibinfo {author} {\bibfnamefont {Q.-H.}\ \bibnamefont {Park}},\ and\ \bibinfo {author} {\bibfnamefont {W.}~\bibnamefont {Choi}},\ }\bibfield  {title} {\bibinfo {title} {Maximal energy transport through disordered media with the implementation of transmission eigenchannels},\ }\href {https://doi.org/10.1038/nphoton.2012.159} {\bibfield  {journal} {\bibinfo  {journal} {Nature Photonics}\ }\textbf {\bibinfo {volume} {6}},\ \bibinfo {pages} {581} (\bibinfo {year} {2012})}\BibitemShut {NoStop}%
\bibitem [{\citenamefont {Shi}\ and\ \citenamefont {Genack}(2012)}]{shi2012}%
  \BibitemOpen
  \bibfield  {author} {\bibinfo {author} {\bibfnamefont {Z.}~\bibnamefont {Shi}}\ and\ \bibinfo {author} {\bibfnamefont {A.~Z.}\ \bibnamefont {Genack}},\ }\bibfield  {title} {\bibinfo {title} {Transmission {{Eigenvalues}} and the {{Bare Conductance}} in the {{Crossover}} to {{Anderson Localization}}},\ }\href {https://doi.org/10.1103/physrevlett.108.043901} {\bibfield  {journal} {\bibinfo  {journal} {Physical Review Letters}\ }\textbf {\bibinfo {volume} {108}},\ \bibinfo {pages} {043901} (\bibinfo {year} {2012})}\BibitemShut {NoStop}%
\bibitem [{\citenamefont {Yu}\ \emph {et~al.}(2013{\natexlab{b}})\citenamefont {Yu}, \citenamefont {Hillman}, \citenamefont {Choi}, \citenamefont {Lee}, \citenamefont {Feld}, \citenamefont {Dasari},\ and\ \citenamefont {Park}}]{yu2013a}%
  \BibitemOpen
  \bibfield  {author} {\bibinfo {author} {\bibfnamefont {H.}~\bibnamefont {Yu}}, \bibinfo {author} {\bibfnamefont {T.~R.}\ \bibnamefont {Hillman}}, \bibinfo {author} {\bibfnamefont {W.}~\bibnamefont {Choi}}, \bibinfo {author} {\bibfnamefont {J.~O.}\ \bibnamefont {Lee}}, \bibinfo {author} {\bibfnamefont {M.~S.}\ \bibnamefont {Feld}}, \bibinfo {author} {\bibfnamefont {R.~R.}\ \bibnamefont {Dasari}},\ and\ \bibinfo {author} {\bibfnamefont {Y.}~\bibnamefont {Park}},\ }\bibfield  {title} {\bibinfo {title} {Measuring {{Large Optical Transmission Matrices}} of {{Disordered Media}}},\ }\href {https://doi.org/10.1103/physrevlett.111.153902} {\bibfield  {journal} {\bibinfo  {journal} {Physical Review Letters}\ }\textbf {\bibinfo {volume} {111}},\ \bibinfo {pages} {153902} (\bibinfo {year} {2013}{\natexlab{b}})}\BibitemShut {NoStop}%
\bibitem [{\citenamefont {G{\'e}rardin}\ \emph {et~al.}(2014)\citenamefont {G{\'e}rardin}, \citenamefont {Laurent}, \citenamefont {Derode}, \citenamefont {Prada},\ and\ \citenamefont {Aubry}}]{gerardin2014}%
  \BibitemOpen
  \bibfield  {author} {\bibinfo {author} {\bibfnamefont {B.}~\bibnamefont {G{\'e}rardin}}, \bibinfo {author} {\bibfnamefont {J.}~\bibnamefont {Laurent}}, \bibinfo {author} {\bibfnamefont {A.}~\bibnamefont {Derode}}, \bibinfo {author} {\bibfnamefont {C.}~\bibnamefont {Prada}},\ and\ \bibinfo {author} {\bibfnamefont {A.}~\bibnamefont {Aubry}},\ }\bibfield  {title} {\bibinfo {title} {Full {{Transmission}} and {{Reflection}} of {{Waves Propagating}} through a {{Maze}} of {{Disorder}}},\ }\href {https://doi.org/10.1103/physrevlett.113.173901} {\bibfield  {journal} {\bibinfo  {journal} {Physical Review Letters}\ }\textbf {\bibinfo {volume} {113}},\ \bibinfo {pages} {173901} (\bibinfo {year} {2014})}\BibitemShut {NoStop}%
\bibitem [{\citenamefont {Pe{\~n}a}\ \emph {et~al.}(2014)\citenamefont {Pe{\~n}a}, \citenamefont {Girschik}, \citenamefont {Libisch}, \citenamefont {Rotter},\ and\ \citenamefont {Chabanov}}]{pena2014a}%
  \BibitemOpen
  \bibfield  {author} {\bibinfo {author} {\bibfnamefont {A.}~\bibnamefont {Pe{\~n}a}}, \bibinfo {author} {\bibfnamefont {A.}~\bibnamefont {Girschik}}, \bibinfo {author} {\bibfnamefont {F.}~\bibnamefont {Libisch}}, \bibinfo {author} {\bibfnamefont {S.}~\bibnamefont {Rotter}},\ and\ \bibinfo {author} {\bibfnamefont {A.~A.}\ \bibnamefont {Chabanov}},\ }\bibfield  {title} {\bibinfo {title} {The single-channel regime of transport through random media},\ }\href {https://doi.org/10.1038/ncomms4488} {\bibfield  {journal} {\bibinfo  {journal} {Nature Communications}\ }\textbf {\bibinfo {volume} {5}},\ \bibinfo {pages} {3488} (\bibinfo {year} {2014})}\BibitemShut {NoStop}%
\bibitem [{\citenamefont {Davy}\ \emph {et~al.}(2015)\citenamefont {Davy}, \citenamefont {Shi}, \citenamefont {Wang}, \citenamefont {Cheng},\ and\ \citenamefont {Genack}}]{davy2015}%
  \BibitemOpen
  \bibfield  {author} {\bibinfo {author} {\bibfnamefont {M.}~\bibnamefont {Davy}}, \bibinfo {author} {\bibfnamefont {Z.}~\bibnamefont {Shi}}, \bibinfo {author} {\bibfnamefont {J.}~\bibnamefont {Wang}}, \bibinfo {author} {\bibfnamefont {X.}~\bibnamefont {Cheng}},\ and\ \bibinfo {author} {\bibfnamefont {A.~Z.}\ \bibnamefont {Genack}},\ }\bibfield  {title} {\bibinfo {title} {Transmission {{Eigenchannels}} and the {{Densities}} of {{States}} of {{Random Media}}},\ }\href {https://doi.org/10.1103/physrevlett.114.033901} {\bibfield  {journal} {\bibinfo  {journal} {Physical Review Letters}\ }\textbf {\bibinfo {volume} {114}},\ \bibinfo {pages} {033901} (\bibinfo {year} {2015})}\BibitemShut {NoStop}%
\bibitem [{\citenamefont {Shi}\ \emph {et~al.}(2015)\citenamefont {Shi}, \citenamefont {Davy},\ and\ \citenamefont {Genack}}]{shi2015a}%
  \BibitemOpen
  \bibfield  {author} {\bibinfo {author} {\bibfnamefont {Z.}~\bibnamefont {Shi}}, \bibinfo {author} {\bibfnamefont {M.}~\bibnamefont {Davy}},\ and\ \bibinfo {author} {\bibfnamefont {A.~Z.}\ \bibnamefont {Genack}},\ }\bibfield  {title} {\bibinfo {title} {Statistics and control of waves in disordered media},\ }\href {https://doi.org/10.1364/oe.23.012293} {\bibfield  {journal} {\bibinfo  {journal} {Optics Express}\ }\textbf {\bibinfo {volume} {23}},\ \bibinfo {pages} {12293} (\bibinfo {year} {2015})}\BibitemShut {NoStop}%
\bibitem [{\citenamefont {Bender}\ \emph {et~al.}(2020)\citenamefont {Bender}, \citenamefont {Yamilov}, \citenamefont {Y{\i}lmaz},\ and\ \citenamefont {Cao}}]{bender2020c}%
  \BibitemOpen
  \bibfield  {author} {\bibinfo {author} {\bibfnamefont {N.}~\bibnamefont {Bender}}, \bibinfo {author} {\bibfnamefont {A.}~\bibnamefont {Yamilov}}, \bibinfo {author} {\bibfnamefont {H.}~\bibnamefont {Y{\i}lmaz}},\ and\ \bibinfo {author} {\bibfnamefont {H.}~\bibnamefont {Cao}},\ }\bibfield  {title} {\bibinfo {title} {Fluctuations and {{Correlations}} of {{Transmission Eigenchannels}} in {{Diffusive Media}}},\ }\href {https://doi.org/10.1103/physrevlett.125.165901} {\bibfield  {journal} {\bibinfo  {journal} {Physical Review Letters}\ }\textbf {\bibinfo {volume} {125}},\ \bibinfo {pages} {165901} (\bibinfo {year} {2020})}\BibitemShut {NoStop}%
\bibitem [{\citenamefont {Guo}\ \emph {et~al.}(2024)\citenamefont {Guo}, \citenamefont {Miller},\ and\ \citenamefont {Fan}}]{guo2024}%
  \BibitemOpen
  \bibfield  {author} {\bibinfo {author} {\bibfnamefont {C.}~\bibnamefont {Guo}}, \bibinfo {author} {\bibfnamefont {D.~A.~B.}\ \bibnamefont {Miller}},\ and\ \bibinfo {author} {\bibfnamefont {S.}~\bibnamefont {Fan}},\ }\bibfield  {title} {\bibinfo {title} {Unitary {{Control}} of {{Multimode Wave Transmission}}},\ }\href@noop {} {\bibfield  {journal} {\bibinfo  {journal} {(Unpublished)}\ } (\bibinfo {year} {2024})}\BibitemShut {NoStop}%
\bibitem [{\citenamefont {Mandel}\ and\ \citenamefont {Wolf}(1995)}]{mandel1995}%
  \BibitemOpen
  \bibfield  {author} {\bibinfo {author} {\bibfnamefont {L.}~\bibnamefont {Mandel}}\ and\ \bibinfo {author} {\bibfnamefont {E.}~\bibnamefont {Wolf}},\ }\href@noop {} {\emph {\bibinfo {title} {Optical Coherence and Quantum Optics}}}\ (\bibinfo  {publisher} {Cambridge University Press},\ \bibinfo {address} {Cambridge ; New York},\ \bibinfo {year} {1995})\BibitemShut {NoStop}%
\bibitem [{\citenamefont {Goodman}(2000)}]{goodman2000}%
  \BibitemOpen
  \bibfield  {author} {\bibinfo {author} {\bibfnamefont {J.~W.}\ \bibnamefont {Goodman}},\ }\href@noop {} {\emph {\bibinfo {title} {Statistical Optics}}},\ \bibinfo {edition} {wiley classics library ed}\ ed.,\ Wiley Classics Library\ (\bibinfo  {publisher} {Wiley},\ \bibinfo {address} {New York},\ \bibinfo {year} {2000})\BibitemShut {NoStop}%
\bibitem [{\citenamefont {Guo}\ and\ \citenamefont {Fan}(2024)}]{guo2024a}%
  \BibitemOpen
  \bibfield  {author} {\bibinfo {author} {\bibfnamefont {C.}~\bibnamefont {Guo}}\ and\ \bibinfo {author} {\bibfnamefont {S.}~\bibnamefont {Fan}},\ }\bibfield  {title} {\bibinfo {title} {Unitary control of partially coherent waves. {{I}}. {{Absorption}}},\ }\href {https://doi.org/10.1103/PhysRevB.110.035430} {\bibfield  {journal} {\bibinfo  {journal} {Physical Review B}\ }\textbf {\bibinfo {volume} {110}},\ \bibinfo {pages} {035430} (\bibinfo {year} {2024})}\BibitemShut {NoStop}%
\bibitem [{\citenamefont {Landau}\ and\ \citenamefont {Lifshitz}(1981)}]{landau1981}%
  \BibitemOpen
  \bibfield  {author} {\bibinfo {author} {\bibfnamefont {L.~D.}\ \bibnamefont {Landau}}\ and\ \bibinfo {author} {\bibfnamefont {L.~M.}\ \bibnamefont {Lifshitz}},\ }\href@noop {} {\emph {\bibinfo {title} {Quantum {{Mechanics}}: {{Non-Relativistic Theory}}}}},\ \bibinfo {edition} {3rd}\ ed.\ (\bibinfo  {publisher} {Butterworth-Heinemann},\ \bibinfo {address} {Singapore},\ \bibinfo {year} {1981})\BibitemShut {NoStop}%
\bibitem [{\citenamefont {O'Neill}(2003)}]{oneill2003}%
  \BibitemOpen
  \bibfield  {author} {\bibinfo {author} {\bibfnamefont {E.~L.}\ \bibnamefont {O'Neill}},\ }\href@noop {} {\emph {\bibinfo {title} {Introduction to {{Statistical Optics}}}}}\ (\bibinfo  {publisher} {Courier Corporation},\ \bibinfo {year} {2003})\BibitemShut {NoStop}%
\bibitem [{\citenamefont {Wolf}(2003)}]{wolf2003}%
  \BibitemOpen
  \bibfield  {author} {\bibinfo {author} {\bibfnamefont {E.}~\bibnamefont {Wolf}},\ }\bibfield  {title} {\bibinfo {title} {Unified theory of coherence and polarization of random electromagnetic beams},\ }\href {https://doi.org/10.1016/S0375-9601(03)00684-4} {\bibfield  {journal} {\bibinfo  {journal} {Physics Letters A}\ }\textbf {\bibinfo {volume} {312}},\ \bibinfo {pages} {263} (\bibinfo {year} {2003})}\BibitemShut {NoStop}%
\bibitem [{\citenamefont {{de Lima Bernardo}}(2017)}]{delimabernardo2017}%
  \BibitemOpen
  \bibfield  {author} {\bibinfo {author} {\bibfnamefont {B.}~\bibnamefont {{de Lima Bernardo}}},\ }\bibfield  {title} {\bibinfo {title} {Unified quantum density matrix description of coherence and polarization},\ }\href {https://doi.org/10.1016/j.physleta.2017.05.018} {\bibfield  {journal} {\bibinfo  {journal} {Physics Letters A}\ }\textbf {\bibinfo {volume} {381}},\ \bibinfo {pages} {2239} (\bibinfo {year} {2017})}\BibitemShut {NoStop}%
\bibitem [{\citenamefont {Zhang}\ \emph {et~al.}(2019)\citenamefont {Zhang}, \citenamefont {Hsu},\ and\ \citenamefont {Miller}}]{zhang2019m}%
  \BibitemOpen
  \bibfield  {author} {\bibinfo {author} {\bibfnamefont {H.}~\bibnamefont {Zhang}}, \bibinfo {author} {\bibfnamefont {C.~W.}\ \bibnamefont {Hsu}},\ and\ \bibinfo {author} {\bibfnamefont {O.~D.}\ \bibnamefont {Miller}},\ }\bibfield  {title} {\bibinfo {title} {Scattering concentration bounds: Brightness theorems for waves},\ }\href {https://doi.org/10.1364/OPTICA.6.001321} {\bibfield  {journal} {\bibinfo  {journal} {Optica}\ }\textbf {\bibinfo {volume} {6}},\ \bibinfo {pages} {1321} (\bibinfo {year} {2019})}\BibitemShut {NoStop}%
\bibitem [{\citenamefont {Korotkova}(2022)}]{korotkova2022}%
  \BibitemOpen
  \bibfield  {author} {\bibinfo {author} {\bibfnamefont {O.}~\bibnamefont {Korotkova}},\ }\href@noop {} {\emph {\bibinfo {title} {Theoretical Statistical Optics}}}\ (\bibinfo  {publisher} {World Scientific},\ \bibinfo {address} {New Jersey London Singapore Beijing Shanghai Hong Kong Taipei Chennai Tokyo},\ \bibinfo {year} {2022})\BibitemShut {NoStop}%
\bibitem [{\citenamefont {Wolf}\ and\ \citenamefont {Maret}(1985)}]{wolf1985}%
  \BibitemOpen
  \bibfield  {author} {\bibinfo {author} {\bibfnamefont {P.-E.}\ \bibnamefont {Wolf}}\ and\ \bibinfo {author} {\bibfnamefont {G.}~\bibnamefont {Maret}},\ }\bibfield  {title} {\bibinfo {title} {Weak {{Localization}} and {{Coherent Backscattering}} of {{Photons}} in {{Disordered Media}}},\ }\href {https://doi.org/10.1103/physrevlett.55.2696} {\bibfield  {journal} {\bibinfo  {journal} {Physical Review Letters}\ }\textbf {\bibinfo {volume} {55}},\ \bibinfo {pages} {2696} (\bibinfo {year} {1985})}\BibitemShut {NoStop}%
\bibitem [{\citenamefont {Yamazoe}(2012)}]{yamazoe2012}%
  \BibitemOpen
  \bibfield  {author} {\bibinfo {author} {\bibfnamefont {K.}~\bibnamefont {Yamazoe}},\ }\bibfield  {title} {\bibinfo {title} {Coherency matrix formulation for partially coherent imaging to evaluate the degree of coherence for image},\ }\href {https://doi.org/10.1364/JOSAA.29.001529} {\bibfield  {journal} {\bibinfo  {journal} {JOSA A}\ }\textbf {\bibinfo {volume} {29}},\ \bibinfo {pages} {1529} (\bibinfo {year} {2012})}\BibitemShut {NoStop}%
\bibitem [{\citenamefont {Okoro}\ \emph {et~al.}(2017)\citenamefont {Okoro}, \citenamefont {Kondakci}, \citenamefont {Abouraddy},\ and\ \citenamefont {Toussaint}}]{okoro2017}%
  \BibitemOpen
  \bibfield  {author} {\bibinfo {author} {\bibfnamefont {C.}~\bibnamefont {Okoro}}, \bibinfo {author} {\bibfnamefont {H.~E.}\ \bibnamefont {Kondakci}}, \bibinfo {author} {\bibfnamefont {A.~F.}\ \bibnamefont {Abouraddy}},\ and\ \bibinfo {author} {\bibfnamefont {K.~C.}\ \bibnamefont {Toussaint}},\ }\bibfield  {title} {\bibinfo {title} {Demonstration of an optical-coherence converter},\ }\href {https://doi.org/10.1364/OPTICA.4.001052} {\bibfield  {journal} {\bibinfo  {journal} {Optica}\ }\textbf {\bibinfo {volume} {4}},\ \bibinfo {pages} {1052} (\bibinfo {year} {2017})}\BibitemShut {NoStop}%
\bibitem [{\citenamefont {Rotter}\ and\ \citenamefont {Gigan}(2017)}]{rotter2017}%
  \BibitemOpen
  \bibfield  {author} {\bibinfo {author} {\bibfnamefont {S.}~\bibnamefont {Rotter}}\ and\ \bibinfo {author} {\bibfnamefont {S.}~\bibnamefont {Gigan}},\ }\bibfield  {title} {\bibinfo {title} {Light fields in complex media: {{Mesoscopic}} scattering meets wave control},\ }\href {https://doi.org/10.1103/RevModPhys.89.015005} {\bibfield  {journal} {\bibinfo  {journal} {Reviews of Modern Physics}\ }\textbf {\bibinfo {volume} {89}},\ \bibinfo {pages} {015005} (\bibinfo {year} {2017})}\BibitemShut {NoStop}%
\bibitem [{\citenamefont {Yamilov}\ \emph {et~al.}(2016)\citenamefont {Yamilov}, \citenamefont {Petrenko}, \citenamefont {Sarma},\ and\ \citenamefont {Cao}}]{yamilov2016}%
  \BibitemOpen
  \bibfield  {author} {\bibinfo {author} {\bibfnamefont {A.}~\bibnamefont {Yamilov}}, \bibinfo {author} {\bibfnamefont {S.}~\bibnamefont {Petrenko}}, \bibinfo {author} {\bibfnamefont {R.}~\bibnamefont {Sarma}},\ and\ \bibinfo {author} {\bibfnamefont {H.}~\bibnamefont {Cao}},\ }\bibfield  {title} {\bibinfo {title} {Shape dependence of transmission, reflection, and absorption eigenvalue densities in disordered waveguides with dissipation},\ }\href {https://doi.org/10.1103/PhysRevB.93.100201} {\bibfield  {journal} {\bibinfo  {journal} {Physical Review B}\ }\textbf {\bibinfo {volume} {93}},\ \bibinfo {pages} {100201} (\bibinfo {year} {2016})}\BibitemShut {NoStop}%
\bibitem [{\citenamefont {Horn}\ and\ \citenamefont {Johnson}(2012)}]{horn2012}%
  \BibitemOpen
  \bibfield  {author} {\bibinfo {author} {\bibfnamefont {R.~A.}\ \bibnamefont {Horn}}\ and\ \bibinfo {author} {\bibfnamefont {C.~R.}\ \bibnamefont {Johnson}},\ }\href@noop {} {\emph {\bibinfo {title} {Matrix Analysis}}},\ \bibinfo {edition} {2nd}\ ed.\ (\bibinfo  {publisher} {{Cambridge University Press}},\ \bibinfo {address} {{Cambridge ; New York}},\ \bibinfo {year} {2012})\BibitemShut {NoStop}%
\bibitem [{\citenamefont {Harris}\ \emph {et~al.}(2020)\citenamefont {Harris}, \citenamefont {Millman}, \citenamefont {{van der Walt}}, \citenamefont {Gommers}, \citenamefont {Virtanen}, \citenamefont {Cournapeau}, \citenamefont {Wieser}, \citenamefont {Taylor}, \citenamefont {Berg}, \citenamefont {Smith}, \citenamefont {Kern}, \citenamefont {Picus}, \citenamefont {Hoyer}, \citenamefont {{van Kerkwijk}}, \citenamefont {Brett}, \citenamefont {Haldane}, \citenamefont {{del R{\'i}o}}, \citenamefont {Wiebe}, \citenamefont {Peterson}, \citenamefont {{G{\'e}rard-Marchant}}, \citenamefont {Sheppard}, \citenamefont {Reddy}, \citenamefont {Weckesser}, \citenamefont {Abbasi}, \citenamefont {Gohlke},\ and\ \citenamefont {Oliphant}}]{harris2020}%
  \BibitemOpen
  \bibfield  {author} {\bibinfo {author} {\bibfnamefont {C.~R.}\ \bibnamefont {Harris}}, \bibinfo {author} {\bibfnamefont {K.~J.}\ \bibnamefont {Millman}}, \bibinfo {author} {\bibfnamefont {S.~J.}\ \bibnamefont {{van der Walt}}}, \bibinfo {author} {\bibfnamefont {R.}~\bibnamefont {Gommers}}, \bibinfo {author} {\bibfnamefont {P.}~\bibnamefont {Virtanen}}, \bibinfo {author} {\bibfnamefont {D.}~\bibnamefont {Cournapeau}}, \bibinfo {author} {\bibfnamefont {E.}~\bibnamefont {Wieser}}, \bibinfo {author} {\bibfnamefont {J.}~\bibnamefont {Taylor}}, \bibinfo {author} {\bibfnamefont {S.}~\bibnamefont {Berg}}, \bibinfo {author} {\bibfnamefont {N.~J.}\ \bibnamefont {Smith}}, \bibinfo {author} {\bibfnamefont {R.}~\bibnamefont {Kern}}, \bibinfo {author} {\bibfnamefont {M.}~\bibnamefont {Picus}}, \bibinfo {author} {\bibfnamefont {S.}~\bibnamefont {Hoyer}}, \bibinfo {author} {\bibfnamefont {M.~H.}\ \bibnamefont {{van Kerkwijk}}}, \bibinfo {author} {\bibfnamefont {M.}~\bibnamefont {Brett}}, \bibinfo {author} {\bibfnamefont
  {A.}~\bibnamefont {Haldane}}, \bibinfo {author} {\bibfnamefont {J.~F.}\ \bibnamefont {{del R{\'i}o}}}, \bibinfo {author} {\bibfnamefont {M.}~\bibnamefont {Wiebe}}, \bibinfo {author} {\bibfnamefont {P.}~\bibnamefont {Peterson}}, \bibinfo {author} {\bibfnamefont {P.}~\bibnamefont {{G{\'e}rard-Marchant}}}, \bibinfo {author} {\bibfnamefont {K.}~\bibnamefont {Sheppard}}, \bibinfo {author} {\bibfnamefont {T.}~\bibnamefont {Reddy}}, \bibinfo {author} {\bibfnamefont {W.}~\bibnamefont {Weckesser}}, \bibinfo {author} {\bibfnamefont {H.}~\bibnamefont {Abbasi}}, \bibinfo {author} {\bibfnamefont {C.}~\bibnamefont {Gohlke}},\ and\ \bibinfo {author} {\bibfnamefont {T.~E.}\ \bibnamefont {Oliphant}},\ }\bibfield  {title} {\bibinfo {title} {Array programming with {{NumPy}}},\ }\href {https://doi.org/10.1038/s41586-020-2649-2} {\bibfield  {journal} {\bibinfo  {journal} {Nature}\ }\textbf {\bibinfo {volume} {585}},\ \bibinfo {pages} {357} (\bibinfo {year} {2020})}\BibitemShut {NoStop}%
\bibitem [{\citenamefont {Mezzadri}(2007)}]{mezzadri2007}%
  \BibitemOpen
  \bibfield  {author} {\bibinfo {author} {\bibfnamefont {F.}~\bibnamefont {Mezzadri}},\ }\bibfield  {title} {\bibinfo {title} {How to generate random matrices from the classical compact groups},\ }\href@noop {} {\bibfield  {journal} {\bibinfo  {journal} {Notices of the American Mathematical Society}\ }\textbf {\bibinfo {volume} {54}},\ \bibinfo {pages} {592} (\bibinfo {year} {2007})},\ \Eprint {https://arxiv.org/abs/math-ph/0609050} {arxiv:math-ph/0609050} \BibitemShut {NoStop}%
\bibitem [{\citenamefont {Yan}\ \emph {et~al.}(2014)\citenamefont {Yan}, \citenamefont {Cui}, \citenamefont {Gu}, \citenamefont {Tian}, \citenamefont {Fu},\ and\ \citenamefont {Wu}}]{yan2014a}%
  \BibitemOpen
  \bibfield  {author} {\bibinfo {author} {\bibfnamefont {X.-B.}\ \bibnamefont {Yan}}, \bibinfo {author} {\bibfnamefont {C.-L.}\ \bibnamefont {Cui}}, \bibinfo {author} {\bibfnamefont {K.-H.}\ \bibnamefont {Gu}}, \bibinfo {author} {\bibfnamefont {X.-D.}\ \bibnamefont {Tian}}, \bibinfo {author} {\bibfnamefont {C.-B.}\ \bibnamefont {Fu}},\ and\ \bibinfo {author} {\bibfnamefont {J.-H.}\ \bibnamefont {Wu}},\ }\bibfield  {title} {\bibinfo {title} {Coherent perfect absorption, transmission, and synthesis in a double-cavity optomechanical system},\ }\href {https://doi.org/10.1364/OE.22.004886} {\bibfield  {journal} {\bibinfo  {journal} {Optics Express}\ }\textbf {\bibinfo {volume} {22}},\ \bibinfo {pages} {4886} (\bibinfo {year} {2014})}\BibitemShut {NoStop}%
\bibitem [{\citenamefont {Wu}\ \emph {et~al.}(2022)\citenamefont {Wu}, \citenamefont {Li},\ and\ \citenamefont {Wu}}]{wu2022d}%
  \BibitemOpen
  \bibfield  {author} {\bibinfo {author} {\bibfnamefont {Z.}~\bibnamefont {Wu}}, \bibinfo {author} {\bibfnamefont {J.}~\bibnamefont {Li}},\ and\ \bibinfo {author} {\bibfnamefont {Y.}~\bibnamefont {Wu}},\ }\bibfield  {title} {\bibinfo {title} {Magnetic-field-engineered coherent perfect absorption and transmission},\ }\href {https://doi.org/10.1103/PhysRevA.106.053525} {\bibfield  {journal} {\bibinfo  {journal} {Physical Review A}\ }\textbf {\bibinfo {volume} {106}},\ \bibinfo {pages} {053525} (\bibinfo {year} {2022})}\BibitemShut {NoStop}%
\bibitem [{Note1()}]{Note1}%
  \BibitemOpen
  \bibinfo {note} {The support of a density matrix $\rho $ is the orthogonal complement of the kernel of $\rho $~\cite {robert2005}.}\BibitemShut {Stop}%
\bibitem [{\citenamefont {Marshall}\ \emph {et~al.}(2011)\citenamefont {Marshall}, \citenamefont {Olkin},\ and\ \citenamefont {Arnold}}]{marshall2011}%
  \BibitemOpen
  \bibfield  {author} {\bibinfo {author} {\bibfnamefont {A.~W.}\ \bibnamefont {Marshall}}, \bibinfo {author} {\bibfnamefont {I.}~\bibnamefont {Olkin}},\ and\ \bibinfo {author} {\bibfnamefont {B.~C.}\ \bibnamefont {Arnold}},\ }\href@noop {} {\emph {\bibinfo {title} {Inequalities: Theory of Majorization and Its Applications}}},\ \bibinfo {edition} {2nd}\ ed.\ (\bibinfo  {publisher} {{Springer Science+Business Media, LLC}},\ \bibinfo {address} {{New York}},\ \bibinfo {year} {2011})\BibitemShut {NoStop}%
\bibitem [{\citenamefont {Nielsen}(1999)}]{nielsen1999}%
  \BibitemOpen
  \bibfield  {author} {\bibinfo {author} {\bibfnamefont {M.~A.}\ \bibnamefont {Nielsen}},\ }\bibfield  {title} {\bibinfo {title} {Conditions for a {{Class}} of {{Entanglement Transformations}}},\ }\href {https://doi.org/10.1103/PhysRevLett.83.436} {\bibfield  {journal} {\bibinfo  {journal} {Physical Review Letters}\ }\textbf {\bibinfo {volume} {83}},\ \bibinfo {pages} {436} (\bibinfo {year} {1999})}\BibitemShut {NoStop}%
\bibitem [{\citenamefont {Gour}\ \emph {et~al.}(2015)\citenamefont {Gour}, \citenamefont {M{\"u}ller}, \citenamefont {Narasimhachar}, \citenamefont {Spekkens},\ and\ \citenamefont {Yunger~Halpern}}]{gour2015}%
  \BibitemOpen
  \bibfield  {author} {\bibinfo {author} {\bibfnamefont {G.}~\bibnamefont {Gour}}, \bibinfo {author} {\bibfnamefont {M.~P.}\ \bibnamefont {M{\"u}ller}}, \bibinfo {author} {\bibfnamefont {V.}~\bibnamefont {Narasimhachar}}, \bibinfo {author} {\bibfnamefont {R.~W.}\ \bibnamefont {Spekkens}},\ and\ \bibinfo {author} {\bibfnamefont {N.}~\bibnamefont {Yunger~Halpern}},\ }\bibfield  {title} {\bibinfo {title} {The resource theory of informational nonequilibrium in thermodynamics},\ }\href {https://doi.org/10.1016/j.physrep.2015.04.003} {\bibfield  {journal} {\bibinfo  {journal} {Physics Reports}\ }\bibinfo {series} {The Resource Theory of Informational Nonequilibrium in Thermodynamics},\ \textbf {\bibinfo {volume} {583}},\ \bibinfo {pages} {1} (\bibinfo {year} {2015})}\BibitemShut {NoStop}%
\bibitem [{\citenamefont {Bengtsson}\ and\ \citenamefont {{\.Z}yczkowski}(2017)}]{bengtsson2017}%
  \BibitemOpen
  \bibfield  {author} {\bibinfo {author} {\bibfnamefont {I.}~\bibnamefont {Bengtsson}}\ and\ \bibinfo {author} {\bibfnamefont {K.}~\bibnamefont {{\.Z}yczkowski}},\ }\href {https://doi.org/10.1017/9781139207010} {\emph {\bibinfo {title} {Geometry of {{Quantum States}}: {{An Introduction}} to {{Quantum Entanglement}}}}},\ \bibinfo {edition} {2nd}\ ed.\ (\bibinfo  {publisher} {Cambridge University Press},\ \bibinfo {year} {2017})\BibitemShut {NoStop}%
\bibitem [{\citenamefont {Gour}\ \emph {et~al.}(2018)\citenamefont {Gour}, \citenamefont {Jennings}, \citenamefont {Buscemi}, \citenamefont {Duan},\ and\ \citenamefont {Marvian}}]{gour2018}%
  \BibitemOpen
  \bibfield  {author} {\bibinfo {author} {\bibfnamefont {G.}~\bibnamefont {Gour}}, \bibinfo {author} {\bibfnamefont {D.}~\bibnamefont {Jennings}}, \bibinfo {author} {\bibfnamefont {F.}~\bibnamefont {Buscemi}}, \bibinfo {author} {\bibfnamefont {R.}~\bibnamefont {Duan}},\ and\ \bibinfo {author} {\bibfnamefont {I.}~\bibnamefont {Marvian}},\ }\bibfield  {title} {\bibinfo {title} {Quantum majorization and a complete set of entropic conditions for quantum thermodynamics},\ }\href {https://doi.org/10.1038/s41467-018-06261-7} {\bibfield  {journal} {\bibinfo  {journal} {Nature Communications}\ }\textbf {\bibinfo {volume} {9}},\ \bibinfo {pages} {5352} (\bibinfo {year} {2018})}\BibitemShut {NoStop}%
\bibitem [{\citenamefont {Luis}(2016)}]{luis2016}%
  \BibitemOpen
  \bibfield  {author} {\bibinfo {author} {\bibfnamefont {A.}~\bibnamefont {Luis}},\ }\bibfield  {title} {\bibinfo {title} {Coherence for vectorial waves and majorization},\ }\href {https://doi.org/10.1364/OL.41.005190} {\bibfield  {journal} {\bibinfo  {journal} {Optics Letters}\ }\textbf {\bibinfo {volume} {41}},\ \bibinfo {pages} {5190} (\bibinfo {year} {2016})}\BibitemShut {NoStop}%
\bibitem [{\citenamefont {Beenakker}(1997)}]{beenakker1997b}%
  \BibitemOpen
  \bibfield  {author} {\bibinfo {author} {\bibfnamefont {C.~W.~J.}\ \bibnamefont {Beenakker}},\ }\bibfield  {title} {\bibinfo {title} {Random-matrix theory of quantum transport},\ }\href {https://doi.org/10.1103/revmodphys.69.731} {\bibfield  {journal} {\bibinfo  {journal} {Reviews of Modern Physics}\ }\textbf {\bibinfo {volume} {69}},\ \bibinfo {pages} {731} (\bibinfo {year} {1997})}\BibitemShut {NoStop}%
\bibitem [{\citenamefont {Zhang}(2011)}]{zhang2011}%
  \BibitemOpen
  \bibfield  {author} {\bibinfo {author} {\bibfnamefont {F.}~\bibnamefont {Zhang}},\ }\href@noop {} {\emph {\bibinfo {title} {Matrix Theory: Basic Results and Techniques}}},\ \bibinfo {edition} {2nd}\ ed.,\ Universitext\ (\bibinfo  {publisher} {Springer},\ \bibinfo {address} {New York},\ \bibinfo {year} {2011})\BibitemShut {NoStop}%
\bibitem [{\citenamefont {Robert}(2005)}]{robert2005}%
  \BibitemOpen
  \bibfield  {author} {\bibinfo {author} {\bibfnamefont {A.}~\bibnamefont {Robert}},\ }\href@noop {} {\emph {\bibinfo {title} {Linear Algebra: Examples and Applications}}}\ (\bibinfo  {publisher} {World Scientific},\ \bibinfo {address} {Hackensack, N.J},\ \bibinfo {year} {2005})\BibitemShut {NoStop}%
\end{thebibliography}%

\end{document}